\documentclass[twocolumn]{aastex62}
\usepackage{amsmath}
\usepackage{amssymb}
\usepackage{mathbbol}
\usepackage{dsfont}
\usepackage{color}
\usepackage{graphicx,longtable,times,threeparttable,multirow}

\DeclareGraphicsExtensions{.eps,.eps.gz,.epsi, .jpg, .png}

\received{April 18, 2018}
\accepted{July, 7, 2018 for publication in ApJ Supplement}

\shorttitle{Disk stellar mass distribution and star formation history}
\shortauthors{Xiang et al.}

\begin{document}
\title
{Stellar mass distribution and star formation history of the Galactic disk revealed by mono-age stellar populations from LAMOST}

\email{msxiang@nao.cas.cn, sjr@bao.cas.cn, x.liu@pku.edu.cn}

\author{Maosheng Xiang}\thanks{LAMOST Fellow}
\affil{Key Laboratory of Optical Astronomy, National Astronomical Observatories, \\
      Chinese Academy of Sciences, Beijing 100012, P. R. China}
\author{Jianrong Shi}
\affil{Key Laboratory of Optical Astronomy, National Astronomical Observatories, \\
      Chinese Academy of Sciences, Beijing 100012, P. R. China}
\affil{University of Chinese Academy of Sciences, Beijing 100049, P. R. China}
\author{Xiaowei Liu}
\affil{South-Western Institute for Astronomy Research, Yunnan University, \\
      Kunming 650500, P. R. China}
\affil{Department of Astronomy, Peking University, Beijing 100871, P. R. China}
\author{Haibo Yuan}
\affil{Department of Astronomy, Beijing Normal University, Beijing 100875, P. R. China}
\author{Bingqiu Chen}
\affil{South-Western Institute for Astronomy Research, Yunnan University, \\
      Kunming 650500, P. R. China}
\author{Yang Huang}\thanks{LAMOST Fellow}
\affil{South-Western Institute for Astronomy Research, Yunnan University, \\
      Kunming 650500, P. R. China}
\author{Chun Wang}
\affil{Department of Astronomy, Peking University, Beijing 100871, P. R. China}
\author{Zhijia Tian}
\affil{Department of Astronomy, Yunnan University, \\
      Kunming 650500, P. R. China}
\author{Zhiying Huo}
\affil{Key Laboratory of Optical Astronomy, National Astronomical Observatories, \\
      Chinese Academy of Sciences, Beijing 100012, P. R. China}
\author{Huawei Zhang}
\affil{Department of Astronomy, Peking University, Beijing 100871, P. R. China}
\author{Meng Zhang}
\affil{Department of Astronomy, Peking University, Beijing 100871, P. R. China}


\begin{abstract}{We present a detailed determination and analysis of 3D stellar mass distribution 
of the Galactic disk for mono-age populations using a sample of 0.93 million main-sequence turn-off 
and subgiant stars from the LAMOST Galactic Surveys. Our results show (1) all stellar populations 
younger than 10\,Gyr exhibit strong disk flaring, which is accompanied with a dumpy vertical density 
profile that is best described by a $sech^n$ function with index depending on both radius and age; 
(2) Asymmetries and wave-like oscillations are presented in both the radial and vertical direction, 
with strength varying with stellar populations; 
(3) As a contribution by the Local spiral arm, the mid-plane stellar mass density at solar radius 
but 400--800\,pc (3--6$^\circ$) away from the Sun in the azimuthal direction has a value of 
$0.0594\pm0.0008$\,$M_\odot$/pc$^3$, which is 0.0164\,$M_\odot$/pc$^3$ higher than 
previous estimates at the solar neighborhood. The result causes doubts on the current estimate 
of local dark matter density;
(4) The radial distribution of surface mass density yields a disk scale length evolving from 
$\sim$4\,kpc for the young to $\sim$2\,kpc for the old populations. The overall population 
exhibits a disk scale length of $2.48\pm0.05$\,kpc, and a total stellar mass of $3.6(\pm0.1)\times10^{10}$\,$M_\odot$ 
assuming $R_{\odot}=8.0$\,kpc, and the value becomes $4.1(\pm0.1)\times10^{10}$\,$M_\odot$ 
if $R_{\odot}=8.3$\,kpc; 
(5) The disk has a peak star formation rate ({\rm SFR}) changing from 6--8\,Gyr at the inner to 4--6\,Gyr 
ago at the outer part, indicating an inside-out assemblage history. 
The 0--1\,Gyr population yields a recent disk total {\rm SFR} of $1.96\pm0.12$\,$M_\odot$/yr. }
\end{abstract}

\section{Introduction}

The Milky Way is the only galaxy for which stellar populations can be characterized star by star 
in full dimensionality -- 3D positions, 3D velocities, mass, age and chemical compositions of their 
photospheres. Therefore it serves as a unique laboratory to understand the matter constitute, 
assemblage and chemo-dynamical evolution history of (spiral) disk galaxies in general 
\citep[e.g.][]{Freeman2002, Rix2013, Bland-Hawthorn2016, Minchev2016}. 
An accurate mapping of the stellar mass distribution in the Milky Way disk, and its variation 
among stellar populations of different ages, are of fundamental importance for Galactic astronomy, 
such as to characterize the disk structure, star formation, assemblage and perturbation history. 
It is also crucial for obtaining proper estimates of the dark matter content, especially the local dark 
matter density \citep[e.g.][]{Read2014}, which provides guidance to the numerous ongoing 
dark matter experiments \citep[e.g.][]{Asztalos2010, Xenon1002012, Kang+2013, Cao2014, ChangJ2014, ChangJ2017, LZ2015}. 
However, due to great challenges encountered in observing the numerous stars spreading 
in the whole sky and covering a huge range of magnitudes and stellar parameters (mass, age and metallicity), 
a detailed map of the completed stellar mass distribution of the Milky Way disk, is still not well-established. 

Since the discovery of the thick disk component by \citet{Gilmore1983} via star counting towards 
the Galactic south pole, it becomes a fashion to describe the stellar (number) density distribution of the 
Galactic disk with a combination of two components, a thin disk and a thick disk \citep[e.g.][]{Chen2001, 
Juric+2008, Chang+2011, Chen+2017}. Whereas there are still large scatters in the derived scale parameters 
of both the thin and the thick disk \citep[e.g.][]{Chang+2011, Jia2014, Lopez-Corredoira+2014, Amores2017}. 
Recent (after 1995) literature reports thin disk scale length of 1 -- 4\,kpc and thick disk scale length of 
2 -- 5\,kpc, while the thin disk scale height has reported values of about 150 -- 350\,pc, and the 
thick disk has reported scale heights of about 600 -- 1300\,pc \citep[e.g.][]{Ojha1996, Ojha2001, Robin1996,
Chen2001, Siegel2002, Du2003, Du2006, Larsen2003, Cabrera-Lavers2005, Karaali2007, Juric+2008, 
Yaz2010, Chang+2011, Jia2014, Chen+2017, Wan2017}. It is likely that a large part of those scatters are 
due to different tracers adopted by those work, which cover different regions of the disk with different 
selection functions in stellar ages \citep{Chang+2011, Amores2017}. There are also debates about the 
relative size of scale length between the thin disk and the thick disk, as photometric stellar density distribution 
generates longer scale length for the geometric thick disk, while spectroscopic sample, which usually 
defines the thick disk in abundance and/or age space, yields shorter scale length for the thick disk 
\citep[e.g.][]{Bovy2012, Cheng+2012, Bovy2016, Mackereth2017}. An explanation of this conflict is likely 
linked to both the time evolution and the flaring structure of the disk. It has been shown that the disk 
scale length may have grown-up significantly with time \citep{Mackereth2017, Amores2017}, which 
is consistent with the concept of an inside-out galaxy assemblage history \citep[e.g.][]{Larson1976, Brook2012}.

Beyond the double-component structure, the disk is also found to be warped and flared in its outskirts 
by young tracers, such as H\,{\sc i} \citep[e.g.][]{Henderson1982, Diplas1991, Nakanishi2003, Levine2006} 
and molecular clouds \citep[e.g.][]{Wouterloot1990, May1997, Nakanishi2006, Watson2017}. 
Warps and flares are also presented for the stellar disk \citep{Lopez-Corredoira2002, Momany2006, Reyle2009, 
Hammersley2011, Lopez-Corredoira+2014, Feast2014}. It is generally believed 
that the flaring is a prominent feature for young stellar disk, whereas it is still unclear to what age  
such structures can survive, and how their strengths evolve with time. The disk is also found to hold asymmetric 
structures and remnants, such as the Monoceros ring \citep{Newberg2002, Rocha-Pinto2003}, 
the Sagittarius Stream \citep{Majewski2003}, the Anti-Center Stream \citep{Crane2003, Rocha-Pinto2003} 
and the Triangulum-Andromeda (TriAnd) stream \citep{Rocha-Pinto2004, Majewski2004}. \citet{Xuyan2015} 
found oscillating asymmetries of stellar number density on two sides of the disk plane in the anti-center direction 
out to a large Galactocentric distance ($\gtrsim$21\,kpc), and the oscillating asymmetries are suggested to be 
results of external perturbations. 
Recently, \citet{Bergemann2018} found that stars in the 
TriAnd at 5\,kpc above the disk mid-plane at a Galactocentric distance of 18\,kpc, as well as stars in the 
A13 over-density at 5\,kpc below the disk mid-plane at a Galactocentric distance of 16\,kpc, exhibit the same 
abundance pattern as the disk stars, suggesting that they belong to the disk and are results of the disk perturbations. 
In the vertical direction, it is found that the stellar number density shows a significant North--South asymmetry, 
exhibiting wave-like disk oscillations \citep{Widrow2012, Yanny2013}. Finally, the most prominent asymmetric 
structures of our Galaxy, as has been known for a long time, are the bar \citep[e.g.][]{McWilliam2010, Nataf2010, Shen2010, 
Wegg2013, Bland-Hawthorn2016, Shen2016} and the spiral arms \citep[e.g.][]{Nakanishi2003, Moitinho2006, Xuye2006}. 
The spiral arms are observed by H\,{\sc i}, molecule clouds and H\,{\sc ii} regions \citep{Nakanishi2003, Xuye2006, Vazquez2008, 
Hou2009, Hou2014, Xuye2013, Griv2017}, and also traced by young stellar associations and open clusters 
\citep{Moitinho2006, Vazquez2008, Griv2017}. Whereas it is still unknown to how old age the spiral arms can survive. 

However, most of the disk structure studies are based on stellar number density for some specific types or colors 
 of stars. They are therefore inevitably affected by selection bias. In order to accurately reveal the underlying 
 disk structures and asymmetries, it is extremely important for such studies to use stellar samples with 
 well-defined selection function, and to properly correct for the sample selection function. 
 We stress that an unbiased characterization of disk structure $should$ be based on stellar mass distribution 
that account for contributions from all underlying populations of stars spreading the full mass function, 
from very low mass below the H-burning limit to the high mass end. This is however, an extremely difficult 
task that has never been carried out in a direct way. 

There have been quite many efforts to estimate the underlying stellar mass distribution of the 
Milky Way disk, either locally or globally. Most of those works are carried out with forward modeling 
via either star counting \citep[e.g.][]{Amores2017, Mackereth2017, Bovy2017} or dynamic method 
\citep[e.g.][]{Bahcall1984a, Bahcall1984b, Pham1997, Bienayme1987, Kuijken1989a, Kuijken1989b, 
Kuijken1989c, Kuijken1991, Bovy2013, Zhang2013, Read2014, Huang+2016, Xia2016, McMillan2017}. 
These forward modeling methods rely on quite a few assumptions, such as the distribution profiles 
of stars (and dark matter), disk star formation history ({\rm SFH}) or stellar dynamics, which usually oversimplify 
the problem. In most cases, one needs also to properly account for the selection bias, although it 
was often omitted. On the other hand, there is a model independent way to determine 
the disk stellar mass density, which constructs the full luminosity function of stars in a given volume 
directly from observations, and converts the luminosity function to the stellar mass function utilizing 
stellar mass--luminosity relation to yield the stellar mass density. This direct method is practicable only 
at the solar neighborhood, where one can obtain approximately a full stellar luminosity function by 
combing observations of various telescopes and instruments, e.g., $Hipparcos$ and HST 
\citep[e.g.][]{Holmberg2000, Chabrier2001, Flynn2006, McKee2015}. 
For both the forward modeling and the direct methods, accurate estimates of stellar distance and proper 
considerations of error propagations are necessary. 

The situation is being improved as precise stellar age and metallicity for large samples of stars 
with well-defined target selection function become available \citep[e.g.][]{Xiang+2015c, Xiang+2017c, Martig2016, Ness2016, Ho2017, Mints2017, Wu2018, Sanders2018}. 
For stellar populations of given age and metallicity, the full stellar mass function can be well 
reconstructed from a subset of stars by using the initial mass function and stellar evolution models, 
both can be considered as, to a large extent, been well-established. With age and metallicity, one 
can thus obtain full stellar mass function to a large distance since the initial mass function is 
suggested approximately uniform in the Milky Way disk \citep{Kroupa2001, Kroupa+2013, Chabrier2003, 
Bastian2010}. With this method, the star formation history is no longer assumption  
but becomes derived quantity. With similar idea, \citet{Mackereth2017} have derived 
the disk stellar mass density distribution for mono-age and mono-abundance populations 
using a sample of 31\,244 APOGEE red giant branch stars, which have age estimates from 
their carbon and nitrogen abundance with typical precision of $\sim$0.2dex (46\%). 
However, they still adopted a forward modeling 
method by inducing assumptions on the disk density profile and star formation history. 

In this work, we present an unprecedented 3D determination of disk stellar mass density for mono-age
populations within a few kilo-parsec of the solar-neighbourhood, utilizing a sample of 0.93 million main-sequence 
turn-off and subgiant (MSTO-SG) stars from the Large sky Area Multi-Object Fiber Spectroscopic Telescope 
\citep[LAMOST;][]{Wang1996, Cui+2012}. The sample stars 
have robust age and mass estimates, with about half of the stars having age uncertainties of only 20--30\% 
and mass uncertainty of a few ($<8$\%) per cent. Such high precision of age estimates 
allows us to distinguish different mono-age stellar populations to a feasible extent. Moreover, the sample 
stars have simple and well-defined target selection function, which allow us to reliably reconstruct 
the underlying stellar populations. We construct a map of 3D disk stellar mass density distribution 
for different age populations, and characterize in detail the local stellar mass density, the radial, azimuthal 
and vertical stellar mass distribution, as well as the disk surface stellar mass density at 
different Galactocentric radii. Our results allow a quantitative study of the global and local 
structures and asymmetries of the disk from stellar mass density derived from complete 
stellar populations. The results also lead to 
a direct measure of the disk star formation history at different Galactocentric annuli. 

The paper is organized as follows. Section\,2 briefly introduces the data sample. Section\,3 
introduces our method for stellar mass density determination, including the correction of 
selection function. Section\,4 presents a test of the method on mock dataset to understand 
the effects of main-sequence star contaminations to our sample stars. 
Section\,5 presents the results and discussions. A summary is presented in Section\,6. 

\section{The data sample}

This work is carried out using the LAMOST MSTO-SG star sample of \citet{Xiang+2017c}, which 
contains mass and age estimates for 0.93 million stars selected in the $T_{\rm eff}$ -- ${\rm M}_V$ diagram
out of 4.5 million stars observed by the LAMOST Galactic surveys \citep{Deng+2012, Zhao+2012, LiuX2015} 
before June 2016. The definition criteria to select the MSTO-SG stars of \citet{Xiang+2017c} is 
\begin{equation}  
T_{\rm eff} > T_{\rm eff}^{\rm bRGB} + \Delta{T_{\rm eff}},
{\rm M}_V < {\rm M}_V^{\rm TO} + \Delta{\rm M}_V,
\end{equation}
where $T_{\rm eff}^{\rm bRGB}$ is the effective temperature of the base-RGB, and is determined 
using the Yonsei-Yale (Y$^2$) stellar isochrones \citep{Demarque2004}. ${\rm M}_V^{\rm TO}$ is 
the $V$-band absolute magnitude of the exact main-sequence turn-off point of the isochrones. 
Both $T_{\rm eff}^{\rm bRGB}$ and ${\rm M}_V^{\rm TO}$ are functions of metallicity. 
For details about the adopted values of $T_{\rm eff}^{\rm bRGB}$, $\Delta{T_{\rm eff}}$, 
${\rm M}_V^{\rm TO}$ and $\Delta{\rm M}_V$, we refer to Tables\,1 and 2 of \citet{Xiang+2017c}. 
To select the MSTO-SG sample stars, a minimum spectral signal-to-noise ratio (S/N) cut 
of 20 (per pixel) is adopted, and about half of the sample stars have a S/N higher than 60. 

Stellar mass and age of the sample stars are determined by matching the stellar parameters (effective temperature $T_{\rm eff}$, 
absolute magnitudes ${\rm M}_V$, metallicity [Fe/H], $\alpha$-element to iron abundance ratio [$\alpha$/Fe]) 
with the Y$^2$ isochrones using a Bayesian method. For details about the age and mass estimation, 
we recommend readers to the comprehensive paper of \citet{Xiang+2017c}. \citet{Xiang+2017c} 
also carried out a variety of tests and examinations to validate the mass and age estimation. 
These tests and examinations include a detailed analysis of results after applying their age (and mass) 
estimation method to mock datasets, comparison of stellar age (and mass) estimates with asteroseismic age  
as well as with age based on the Gaia TGAS parallax, robustness examinations of age estimates with duplicate 
observations, and validations with member stars of open clusters. 
These tests and examinations validate that not only the age (and mass) estimates but also 
their error estimates are reliable. About half of the sample stars have age errors 
of about 20--30\%, while the other half have larger errors. The mass estimates have a medium 
error of 8\%. The amount of errors are largely determined by the spectral S/N (thus the parameter errors). 
Fig.\,1 shows the age distribution of the MSTO-SG sample stars in the Galactic coordinate 
centered on the Galactic anti-center. 
\begin{figure}
\centering
\includegraphics[width=90mm]{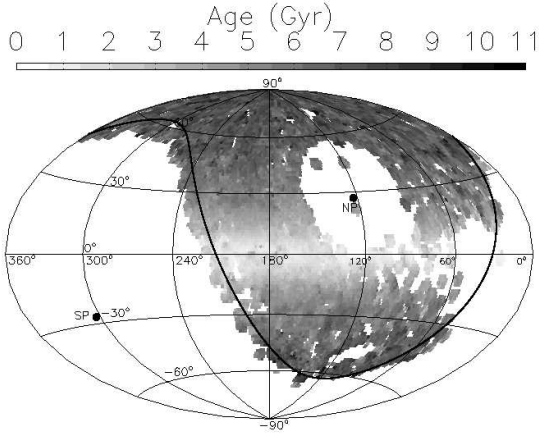}
\caption{Color-coded distribution of medium stellar ages in Galactic coordinates ($l$, $b$). The image center 
shows the Galactic anti-center ($l=180^\circ$, $b=0^\circ$). Solid lines show the Galactic longitudes, latitudes 
as well as the celestial equator. 
Throughout this paper, the Hammer-Aitoff projection is adopted.}
\label{Fig1}
\end{figure}

Stellar parameters, including $T_{\rm eff}$, ${\rm M}_V$, log\,$g$, [Fe/H] and [$\alpha$/Fe], 
for the MSTO-SG sample stars and for the whole 4.5 million LAMOST stars 
are determined with the LAMOST Stellar Parameter Pipeline at Peking University \citep[LSP3;][]{Xiang+2015a, LiJ+2016, Xiang+2017a}, 
using the same version as adopted for the LSS-GAC DR2 \citep{Xiang+2017b}, the second data release of value-added catalogues 
for the LAMOST Spectroscopic Survey of the Galactic Anti-center \citep[LSS-GAC;][]{LiuX2014}.  
Note that the ${\rm M}_V$ is derived directly from the spectra with a multivariate regression 
method based on kernel-based principal component analysis (KPCA), utilizing the LAMOST and $Hipparcos$ \citep{Perryman1997} 
common stars as training dataset \citep{Xiang+2017a, Xiang+2017b}. With the absolute magnitudes, 
stellar distance is deduced from the distance modulus, utilizing interstellar extinction 
derived with the `star pair' method \citep{Yuan2013, Yuan+2015, Xiang+2017b}. 
A comparison with distance inferred from the Gaia TGAS parallax \citep{Brown2016} indicates that 
our distance estimates reach a precision of 12\% given the relatively high spectral S/N of the 
LAMOST-TGAS common stars, and the systematic error is negligible. 
A comparison with distance inferred from the Gaia DR2 parallax gives comparable results (see Section 4).
The overall MSTO-SG sample stars have a median distance error of 17\%. 

\section{method}

\subsection{Methodology overview }

The method aims to derive a three dimensional distribution of stellar mass density of the Galactic 
disk for mono-age stellar populations by counting the MSTO-SG stars. 
Here by using the MSTO-SG stars as tracers, we intend to derive the stellar mass density of the whole 
populations, i.e., populations across of the whole stellar mass function, from the low- to the 
high-mass end. This is not a straightforward task as it appears, and can only be carried out with mono-age 
and mono-metallicity populations if we do not impose strong assumptions on star formation 
history and stellar migrations.

In principle, the stellar mass function in a given volume of the Galactic disk is a combined result
of in-situ star formation, stellar evolution and stellar migration.
Mathematically, we can describe the number distribution of in-situ stars 
at any given position (with limited volume) of the disk $P(l, b, d)$ as a function of mass $M$, 
metallicity $Z$ and age $\tau$ by
 \begin{equation}
 \begin{aligned}
 N(M, Z, \tau)  = N_{\rm in}(M, Z, \tau) + N_{\rm k}(M, Z, \tau),
  \end{aligned}
 \end{equation}
 where $N_{\rm in}$ represents stars formed in-situ, $N_{\rm k}$ represents stars 
 migrated to their current position due to kinematic process. 
 For stars formed in-situ, the stellar mass distribution is described by  
 \begin{equation}
 \begin{aligned}
 N_{\rm in}(M, Z, \tau) = & \psi(\tau)\phi(Z\mid\tau)\xi(M_{\rm ini}\mid Z,\tau)F(M\mid M_{\rm ini},Z,\tau),  
 \end{aligned}
 \end{equation}
 where $\psi$, $\phi$ and $\xi$ are respectively
 the star formation history, the chemical enrichment history and the stellar initial mass function ({\rm IMF}).
 The $F$ converts the initial mass ($M_{\rm ini}$) of stars with given age and metallicity to the present stellar mass 
 ($M$) by considering mass loss due to stellar evolution. 
 For stars migrated to the current position, their mass distribution depends also on the details 
 of the migration process, which might be a function of mass, age and metallicity. So that, 
 \begin{equation}
 \begin{aligned}
 N_{\rm k}(M, Z, \tau) = & \int\int\int\psi^\prime(\tau)\phi^\prime(Z\mid\tau)\xi^\prime(M_{\rm ini}\mid Z,\tau)\\
         & F(M\mid M_{\rm ini},Z,\tau)K_{P^\prime\rightarrow P}(M_{\rm ini},Z,\tau){\rm d}l^\prime b^\prime d^\prime, 
 \end{aligned}
 \end{equation}
where $\psi^\prime$, $\phi^\prime$ and $\xi^\prime$ are respectively the star formation history, 
the chemical enrichment history and the stellar initial mass function at any given position 
$P^\prime(l^\prime, b^\prime, d^\prime)$, and $K_{P^\prime\rightarrow P}(M_{\rm ini},Z,\tau)$ 
is a function to describe the probability that stars migrated from $P^\prime(l^\prime, b^\prime, d^\prime)$ 
to $P(l, b, d)$. 
Note that here we have ignored the change of stellar metallicity due to stellar 
evolution, i.e., we assume that metallicity $Z$ is the same as the initial stellar metallicity. 
 In reality, the Milky Way may have experienced a complex assemblage history. As a consequence, 
 the star formation history, the chemical history as well as the migration term must vary with
 positions across the disk with probably complex form.

Nevertheless, it is possible to make two reasonable assumptions to simplify the issue.
One is that the {\rm IMF} is universal across the Galactic disk, and it is not sensitively
depending on $\tau$ and $Z$ in the disk volume concerned by this work.
Note that a universal IMF of the Galactic disk has been supported by previous studies 
\citep[e.g.][]{Kroupa2001, Kroupa+2013, Chabrier2003, Bastian2010}.
Another assumption is that stellar migration due to kinematic process does not
prefer special stellar mass, so that the kinematic
term in Equation\,4 is a constant function of stellar mass. 
With these assumptions, for stellar population of given $\tau$ and $Z$ in a given
volume, the number (and of course mass) distributions of both the in-situ formed 
and the migrated stars are determined by only the {\rm IMF} and the stellar evolution process, 
both of which are universal. 
For mono-age and mono-metallicity populations, there is no need to impose assumptions 
on the star formation history, the chemical enrichment history 
and the kinematical/dynamical history to derive the full stellar mass function from the 
MSTO-SG sample stars.

To derive the stellar mass density, we group the MSTO-SG sample stars into 
$3^\circ\times3^\circ$ line of sights in ($l$, $b$) space (see \S{3.2}). In each line of sight, the stars
are divided into distance bins with a constant bin width of $0.2$ in logarithmic scale, and with a lower
and upper limiting bin size of 100\,pc and 1000\,pc, respectively. In each distance bin,
the stellar mass density is calculated by
\begin{equation}
\rho = \frac{\sum\limits_{i=1}^n M_i \centerdot W_{\rm CMD}^i \centerdot W_D^i  \centerdot W_{\rm IMF}^i}{4\pi \centerdot V}, \\
\end{equation}
\begin{equation}
V = \frac{A}{3}(D_2^3 - D_1^3),   
\end{equation}
where $M_i$ is the mass of the $i_{\rm th}$ MSTO-SG star, $n$ the number of MSTO-SG stars 
in the distance bin of concern. $W_{\rm CMD}$ is the weight assigned to each MSTO-SG star to account 
for selection function of the survey in the color-magnitude diagram (CMD; \S{3.2}), $W_D$ the weight 
to account for volume completeness, which is defined in \S{3.3}, and $W_{\rm IMF}$ the weight 
to convert the mass density of MSTO-SG stars to mass density of stellar populations of all masses (\S{3.4}).  
To compute the volume ($V$) of the distance bin, $A$ is the sky area of the line of sights, $D_1$ and $D_2$ 
are respectively the lower and upper boundary of the distance bin. Note that the lower 
boundary of the first distance bin and the upper boundary of the last distance bin in each 
line of sight are jointly determined by the limiting apparent magnitude of the survey and the limiting 
absolute magnitude of the MSTO-SG stars (\S{3.3}), and are independent of our binning strategy.

To obtain an error estimate of the derived stellar mass density, we adopt a Monte-Carlo 
approach. Specifically, we calculate the density for many times, in each time we retrieve a new set 
of values for distance, age, mass and $W_{\rm CMD}$ for all the MSTO-SG sample stars from a 
Gaussian function characterized by the measured values and their errors, and repeat the process 
to derive the stellar mass density. The standard deviation of the measured densities in each distance 
bin is adopted as an error estimate of the derived stellar mass density. The latter is adopted as 
that derived with the original (measured) set of parameters.   
Considering the time cost, the number of realizations is adopted to be 21. 
To increase the sampling density, we have also opted to double the number of bins 
in each line of sight by shifting the bins by half of the bin width. 

\subsection{Selection function in the CMD} 
Several selection processes have been incorporated subsequently to generate the MSTO-SG star sample: 
1) the photometric catalogs which afford input stars for LAMOST surveys are magnitude limited ones. 
Stars brighter or fainter than the limit magnitudes are not observed; 2) Only a part of stars in the photometric catalogs 
are targeted by LAMOST via target selection in the CMD; 
3) As mentioned in Section\,2, not all stars targeted by LAMOST got spectra with 
enough S/N and stellar parameters successfully; 
4) A few criteria have been used to select the MSTO-SG sample from stars targeted by LAMOST 
and have stellar parameter determinations. 
All these processes cause incompleteness of the MSTO-SG stars in the CMD. 

\begin{figure}
\includegraphics[width=90mm]{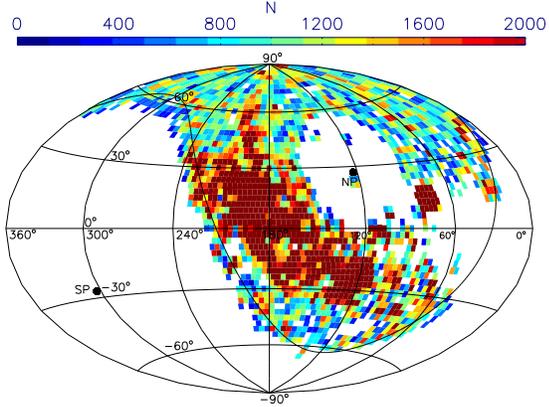}
\caption{Color-coded distribution of number of stars that have spectroscopic stellar parameters  
and with spectral ${\rm S/N}>20$ in subfield of $3^\circ\times3^\circ$ in $(l,b)$ plane.}
\centering
\label{Fig2}
\end{figure}
The LAMOST Galactic surveys select input targets from the photometric catalogs
uniformly and/or randomly in the CMDs \citep{Carlin+2012, LiuX2014, Yuan+2015}. 
Such simple yet non-trivial target selection strategies allow us to reconstruct 
the photometric catalog from a selected spectroscopic sample \citep{Chen2018}.
\citet{Chen2018} present a detailed example to demonstrate how to correct for 
the LSS-GAC selection function in the CMD rigorously. 
The method is also appropriate for the whole LAMOST Galactic spectroscopic surveys.
In most cases, there are more than one LAMOST plate observed for a given field on the sky. 
These plates are usually observed under different weather conditions. 
\citet{Chen2018} thus derive the selection function plate by plate.
In addition, for each plate, there are 16 spectrographs with different instrument performance
thus different selection functions. \citet{Chen2018} thus derive the selection function for 
different subfields with sky area similar to that of a spectrograph. 

For the specific purpose of star counts with MSTO-SG stars, here we adopt similar
but slightly different strategy with respect to the method of \citet{Chen2018}. 
We consider the selection functions at different pencil beams (or line of sights) on the sky 
defined with $3^\circ\times3^\circ$ in ($l$, $b$) plane, and we combine spectroscopic stars 
observed by all plates in each line of sight. 
In order to distinguish with the LAMOST field, below we will use the `subfield' to 
describe each line of sight. 
In addition, we adopt larger bin size when dividing the stars into bins in the CMD. 
All these efforts are intend to reduce the fluctuation of selection function on the CMD 
by encompassing more stars in each CMD cell. These adjustments, in the majority cases, 
improve the selection function small but important for the purpose of star counts 
for mono-age populations.

In total, there are 2144 subfields that each contains more than 20 unique stars observed
by LAMOST and have a spectral S/N higher than 20, the SNR cut adopted for
the MSTO-SG star sample. After a careful subfield by subfield inspect on the spatial 
coverage and CMD for both the spectroscopic and the photometric stars, we exclude 
307 subfields for which either the photometric catalog is incomplete or the spectroscopic 
stars (observed by LAMOST with a S/N higher than 20) have poor coverage on the CMD. 
For the remaining 1837 subfields adopted by this work, the median 
number of spectroscopic stars per subfield is 1149, and the minimum number is 123.  
Fig.\,2 plots a color-coded distribution of the number of spectroscopic stars in each subfield 
in the $(l, b)$ plane.
Note that a small number of stars in a given subfield does not necessarily mean a poor 
sampling rate. This is because the LAMOST surveys categorize stars into very bright ($9\lesssim r \lesssim14$\,mag),
bright ($14<r\lesssim16.3$\,mag), medium bright ($16.3\lesssim r<17.8$\,mag) and
faint ($17.8\lesssim r<18.5$\,mag) plates according to the apparent magnitudes to optimize the
survey strategy \citep{Deng+2012, LiuX2014, Yuan+2015}, and not all of the LAMOST fields 
having all these categories of plates observed. The number of stars in each subfield 
is thus largely determined by the survey depth, and also 
depends on the Galactic latitude, as fields at low Galactic latitudes having  
more stars than those at high Galactic latitudes. 

\begin{figure}
\centering
\includegraphics[width=85mm]{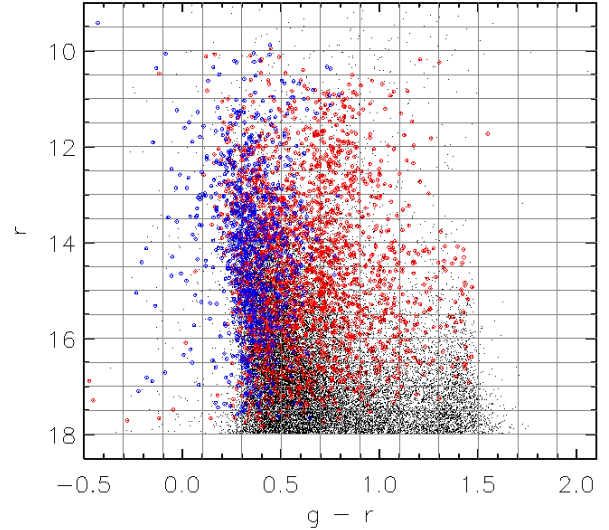}
\caption{Color-magnitude diagram for one subfield centered on $l=180^\circ$ and $b=21^\circ$.
Black dots are photometric stars that combining the XSTPS-GAC and APASS catalogs,
blue and red squares are respectively MSTO-SG stars and other types of stars
that have LAMOST stellar parameters and with spectral ${\rm S/N}>20$.}
\label{Fig3}
\end{figure}
\begin{figure}
\centering
\includegraphics[width=90mm]{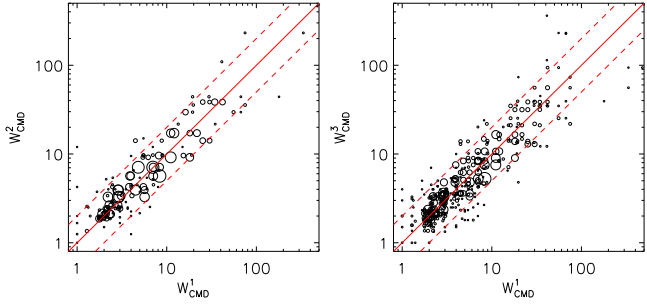}
\caption{Comparison of CMD weights derived with different binning configurations 
for stars in subfield ($l=180^\circ$, $b=21^\circ$).
In the left panel, both sets of CMD weights are derived with a bin size of $0.2\times0.5$\,mag 
in the $(g-r, r)$ diagram but the grids are offset by half bin size. In the right panel, 
$W_{\rm CMD}^1$ is the same as those in the left panel, while $W_{\rm CMD}^3$ is derived with 
a bin size of $0.3\times1.0$\,mag in the $(g-r, r)$ diagram. Size of the symbols indicate the 
number of stars that have the same values of CMD weights (based on the definition, 
all stars in a same CMD bin have the same value of CMD weight). The solid line  
indicates the 1:1 line, while the dashed lines indicate 1:2 and 2:1 lines.} 
\label{Fig4}
\end{figure}
For each subfield, we correct for the selection function in the ($g-r$, $r$)
diagram of the photometric catalog by assigning weights ($W_{\rm CMD}$) to individual stars.
The weight is defined as 
\begin{equation}
W_{\rm CMD}=N_{\rm {ph, CMD}} / N_{\rm {sp, CMD}},
\end{equation}
 where $N_{\rm {ph, CMD}}$ is the number of stars from the photometric catalog in a given CMD cell, 
while $N_{\rm {sp, CMD}}$ is the number of stars in that CMD cell but also have stellar parameters 
from LAMOST spectra with ${\rm S/N}>20$. For convenience, here we give the inverse 
of this CMD weight (i.e., $N_{\rm {sp, CMD}}$/$N_{\rm {ph, CMD}}$) a name as `sampling rate', 
as it means the fraction of photometric stars that are successfully observed by the spectroscopic survey.
Results are deduced using two sets of CMD cells with different sizes,
namely, 0.2$\times$0.5\,mag and 0.3$\times$1.0\,mag. For each set of cell size, 
two sets of weights are derived by offsetting the cells by half length of the cell size.
The final CMD weight is adopted as the average of the four sets of values,
and the standard deviation is adopted as an error estimate of the mean weight.
As an example, Fig.\,3 plots the CMD for one subfield ($l=180^\circ$, $b=21^{\circ}$).
The figure shows that the LAMOST stars have a good coverage on the CMD, which is necessary
to properly recover the photometric sample. Fig.\,4 plots the comparison of weights 
derived with different binning configurations. It shows considerable scatters of CMD weights 
among different binning configurations for a given star. The median value of 
relative errors of the CMD weights for all stars in this line of sight is 18 per cent. 
Note that the LAMOST target selection is in fact based on both ($g-r$, $r$) and ($r-i$, $r$)
diagrams \citep{Carlin+2012, LiuX2014, Yuan+2015}, while here we have used only
the ($g-r$, $r$) diagram to derive the selection function. Such a simplification is
not expected to induce significant bias given that stellar locus in the $(g-r)$ versus $(r-i)$
diagram is quite tight \citep[e.g.][]{Covey+2007, Yuan+2015b},
and that we have adopted a large cell size in the CMD.
Note also that although the selection of VB targets is not carried out in the ($g-r$, $r$)
diagram but according to only the magnitudes of the targets \citep{Yuan+2015},
our approach is expected to be still valid as it does not drop information.

\begin{figure}
\centering
\includegraphics[width=85mm]{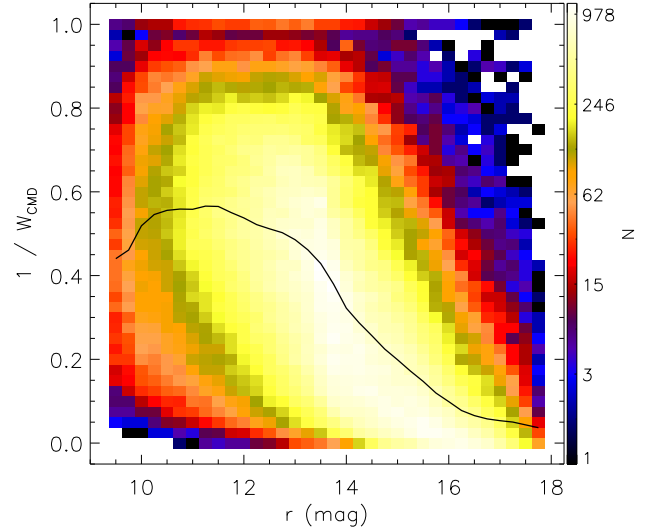} 
\caption{Color-coded distribution of star numbers in the plane of the $r$-band magnitude
versus the sampling rate, i.e. the inverse of the derived CMD weight. The black curve delineates the
median value of sampling rates of individual stars as a function of $r$ magnitude.}
\label{Fig5}
\end{figure}
Fig.\,5 plots the number density of stars in plane of the $r$-band magnitude and the inverse of
the derived CMD weight (i.e. the sampling rate) for all the MSTO-SG sample stars.
As expected, the sampling rate is shown to decrease with increasing magnitude, because the number
of faint stars in the photometric catalog increases steeply with magnitude while the LAMOST
targets have a much flatter distribution as a function of magnitude.
Nevertheless, more than 86\% of the stars have a sampling rate larger than
0.1. The values are even higher for very bright ($r<14$\,mag) stars, as the sampling 
rate computed for individual stars yields a median value of $\sim$0.5, indicating that half of the
very bright stars in the sky area of concern have been successfully observed by the 
LAMOST surveys.

The $g$ and $r$-band photometry are from a combination of different
surveys, namely the Xuyi Schmidt Telescope Photometric Survey of the Galactic Anti-center
\citep[XSTPS-GAC;][]{Zhang+2014, LiuX2014}, the Sloan Digital Sky Survey
\citep[SDSS;][]{York+2000, Ahn+2012}, and the AAVSO Photometric All-Sky
Survey \citep[APASS;][]{Munari+2014}.
The complete magnitude range in $r$-band is $\sim$13--19\,mag for the XSTPS-GAC,
$\sim$14--22\,mag for the SDSS, and $\sim$9-14.5\,mag for the APASS photometric catalog.
A combination of them therefore provides a complete photometric catalog from 9 to 19\,mag,
which covers well the magnitude range of the LAMOST Galactic surveys.
As mentioned above, we have inspected the CMD of all the original 2144 subfields by eye, and excluded 
307 of them. In addition, for some subfields that only the APASS
catalog is available, we set an upper magnitude limit of 14.5\,mag in $r$-band by excluding
fainter stars.

\subsection{Determination of complete volume}
\begin{figure}
\centering
\includegraphics[width=90mm]{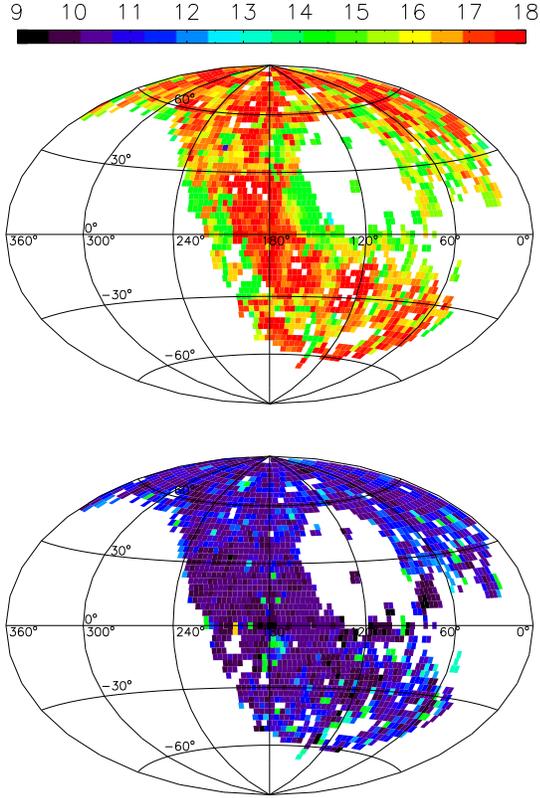}
\caption{Faint (upper) and bright (lower) limiting apparent magnitudes in $r$ band for individual subfields.}
\label{Fig6}
\end{figure}
Applying the CMD weight to individual MSTO-SG sample stars leads to a 
complete sample in magnitude rather than volume. Moreover, the limiting magnitudes 
vary from one subfield to another due to different observation progress. 
We define the bright and faint limiting magnitudes subfield by subfield via inspecting the 
CMD.  
Fig.\,6 shows the limiting magnitudes at both the bright and the faint ends for the individual
subfields. For most of the subfields, the bright limiting magnitudes are $\sim$10\,mag,
while some subfields have a bright limiting magnitude fainter than 14\,mag as there
are no very bright plates observed. At the faint end, more than one third of the subfields 
have a limiting magnitude fainter than 17\,mag, and about 15\% of the subfields 
have a limiting magnitude brighter than 14\,mag as only very bright plates are observed.

\begin{figure}
\centering
\includegraphics[width=80mm]{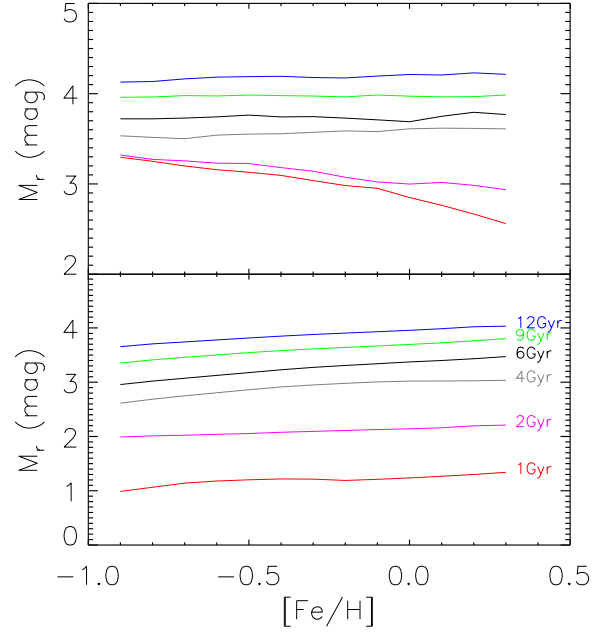}
\caption{Faint (upper) and bright (lower) limiting absolute magnitudes of the MSTO-SG stars
as a function of [Fe/H] for different ages. Values of the absolute magnitudes are directly 
from the definition of MSTO-SG stars in the $T_{\rm eff}$--${\rm M}_V$ diagram.}
\label{Fig7}
\end{figure}
According to our definition criteria to select the MSTO-SG sample stars, 
the absolute magnitudes of the MSTO-SG stars span a wide range of values depending on mass, age and metallicity. 
We use the following equation to define a complete volume,
\begin{equation}
 m_r^{\rm B}-{\rm min}\{M_r^{\rm B}\}-A_r^{\rm B} < 5\log D-5 < m_r^{\rm F}-{\rm max}\{M_r^{\rm F}\}-A_r^{\rm F},
\end{equation}
where $D$ is the distance of the star, $m_r^{\rm B}$ and $m_r^{\rm F}$ are respectively the bright
and the faint limiting apparent magnitudes, ${\rm min}\{M_r^{\rm B}\}$ and ${\rm max}\{M_r^{\rm F}\}$
are respectively the minimal bright and the maximal faint limiting absolute
magnitude for stars of all populations (age and metallicity) of concern, $A_r^{\rm B}$ and $A_r^{\rm F}$
are the $r$-band interstellar extinction at respectively the near and the farther side of the complete
distance, and they are determined iteratively using the LAMOST stars whose E(B-V) are determined with
the `star-pair' method with typical uncertainty of $\sim$0.04\,mag \citep{Yuan+2015, Xiang+2017b}.
The selection function in distance defining the complete volume thus can be written as, 
\begin{eqnarray}
W_D = 
       \left\{
          \begin{array}{ll} 
             1,  {\rm if} ~D_{\rm min} < D < D_{\rm max}, \\
             0,  {\rm if} ~D < D_{\rm min}  ~ {\rm or} ~ D > D_{\rm max}, 
          \end{array}
           \right.
\end{eqnarray}
where
\begin{equation}
  D_{\rm min}= 10^{(m_r^{\rm B}-{\rm min}\{M_r^{\rm B}\}-A_r^{\rm B} + 5)/5}, 
\end{equation}
\begin{equation} 
  D_{\rm max} = 10^{(m_r^{\rm F}-{\rm max}\{M_r^{\rm F}\}-A_r^{\rm F} + 5)/5}.
\end{equation}

It is clear that the complete volume (distance) for each subfield varies with stellar populations of
different age and metallicity as they have different absolute magnitudes.
Fig.\,7 plots the bright and the faint limiting absolute magnitudes of the MSTO-SG stars
as a function of [Fe/H] for different ages. Note that those values are directly from 
the definition criteria based on the isochrones, and are independent of the absolute 
magnitude estimates of the sample stars. The figure shows that from 1 to 12\,Gyr, the limiting
absolute magnitudes vary more than 1 and 3\,mag respectively at the fainter and the brighter end.
For a given age, the limiting absolute magnitudes depend marginally on the metallicity except
for the very young ($<2$\,Gyr) stars, which exhibit a variation of $\sim$0.5\,mag
from a [Fe/H] value of $-1.0$ to 0.3\,dex.
In each subfield, we define a complete volume for each population of age 0--1,
1--2, 2--4, 4--6, 6--8, 8--10, 10--14 and 1--14\,Gyr based on Equation\,8. For each population,
only stars within the complete volume are used for star counts, while stars outside the complete
volume are discarded from the sample. 
Here we define the 1--14\,Gyr rather than the whole stellar 
population of 0--14\,Gyr because the latter has a significantly larger dynamic 
range of absolute magnitude at the brighter end, thus much smaller complete volume. 

\subsection{The IMF weight}
\begin{figure}
\centering
\includegraphics[width=85mm]{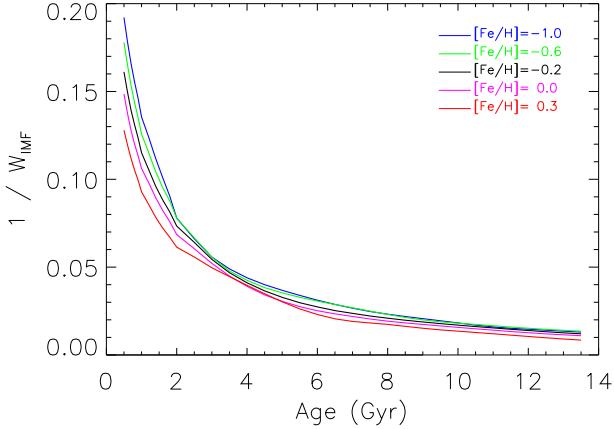}
\caption{Mass fraction of the MSTO-SG stars respect to the whole stellar
population of all masses, i.e., the inverse of IMF weight, as a function of age for different
metallicities.}
\label{Fig8}
\end{figure}
For each age and each metallicity, mass of the MSTO-SG stars is converted to that of the
whole stellar population of all masses with $W_{\rm IMF}$ derived utilizing the {\rm IMF} of \citet{Kroupa2001} and
the Y$^2$ isochrones, the isochrones used to define the trajectories 
of the MSTO stars \citep{Xiang+2017c}. For a mono-age and mono-metallicity population, 
the $W_{\rm IMF}$ is defined as 
\begin{equation}
W_{\rm IMF} = \frac{\int_{M_1}^{M_2}\zeta(M){\rm d}M} {\int_{0.08M_\odot}^{110M_\odot}\zeta(M){\rm d}M}, 
\end{equation}
where 
\begin{equation}
\zeta(M) = \xi(M)F(M\mid M_{\rm ini})
\end{equation}
 is the joint product of the initial stellar mass function and 
the function account for stellar evolution. The $M_1$ and $M_2$ are respectively the lower and upper boundary of 
the MSTO-SG stars, and are determined by the sample selection criteria. 
Here the total stellar mass for the whole population is calculated by 
imposing a lower mass cut of 0.08\,$M_{\odot}$ and a higher mass cut of 110\,$M_{\odot}$. 

To account for mass loss due to stellar evolution, stars with initial mass more massive than the Tip-RGB and
smaller than 10\,$M_{\odot}$ are assumed to have had
become white dwarfs (WD), which have a fixed mass of 0.6\,$M_{\odot}$ \citep[e.g.][]{Rebassa-Mansergas2015}.
Since some stars more massive than the Tip-RGB must have become HB or AGB stars, which are
probably more massive than WDs, the current treatment thus may have slightly underestimated the mass
of the whole stellar population.
Stars with initial mass of 10--29\,$M_{\odot}$ are assumed to have had become neutron stars (NS), which have a fixed mass
of 2.0\,$M_{\odot}$, while stars with initial mass more massive than 29\,$M_{\odot}$ are assumed to have had become
black holes (BH), which have a fixed mass of 10\,$M_{\odot}$.
The mass fraction of NS and BH respect to the whole stellar population is found to be $\sim$5 per cent.
Fig.\,8 plots the inverse of $W_{\rm IMF}$, i.e., mass ratio of the MSTO-SG stars to the whole stellar population,
as a function of age for different metallicities.
The figure shows that the mass ratio decreases from 10--20\% for young ($\lesssim1$\,Gyr) stars
to 1--2\% for old ($\gtrsim8$\,Gyr) stars.
The log-normal {\rm IMF} of \citet{Chabrier2003} is found to yield a stellar mass density lower
than that of the Kroupa {\rm IMF} by $\sim$10\%. While the \citet{Salpeter1955} {\rm IMF} is found to yield a stellar mass
density higher than that of the Kroupa {\rm IMF} by about 75\% as it predicts much more low mass stars.
Note that since the {\rm IMF} weight is derived for stellar mass range of 0.08--110\,$M_{\odot}$, we 
thus have not account for contributions of substellar objects (e.g. brown dwarfs) to the total mass.
Brown dwarfs were suggested to contribute a local density of 0.0015--0.004\,$M_{\odot}$/pc$^3$ \citep{Chabrier2003, Flynn2006, McKee2015}
and a surface mass density of about 1--2\,M$_{\odot}$/pc$^2$ \citep{Flynn2006, McKee2015} at the solar neighbourhood.
In addition, there could be also more low mass ($<0.5$\,$M_{\odot}$) stars than prediction of the
Kroupa {\rm IMF} due to possibly undetected binaries in the sample used to derive the IMF \citep{Kroupa+2013}.
This may also lead to an underestimation of the current mass density.


\section{Contaminations of main-sequence stars}
\begin{figure*}
\centering
\includegraphics[width=160mm]{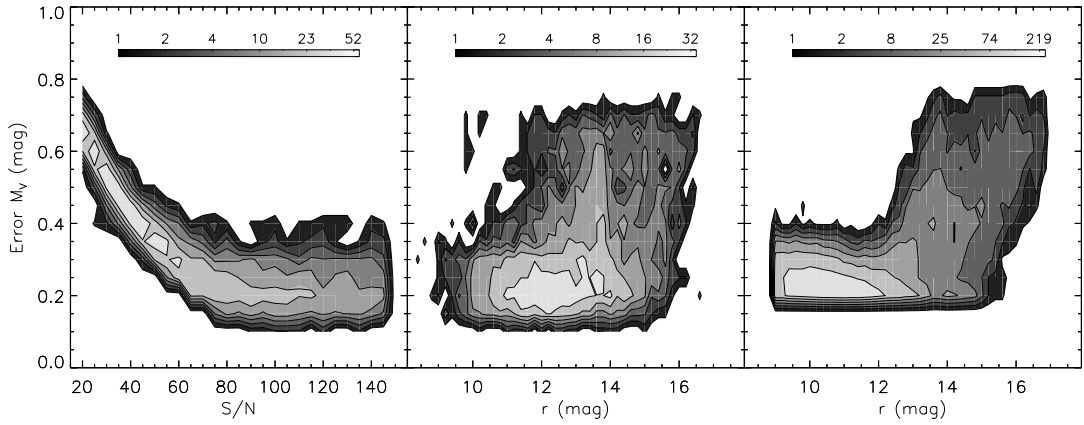}
\caption{${\rm M}_V$ errors as a function of spectral S/N ({\em left}) and $r$-band magnitudes
({\em middle}) for the MSTO-SG sample stars. Colors indicate the number density of stars. 
The {\em right} panel shows the simulated ${\rm M}_V$ errors
as a function of $r$-band magnitude. An error limit of 0.2\,mag is set at the lower end.}
\label{Fig9}
\end{figure*}

\begin{figure*}
\centering
\includegraphics[width=160mm]{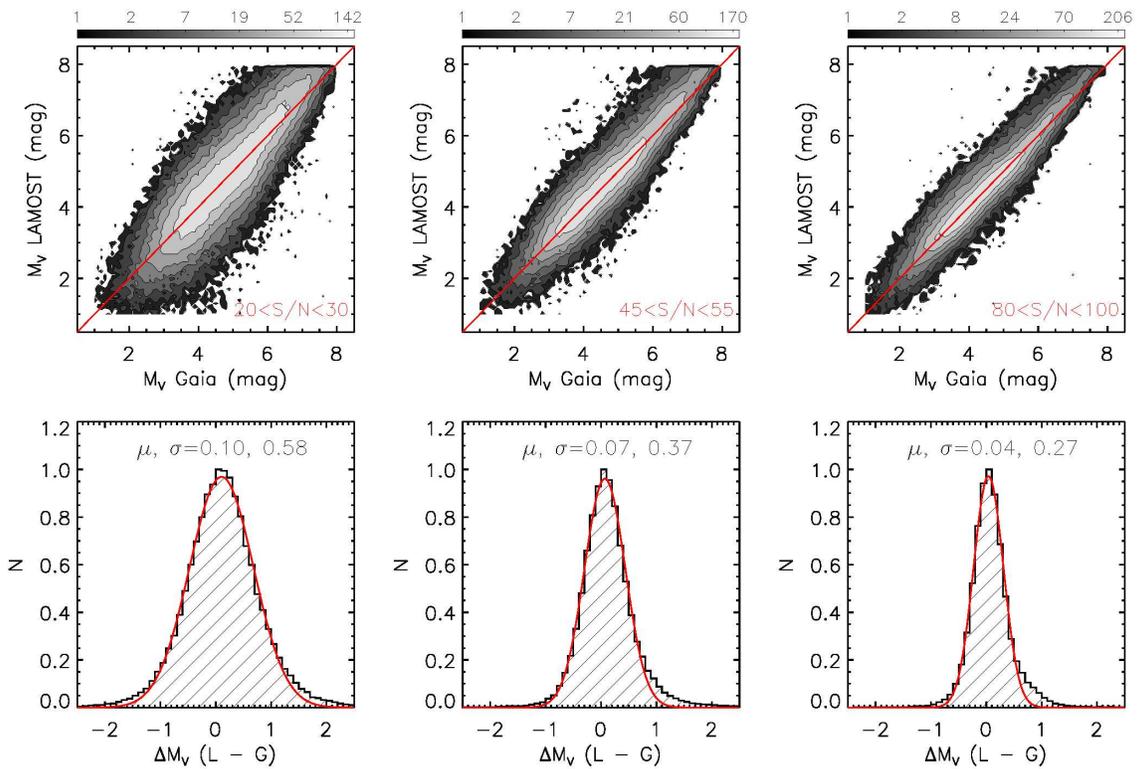}
\caption{The upper row shows comparison of LAMOST ${\rm M}_V$ with values inferred from Gaia DR2 parallax.
Colors indicate the number density of stars. From left to right panels are results for different spectral S/N bins, 
as marked in the plots.
The bottom row plots histograms of the differences, as well as Gaussian fits to the histograms.
The mean and 1$\sigma$ value of the differences, are marked in the figure.
All stars are required to have a ${\rm M}_V$ error in the Gaia DR2 values smaller than 0.1\,mag.}
\label{Fig10}
\end{figure*}

\begin{figure*}
\centering
\includegraphics[width=160mm]{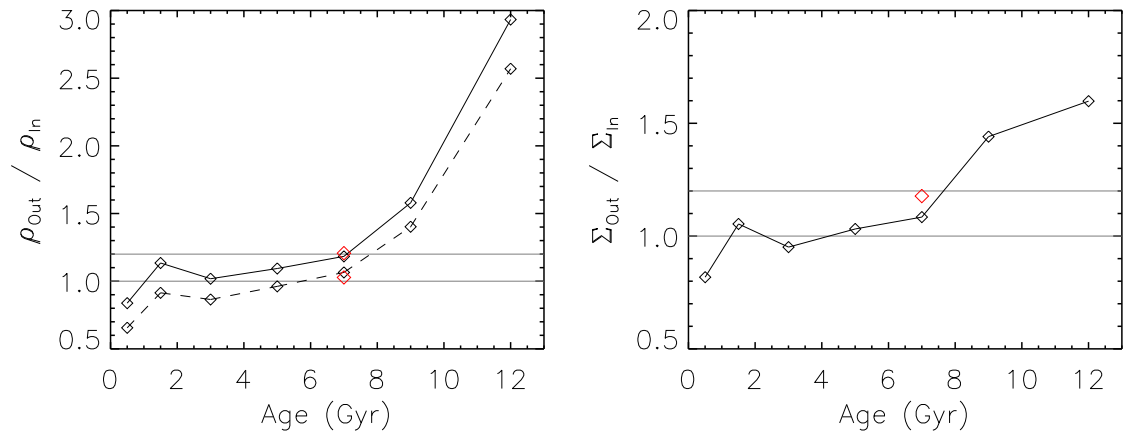}
\caption{{\em Left}: Ratios between the derived mean density within $|Z|<50$\,pc from the mock data 
and the input mean stellar mass density within $|Z|<50$\,pc (solid line) or the input mid-plane density (dashed line). 
{\em Right:} Ratios between the derived and the model input 
surface mass density. For both the left and the right panels, the red symbol shows the result of 
the whole stellar populations of 0--14\,Gyr. The horizontal lines delineate the constant values of 1.0 and 1.2. }
\label{Fig11}
\end{figure*}
\begin{table}
\caption{Parameters for mock disk populations.}
\label{}
\begin{tabular}{ccccc}
\hline
 Age range (Gyr)  &  [Fe/H]    &  [$\alpha$/Fe]    &    $\rho_0$ ($M_{\odot}$/pc$^3$)  & $H_Z$ (kpc)  \\
\hline                                                                                                                                                
 0 -- 1 &  0.1 & 0.0 & 0.0120  & 0.08 \\
 1 -- 2 &  0.1 & 0.0 & 0.0120  & 0.11 \\
 2 -- 3 &  0.0 & 0.0 & 0.0100  & 0.14 \\
 3 -- 4  &  0.0 & 0.0 & 0.0100  & 0.17 \\
 4 -- 5 &  $-0.1$ & 0.0 & 0.0077  & 0.19 \\
 5 -- 6 &  $-0.1$ & 0.0 & 0.0077  & 0.21 \\
 6 -- 7 & $-0.2$ & 0.1 & 0.0056  & 0.26 \\
 7 -- 8 & $-0.2$ & 0.1 & 0.0056  & 0.30 \\
 8 -- 9 & $-0.3$ & 0.1 & 0.0030  & 0.37 \\
 9 -- 10 & $-0.3$ & 0.1 & 0.0030  & 0.43 \\
 10 -- 11 & $-0.4$ & 0.3 & 0.0014  & 0.80 \\
 11 -- 12 & $-0.5$ & 0.3 & 0.0014  & 0.80 \\
 12 -- 13 & $-0.6$ & 0.3 & 0.0014  & 0.80 \\
 \hline
\end{tabular}
\end{table}

There are more main-sequence stars than MSTO-SG stars due to the nature of
{\rm IMF}, therefore the random errors of stellar parameters (particularly ${\rm M}_V$),
may cause a net contamination from main-sequence stars to the MSTO-SG star sample. 
The contaminations are expected to cause overestimate 
of the stellar mass density, especially for the old stellar populations due to their closer positions  
to the bulk main-sequence in the H-R ($T_{\rm eff}$ -- ${\rm M}_V$) diagram. 
The percentile value of the main-sequence contaminations to the underlying 
MSTO-SG stars are mainly determined by the amount of random errors of parameter 
estimates, and also moderately depending on the local star formation history (i.e. the 
relative amount of stars among different age populations).

A series of tests and examinations have been carried out to validate the estimates 
of stellar parameters and their errors \citep{Xiang+2015a, Xiang+2017a, Xiang+2017b, Xiang+2017c}.  
The amount of parameter errors are found to depend sensitively on the spectral S/N. 
For the MSTO-SG sample stars, it is found that as the S/N increases from 20 to 80,   
typical random errors in $T_{\rm eff}$ decrease from 100\,K to 65\,K, random errors 
in ${\rm M}_V$ decrease from $\sim$0.7\,mag to 0.3\,mag, while random errors in [Fe/H] 
decrease from 0.16\,dex to 0.08\,dex. The errors in $T_{\rm eff}$ and [Fe/H] have also 
moderate dependence on the spectral type, as the early type stars having larger random 
errors in general. 
The left and middle panel of Fig.\,9 plots the errors of ${\rm M}_V$ for the MSTO-SG
sample stars with $7950<R<8050$\,pc as a function of S/N and $r$-band magnitude, respectively.
The figure shows that errors of ${\rm M}_V$ decrease from $\sim$0.7\,mag at a S/N of 20 
to about 0.2--0.3\,mag at a S/N higher than $\sim$\,80. Note that 60\% of our  
MSTO-SG sample stars have a S/N higher than 50. 
For nearby MSTO-SG stars, the spectral S/N's are even higher because the stars are brighter. 
At $7950<R<8050$\,pc, the MSTO-SG stars have a median S/N value of 90, and 76\% of the stars
have a S/N higher than 50. 

To further validate the error estimates with Gaia DR2, in Fig.\,10 we plot a comparison of ${\rm M}_V$ with values inferred
from the Gaia DR2 parallax \citep{Brown2018, Luri2018} for different S/N's.
The figure shows good agreements in general, and the differences are described well
by Gaussian distribution, with 1$\sigma$ value consistent well with our error estimates for
the intermediate and high S/N bins. For the lowest S/N bin of 20--30, the Gaussian 1$\sigma$ value
is lower than the error estimates by $\sim$0.1\,mag, indicating that ${\rm M}_V$ errors
for our MSTO-SG stars with low S/N's may have been slightly overestimated. However, this
does not have a negative impact on our conclusions since we then obtain a more conservative
estimate of the mass excess. At the fainter side of ${\rm M}_V\gtrsim5$\,mag,
the ${\rm M}_V$ from LAMOST spectra may be slightly overestimated (by $\sim$0.1\,mag at ${\rm M}_V\sim5$\,mag).
Again, this will not have a negative impact on our results since the overestimation tends
to reduce contaminations of main-sequence stars to the MSTO-SG sample.

Given the good knowledge of the stellar parameter errors, as well as the fact 
that, as the basis of this work, the underlying stellar mass function for mono-age and 
mono-metallicity population is well characterized by stellar initial mass function and stellar evolution, 
the amount of contamination can be practicably estimated from a mock dataset.
We thus use mock data to assess effects on the stellar mass density determination 
caused by the inevitably happened contamination of main-sequence stars. 
Our mock data are composed of a set of single-age exponential-disk populations in 
$7950<R<8050$\,pc created by Monte-Carlo sampling.
Parameters of the mock populations are shown in Table\,1.
The adopted parameters have a trend with age comparable to the measured ones
utilizing the MSTO-SG sample stars.
Random errors of parameters are incorporated into the generated parameters of individual mock stars.
Note that a realistic modeling of the parameter errors considering the S/N
effect is very complex since it requires a priori knowledge of the S/N distribution of all the
individual plates and spectrographs of the surveys. To simplify the problem, we use
the $r$-band magnitude as an indirect indicator of the the S/N, considering that fainter stars
generally have lower S/N's, and then assign the parameter errors based on the $r$-band magnitude
and effective temperature.
The right panel of Fig.\,9 shows the adopted ${\rm M}_V$ errors for the mock data. 

Fig.\,11 plots the ratios between the derived density within $|Z|<50$\,pc 
and the model inputs for different populations. Here the bin size (50\,pc) is adopted as the same 
as that for the real data (Section\,5). 
The figure shows that the derived stellar mass density is significantly higher than the model 
input for the old ($\gtrsim$8\,Gyr) populations, while lower than the model input for the 
youngest ($<1$\,Gyr) population. Similar patterns are seen also for the surface mass density. 
The lower derived mass density respect to the model input for the youngest populations
is mainly due to a systematic overestimation of age for those youngest stars. 
Such a systematic overestimation of age for those youngest stars has been found by 
\citet{Xiang+2017c} via validation with member stars of open clusters (see their Fig.\,13).
This actually also leads to higher derived stellar mass density respect to model input for the 1--2\,Gyr population. 
For the oldest population of 10--14\,Gyr, the derived density is higher than model input  
by amount of $\sim$190\% due to severe contamination from main-sequence stars. 
The derived surface mass density for the oldest population is 60\% higher than the model input, 
much smaller than that of the volume density within $|Z|<50$\,pc. This is because most of 
the main-sequence contamination stars have smaller scale heights than the underlying 
oldest population. For the whole stellar population of 0--14\,Gyr, the derived stellar mass 
density is $\sim$20\% higher than the model input, while the derived surface mass density 
is $\sim$18\% higher than the model input. 
Note that for the whole stellar population of 0--14\,Gyr, the measured stellar mass density 
within $|Z|<50$\,pc is found to be very close to (only 3\% higher) the model input mid-plane 
stellar mass density. 

Finally, we mention that since the true star formation history (or the age -- disk scale height relation) 
maybe different from the one adopted here, our estimate of contamination rate may suffer some uncertainties. 
However, such uncertainties for the overall populations are found to be 
small by varying the star formation history in reasonable range. This is largely because the 
contaminations mainly affect the stellar mass density estimates for old populations, which 
occupy only a limited part of the total stellar mass density. 
To better assess the contaminations, examinations with respect to independent, high accuracy 
set of observation data are also desired. During the review of this manuscript, the Gaia DR2  
become available, which provides the possibility for an independent check of the contaminations 
since the Gaia DR2 provides much more precise absolute magnitudes for bright ($r\lesssim16$\,mag)
 stars. A careful work for the same purpose of this paper based on Gaia DR2 parallax is ongoing.  
 As a preliminary result, we find that in $7.8<R<8.2$\,kpc, the Gaia DR2 
 yields a disk mid-plane total stellar mass density in good agreement with the current estimate (\S{5.3}) 
 after considering the 20\% contamination (with a difference of $\lesssim$0.002\,$M_{\odot}$/pc$^3$), 
 implying that the current estimate of contamination rate is reasonable. 
In addition, we may also validate the results with other advanced and independent mock data  
sets, such as those from the Galaxia \citep{Sharma2011}, the Galmod \citep{Pasetto2018} and that of \citet{Rybizki2018}.

\section{Results and Discussion}
In this section, we present the disk stellar mass distribution and star formation history derived from 
the LAMOST MSTO-SG stars. We will present the mass distribution for stellar populations in 
age bins of 1--14, 0--1, 1--2, 2--4, 4--6, 6--8, 8--10, 10--14\,Gyr. Here we present results of 
the 1--14\,Gyr rather than the whole stellar population of 0--14\,Gyr because, as mentioned in 
\S{3.3}, the latter has poor complete volume due to the large dynamic range of absolute magnitude. 
We describe the 3D mass distribution using the cylindrical coordinate ($R$, $\phi$, $Z$). 
The Sun is assumed to be located at $R=8.0$\,kpc, $\phi$=180$^\circ$ and $Z=0$.

\subsection{Stellar mass distribution in the disk $R$-$Z$ plane}
\begin{figure*}
\centering
\includegraphics[width=180mm]{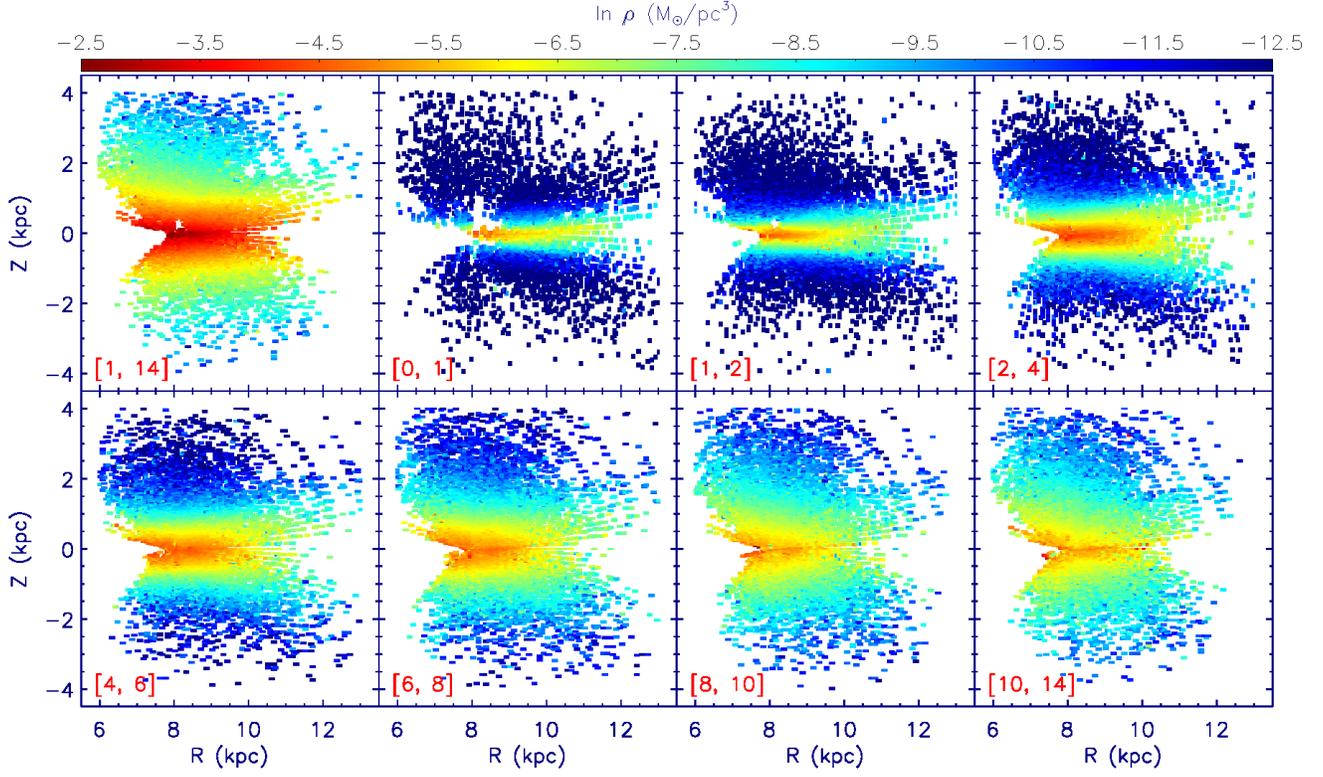}
\caption{Distribution of stellar mass density in the disk $R-Z$ plane.
Different panels are results for different age populations, as marked at the
bottom-left corner. The density map is generated by adopting a bin size of $0.1\times0.05$\,kpc.}
\label{Fig12}
\end{figure*}
\begin{figure*}
\centering
\includegraphics[width=180mm]{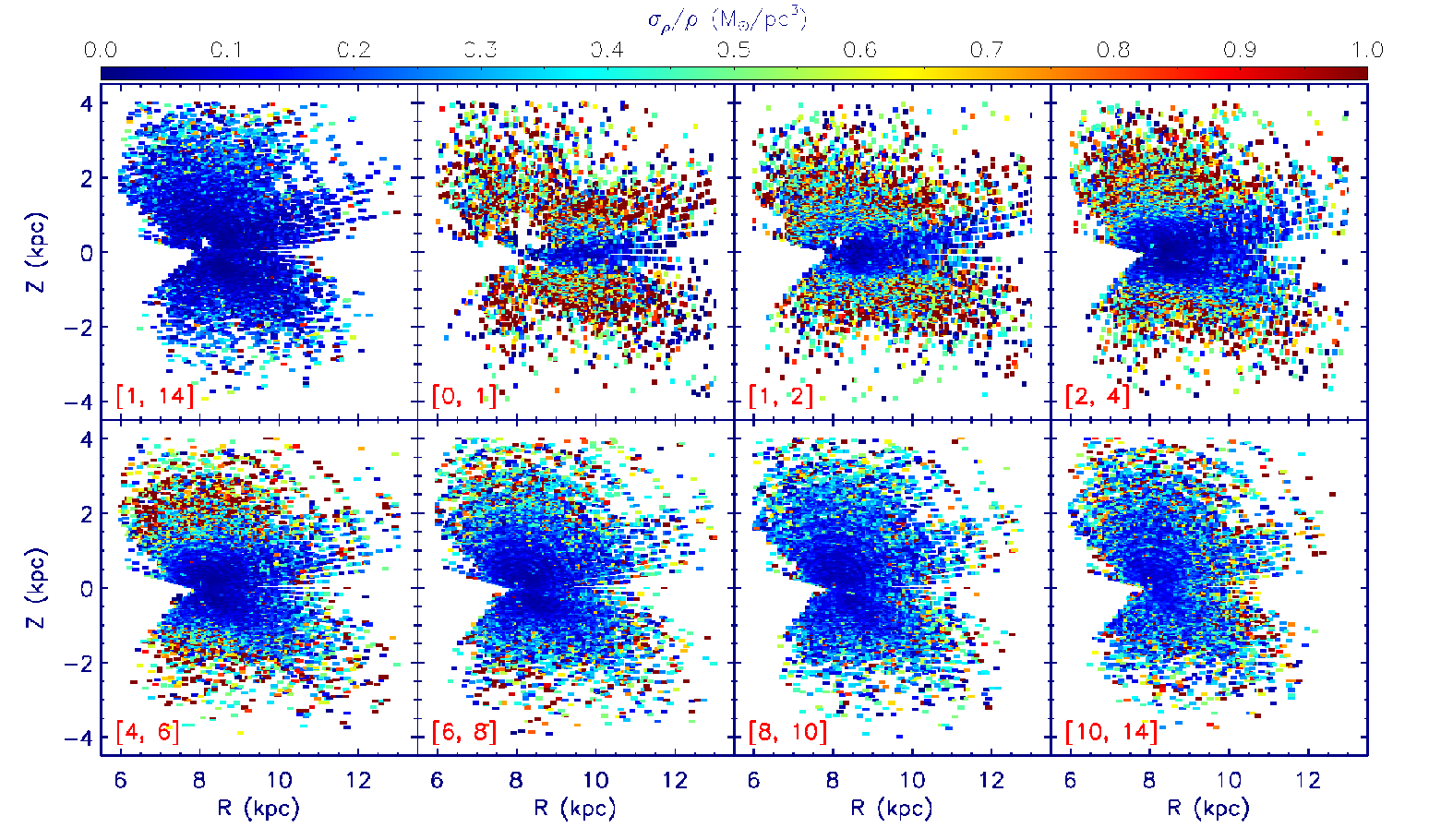}
\caption{Same as Fig.\,12 but for relative errors of the stellar mass density estimates.}
\label{Fig13}
\end{figure*}
We create a 2D density map in the disk $R$-$Z$ plane by dividing the measurements into bins of 
0.1$\times$0.05\,kpc. In each bin, all measurements in the azimuthal direction are averaged by taking 
the volume as a weight. To increase the sampling density, we have also opted to create a dense
grid with steps of 0.05 and 0.025\,kpc, in the radial and vertical direction, which means that there are 50\% overlaps between 
the adjacent bins. The results are shown in Fig.\,12.
The figure presents clear temporal evolution of the disk morphology. 
Younger stellar populations are more concentrated to the disk mid-plane and exhibit strong 
flaring phenomenon. For populations older than 8\,Gyr, the disk morphology become 
outward folding, which shows a decrease of density with increasing radius, although 
a quantitative description suggests that there are also flaring phenomenon. 
At $8\lesssim R\lesssim9$\,kpc and $Z\sim0$\,kpc, the maps for the 1--14\,Gyr and the relatively 
young populations ($\lesssim4$) present an over-density, which is particularly 
clear in the 1-14\,Gyr bin due to the high contrast of color scale. 
This over-density, as will be discussed below, is contributed by the Local stellar arm. 
Although with very low density, there are stars with very young age ($<2$\,Gyr) at 
unexpected large heights (e.g. $>2$\,kpc). We suspect they are contaminations 
of halo stars or blue strugglers of the (old) thick disk whose ages are wrongly estimated 
\citep{Xiang+2017c}.   
Note that at the outer boundary, the distribution of the data points shows
some arc-like structures. They are artifacts due to the binning strategy to measure
the density. These structures have however, no significant impact on the
overall stellar mass density distribution.
Fig.\,13 shows the error estimates of the stellar mass density determinations. 
The relative errors of the mass density increase with vertical height above the 
disk plane, mainly due to decrease of stellar number density at larger heights. 
For the 1--14\,Gyr population, the median value of relative errors for individual $R$--$Z$ bins 
is 10\%, while at small heights (e.g. $|Z|<200$\,pc), the relative errors are smaller than 5\%. 
Note that for this plot, as well as for determining the disk structure, we have imposed 
a minimum error limit of 5\% by setting all smaller values to be 5\%.  
For young stellar populations at large heights above the disk plane, the mass 
density estimates have large relative errors which may reach 100\%.

\begin{table}
\caption{The range of parameters used for the MCMC fitting.}
\label{}
\begin{tabular}{lll}
\hline
$\rho_{1, R_\odot}$ ($M_{\odot}$ pc$^{-3}$)  &  [0.0001, 0.1]    & mid-plane density at $R_0$ \\
$\rho_{2, R_\odot}$ ($M_{\odot}$ pc$^{-3}$)  &  [0.00001, 0.1]  &  \\
 $L_1$ (kpc)                                       &  [0.1, 10]     & scale length \\
$L_2$ (kpc)  &   [0.1, 10]        &   \\
$Z_{0,1}$ (kpc)   &   [$-$0.2, 0.2]     & height of the disk mid-plane   \\
$Z_{0,2}$ (kpc)    &  [$-$0.2, 0.2]     &  \\
$H_1$ (kpc)   &  [0.01, 3.0]     & scale height  \\
$H_2$ (kpc)   &  [0.01, 3.0]    &  \\
$\beta_1$           &   [0, 3]     &  slope of scale heights with $R$ \\
$\beta_2$           &   [0, 3]     & \\
$n_1$                &   [0, 20]   & index of the $sech^n$ function \\
$n_2$                &   [0, 20]   &  \\
\hline
\end{tabular}
\end{table}

\begin{figure*}
\centering
\includegraphics[width=180mm]{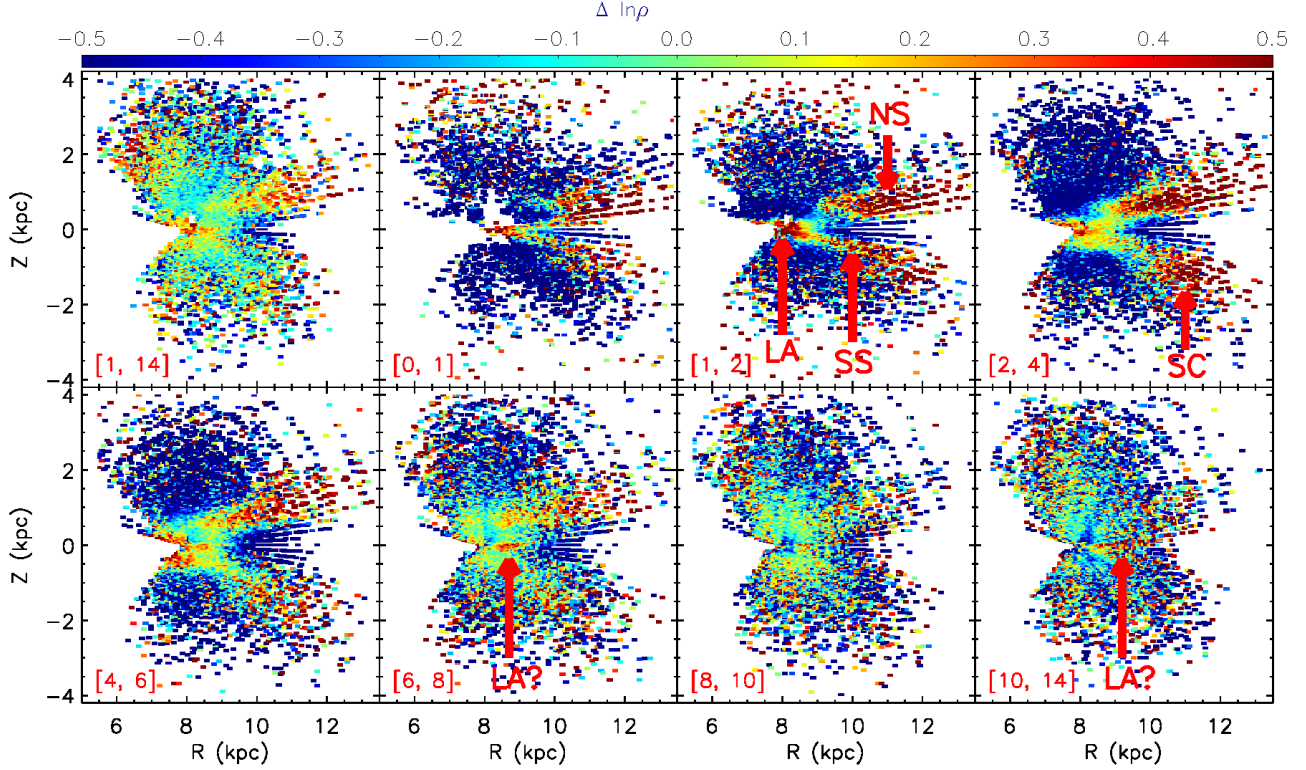}
\caption{Residuals of the stellar mass density distribution in the disk $R-Z$ plane, after subtracting 
the double-disk component fits with constant scale heights. Arrows in red mark several prominent 
 over-density structures, namely the Local arm (`LA'), the northern stream (`NS'), the southern stream 
 (`SS') and the southern clump (`SC'). `LA?' in the bottom panels indicates that the over-density  
structure is suspected to be associated with the `LA' structure presented for younger populations.}
\label{Fig14}
\end{figure*}
\begin{figure*}
\centering
\includegraphics[width=180mm]{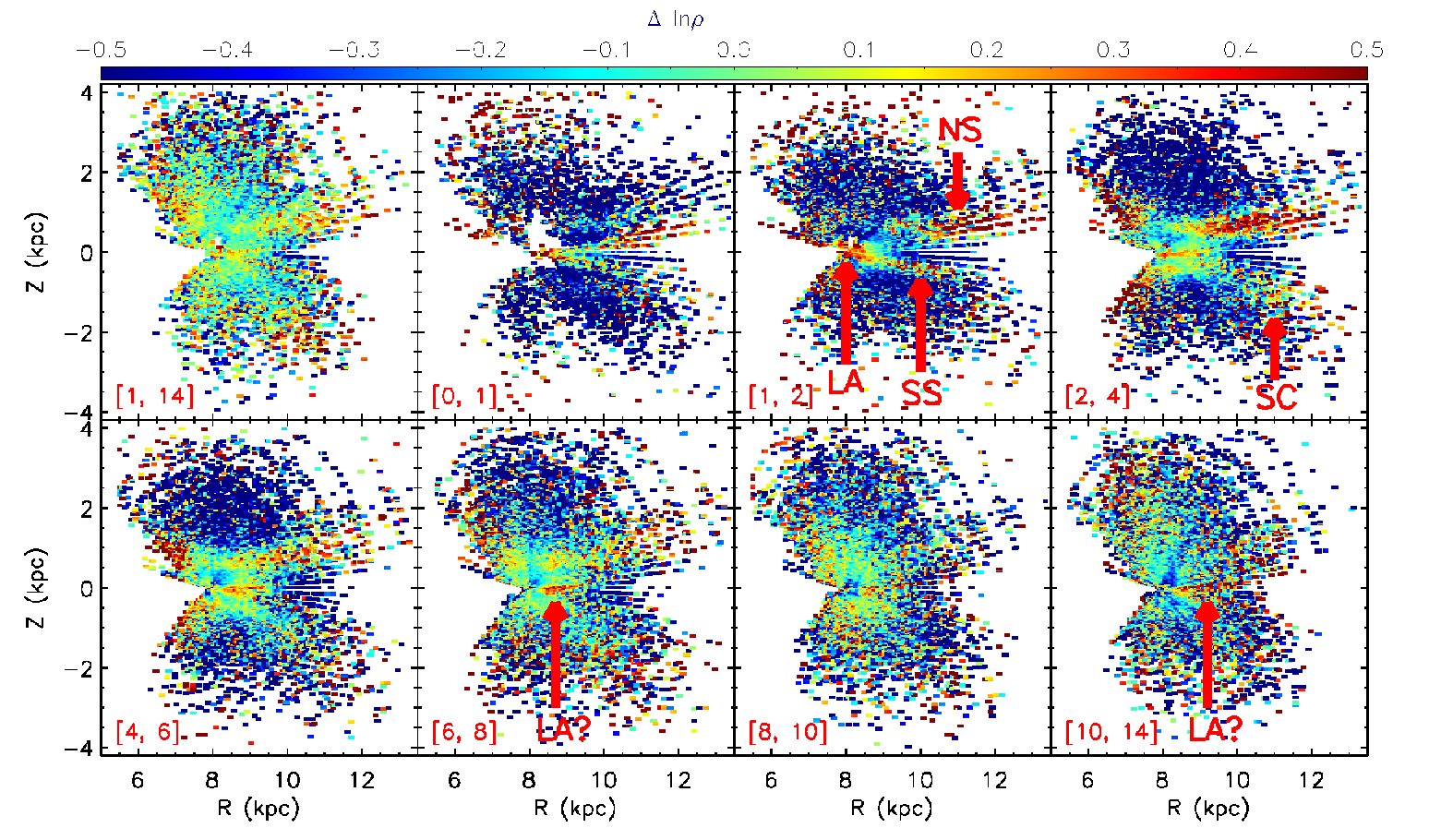}
\caption{Same as Fig.\,14 but for residuals after subtracting
the double-disk component fits with flared scale heights.}
\label{Fig15}
\end{figure*}

It is suggested that the radial luminosity (and mass) profiles of galactic disks are well described by 
exponential functions, while the vertical profiles are better described by $sech^n$ functions \citep{vanderKruit1988, vanderKruit2011}. 
We therefore fit the mass distribution with a sum of two $sech^n$ functions 
with flared disk scale heights, 
\begin{equation}  
\begin{aligned}
{\rho}(R,Z) = & {\rho}_{1, R_{\odot}}\exp\left(-\frac{R-R_{\odot}}{L_1}\right){\rm sech}^{2/n_1}\left(-\frac{n_1|Z-Z_{0,1}|}{2H_1^{\prime}}\right)\\
& + \rho_{2, R_{\odot}}\exp\left(-\frac{R-R_{\odot}}{L_2}\right){\rm sech}^{2/n_2}\left(-\frac{n_2|Z-Z_{0,2}|}{2H_2^{\prime}}\right), 
\end{aligned}
\end{equation}
\begin{equation}  
\begin{aligned}
&H_i^{\prime} = H_i\times(1.0+\beta_i(R-R_0)),
\end{aligned}
\end{equation}
where ${\rho}_{i, R_{\odot}}$ is the volume density of the $i_{\rm th}$ component at solar radius.  
$L_i$ and $H_i$ are respectively the scale length and height of the $i_{\rm th}$ component. 
$Z_{0, i}$ is the position of the mass-weighted mid-plane of the disk, which is 
a free parameter in the fitting. The index $n_i$ a free parameter, and the vertical profile becomes the 
isothermal distribution when $n=1$, and becomes the exponential function when $n=\infty$.
The disk flaring is described by a linear outward increase of the scale height, 
and $\beta_i$ is the increasing rate of scale height for the $i_{\rm th}$ component, 
which describes the strength of the flaring. Fixing $\beta_i=0$ corresponds to a 
constant scale height model.  
The fitting is implemented by searching for the best set of parameters with 
a Markov Chain Monte Carlo (MCMC) method. The best-fit parameters are taken as those 
yields the minimum $\chi^2$, which is defined as 
\begin{equation}
\chi^2 = \sum\limits_{i=1}^n\frac{(\rho_{\rm measure}^i - \rho_{\rm model}^i)^2}{\sigma_i^2}, 
\end{equation}
where $\rho_{\rm measure}^i$ and $\rho_{\rm model}^i$ are respectively the measured 
and the model-predicted stellar mass density for the $i_{\rm th}$ $R$--$Z$ bin,  $\sigma_i$ 
is the error estimate for the measured stellar mass density. 
Errors of the best-fit parameters are adopted as the standard deviations of the 
individual sets of parameters generated by the MCMC method.
Here we have adopted the MCMC code written by Ankur Desai (v1.0) in IDL environment. 
The allowed range of parameters for the MCMC fitting are presented in Table\,2. 

Table\,3 presents the results of the fitting for disk models with both constant scale heights, i.e. fixing $\beta_i=0$, 
and flared scale heights. 
It is found that for the disk model with constant scale heights, the population 
of 1--14\,Gyr yields a scale length of 3677$\pm$57 and 4457$\pm$80\,kpc, and a scale height of 300$\pm$2 and 981$\pm$12\,pc for the 
thin and the thick disk component, respectively. While when accounting for the flaring, the values become 
2216$\pm$30 and $1405\pm25$\,kpc for the scale length, 265$\pm$2 and 920$\pm$8\,pc for the scale height, with a $\beta$ value of 
0.178 and 0.124 for the thin and thick disk, respectively.   
We expect that these scale parameters derived from the 1--14\,Gyr population 
are good approximates to those of the whole stellar population of 0--14\,Gyr, as the youngest 
($<1$\,Gyr) population contribute only a minor amount ($<$1/10) of stellar mass. 
Models with constant scale heights have failed to yield converged values of scale lengths 
for young populations, as the derived values reach the upper boundary set for the fitting. 
This is because the density distributions in the $R$--$Z$ plane for those populations are 
significantly flared. While Table\,3 shows that, in most cases, the flared disk model can yield 
reasonable description to the density distributions. Also, in some cases for both the constant 
height model and the flared model, the index value of the $sech^n$ function reach the upper 
limit set for the fitting, which suggests that the realistic vertical density distribution is more 
resemble to an exponential profile.

The mass-weighted disk mid-plane is found to be $10\pm1$\,pc below the Sun. The value is 
smaller than many of the previous estimates, which give values of about 15--27\,pc \citep[e.g.][]{Chen2001, Juric+2008, Widmark2017}. 
However, our results show that positions of the mass-weighted disk mid-plane evolve with age, 
with younger populations have smaller offsets respective to the Sun, from about 1$\pm$2\,pc 
for the youngest population to $\sim$30\,pc for the oldest population. It is likely that the higher 
values in literature are caused by bias of their sample stars toward old populations. 
In fact, it has been shown that tracers with young ages, such as open clusters and A/F dwarfs 
generate disk mid-plane positions with small offset (a few parsec) respect to the Sun \citep[e.g.][]{Joshi2016, Bovy2017}, 
which are consistent with our estimates of young stellar populations. 

\begin{table*}
\centering
\caption{Derived parameters for stellar mass distribution in the disk $R$-$Z$ plane.}
\tiny 
\begin{tabular}{l}
(a) Fitting the mass distribution using double-component disk models with constant scale heights.   \\
\end{tabular}
\begin{tabular}{lllllllll}
\hline
Age   (Gyr)                            & 1-14                             &      0-1                                       & 1-2                                      & 2-4                            & 4-6                  &  6-8                                           & 8-10                            & 10-14 \\
$\rho_{1, R_\odot}$   ($M_{\odot} pc^{-3}$) &  0.0477$\pm$0.0004 &  0.0040$\pm$0.0002  &  0.0058$\pm$0.0001 &  0.0116$\pm$0.0002 &  0.0086$\pm$0.0001  &  0.0069$\pm$0.0001  &  0.0056$\pm$0.0001 &  0.0056$\pm$0.0001  \\
$\rho_{2, R_\odot}$   ($M_{\odot} pc^{-3}$) &  0.0028$\pm$0.0002   &  2.1865e-5$\pm$1.4881e-6 &  2.5470e-5$\pm$2.3762e-6 &  3.2760e-5$\pm$2.3147e-6  &  0.0001$\pm$2.5127e-5  &  0.0004$\pm$3.7328e-5 &  0.0007$\pm$0.0001 &  0.0002$\pm$1.1773e-5  \\
$L_1$ (pc)   &  3677$\pm$57 &  9997$\pm$38$^{n}$  &  9994$\pm$26$^{n}$  &  9999$\pm$13$^{n}$  &  9371$\pm$361$^{n}$ &  2842$\pm$72   &  1906$\pm$51   &  2285$\pm$34     \\
$L_2$ (pc)   &  4457$\pm$80   &  9996$\pm$22$^{n}$   &  9998$\pm$19$^{n}$ &  9999$\pm$21$^{n}$ &  9999$\pm$14$^{n}$ &  9999$\pm$63$^{n}$  &  5737$\pm$259    &  9991$\pm$146$^{n}$ \\
$Z_{0, 1}$ (pc)  &    $-$9$\pm$1   &    $-$3$\pm$2  &    $-$5$\pm$1   &   $-$15$\pm$1  &   $-$19$\pm$1   &   $-$32$\pm$2  &   $-$22$\pm$3  &   $-$36$\pm$3 \\
$Z_{0, 2}$ (pc)  &  $-$119$\pm$6  &  $-$168$\pm$12 &   $-$28$\pm$10  &  $-$197$\pm$11$^{n}$  &  $-$200$\pm$4$^{n}$   &   $-$88$\pm$8        &   $-$47$\pm$9    &  $-$200$\pm$1$^{n}$  \\
$H_1$ (pc)   &   300$\pm$2   &   124$\pm$2   &   152$\pm$1  &   220$\pm$1 &   263$\pm$2    &   289$\pm$6         &   388$\pm$7   &   488$\pm$4  \\
$H_2$ (pc)   &   981$\pm$12  &   818$\pm$13   &   804$\pm$15  &  2020$\pm$100    &  1174$\pm$34    &   954$\pm$23     &  1016$\pm$23   &  1371$\pm$142 \\
$n_1$    &  19.92$\pm$0.86$^{n}$      &   3.99$\pm$0.76 &  10.17$\pm$2.09   &   9.36$\pm$1.67   &   2.71$\pm$0.15    &   1.66$\pm$0.11    &  15.40$\pm$2.91   &  19.90$\pm$0.71$^{n}$ \\
$n_2$    &  14.36$\pm$4.67  &  17.24$\pm$4.56$^{n}$     &   4.40$\pm$4.74   &   9.38$\pm$4.17   &   3.13$\pm$5.94  &   3.37$\pm$0.49  &   5.41$\pm$3.49     &   1.00$\pm$0.45   \\
$\chi_{\rm red}^2$  &   2.98 &   4.90   &   3.92  &   4.06   &   3.45   &   3.30   &   3.45   &   2.60   \\
\hline
\end{tabular}
\begin{tabular}{l}
(b) Fitting the mass distribution using double-component disk models with flared scale heights.  \\
\end{tabular}
\begin{tabular}{lllllllll}
\hline
Age   (Gyr)                             & 1-14                             &      0-1                                       & 1-2                                      & 2-4                            & 4-6                  &  6-8                                           & 8-10             & 10-14 \\
$\rho_{1, R_\odot}$   ($M_{\odot} pc^{-3}$)            & 0.0563$\pm$0.0005    &   0.0054$\pm$0.0003            &  0.0077$\pm$0.0002           &  0.0126$\pm$0.0002   &  0.0099$\pm$0.0001    &  0.0078$\pm$0.0001  &  0.0056$\pm$0.0001   &  0.0054$\pm$0.0001 \\
$\rho_{2, R_\odot}$   ($M_{\odot} pc^{-3}$)             & 0.0037$\pm$0.0001    &  2.0617e-5$\pm$1.2001e-6   &  3.9331e-5$\pm$2.3209e-6   &  0.0001$\pm$3.3480e-6   &  0.0004$\pm$2.5153e-5  &  0.0003$\pm$4.9607e-5 &  0.0006$\pm$4.6273e-5  &  0.0003$\pm$2.3280e-5 \\
$L_1$ (pc)                     & 2216$\pm$30              &  3270$\pm$192                     &  2284$\pm$44       &  2670$\pm$67   &  2025$\pm$32    &  2050$\pm$36     &  2039$\pm$38  &  2248$\pm$35 \\
$L_2$ (pc)                     & 1405$\pm$25              &  9949$\pm$177$^{n}$           &  2331$\pm$79      &  3480$\pm$209   &  3326$\pm$263   &  2803$\pm$184   &   882$\pm$18   &  7490$\pm$887 \\
$Z_{0, 1}$ (pc)               &  $-$10$\pm$1             &    $-$1$\pm$2                         &    $-$5$\pm$1      &   $-$14$\pm$1     &   $-$19$\pm$1    &   $-$34$\pm$2    &   $-$17$\pm$3      &   $-$32$\pm$3  \\
$Z_{0, 2}$ (pc)               & $-$114$\pm$5            &   $-$51$\pm$11                        &   $-$61$\pm$9    &  $-$126$\pm$21   &   $-$88$\pm$10   &   $-$81$\pm$13  &   $-$96$\pm$14  &  $-$200$\pm$1$^{n}$ \\
$H_1$ (pc)                    &   265$\pm$2                &    91$\pm$2                             &   117$\pm$1      &   166$\pm$1         &   198$\pm$3       &   306$\pm$5    &   405$\pm$6    &   498$\pm$5 \\
$H_2$ (pc)                    &    920$\pm$8               &   777$\pm$11                          &   758$\pm$10    &  1466$\pm$45       &   853$\pm$18    &  1202$\pm$43   &   1208$\pm$34   &  1907$\pm$74   \\
$\beta_1$                      & 0.178$\pm$0.005       &   0.222$\pm$0.009                &   0.270$\pm$0.005   &   0.212$\pm$0.006  &   0.222$\pm$0.007    &   0.127$\pm$0.005   &   0.233$\pm$0.009   &   2.141e-5$\pm$1.836e-4  \\
$\beta_2$                      & 0.123$\pm$0.004       &   0.050$\pm$0.003    &   0.107$\pm$0.005    &   0.105$\pm$0.009     &   0.078$\pm$0.006     &   0.055$\pm$0.008   &   0.222$\pm$0.005 &   0.058$\pm$0.009 \\
$n_1$                            & 19.97$\pm$1.09$^{n}$         &   3.36$\pm$0.50    &   5.25$\pm$0.62      &   2.42$\pm$0.14     &   1.34$\pm$0.07       &   2.18$\pm$0.15    &  19.40$\pm$1.61$^{n}$  &  19.68$\pm$0.60$^{n}$  \\
$n_2$                            & 18.71$\pm$2.49$^{n}$         & 11.45$\pm$4.30  &  15.72$\pm$4.60      &  16.26$\pm$4.06     &  19.10$\pm$3.92$^{n}$      &   4.08$\pm$6.17   &  17.76$\pm$3.10$^{n}$   &  19.73$\pm$2.49$^{n}$  \\
$\chi_{\rm red}^2$        &  2.84                           &   4.45      &   3.34                &   3.53     &   3.13     &   3.22   &   3.45   &   2.60  \\
\hline
\end{tabular}     
\begin{tablenotes}
\item[1] $n$:  parameter value reaches the boundary due to convergence failure of the fitting. 
\end{tablenotes}                                                
\end{table*}

Figs.\,14 and 15 plots the residual map of the fits for disk models with constant and flared scale heights, respectively.  
The figures illustrate that the mass distribution is much more complex than the double 
exponential plus $sech^n$ functions in terms of that there exists prominent patterns and asymmetric structures. 
For young ($\lesssim4$\,Gyr) populations, as well as the population of 1--14\,Gyr, 
there is an over-density at the solar radius near the disk mid-plane (see `LA' in the figure). Such an over-density 
has also been seen in Fig.\,12, as mentioned above. The azimuthal distribution of this over-density 
suggests that it is actually a Local stellar arm (see Section\,5.2). 
Fig.\,14 shows that for populations of $\lesssim8$\,Gyr, the outer disk 
exhibits strong stripes of mass excess at both the northern and the southern side.
Those stripe-like structures have a large extension in the radial direction. 
The northern stripe (see `NS' in the figure) becomes prominent from $R\sim9$\,kpc, and reach beyond $R=13$\,kpc, 
the limit of our sample stars. 
For young ($\lesssim2$\,Gyr) populations, the southern stripe (see `SS' in the figure) extends from about the solar radius 
to a large distance, while for older populations, the most prominent feature in the southern 
disk is a clump of mass excess at $R\sim11$\,kpc (see `SC' in the figure). Given the distance limit of our 
sample stars, it is not sure if the southern clump is a stripe-like structure that extends 
to large distance or a local clump-like structure. 
Near the disk mid-plane, the outer disk of $R>9$\,kpc shows a significant under-density. 
These under-densities near the disk mid-plane, as well as the over-densities above the
disk mid-plane, lead to a rather dumpy vertical density profile (see Section\,5.4). 
For old populations of $\gtrsim8$\,Gyr, the patterns become sparse and weak, but it seems that 
there are some mass excesses near the disk mid-plane of $8.5\lesssim R\lesssim10$\,kpc 
(see `LA?' in the figure), which is the opposite case to the younger populations. 
Although with less strength, Fig.\,15 shows almost all the patterns and structures  
seen in Fig.\,14 -- namely, the over-densities near the disk mid-plane, for which the 
positions in the radial direction move from the solar radius for young populations 
to the outer disk of $R\gtrsim9$\,kpc for old populations, the northern stripe of mass excess 
at $9\lesssim R\lesssim13$\,kpc for young and intermediate populations, the southern 
stripe of mass excess at $8\lesssim R\lesssim10$\,kpc for young populations, and 
the southern clump of mass excess at $R\sim11$\,kpc for intermediate populations (2--4\,Gyr). 
Fig.\,15 thus illustrates that structures shown in Fig.\,14 can not be fully explained by the 
(symmetric) disk flaring since they remain in residuals derived by subtracting models 
taking the flaring into consideration.    
The reason is largely because of the asymmetric nature of the structures at the northern and 
southern parts of the disk. 
Fig.\,15 shows also prominent over-densities at both the northern and southern sides 
above the disk mid-plane at the inner disk ($R<8$\,kpc), which are particularly 
strong for young and intermediate populations. Those over-densities are not presented 
in Fig.\,14. We suspect that those over-densities are probably caused by an imperfect disk flaring model. 
It is probably that the flaring starts at a Galactocentric distance beyond the solar radius, and 
the inner disk needs to be described by constant scale heights.

Using the SDSS photometry, \citet{Xuyan2015} found that stellar number density in the disk 
anti-center direction exhibits significant oscillation, and there are more stars in the northern disk 
at a distance of $\sim$2\,kpc from the Sun. This is consistent 
with our results, as we see strong mass excess at the northern disk at $R\sim10$\,kpc.  
\citet{Xuyan2015} found that the oscillation extends to large distance 
($\gtrsim$15\,kpc from the Sun) in the outer disk, it is thus natural to believe that patterns 
shown in Fig.\,15 are parts of a global oscillation structure in larger scale. Although 
\citet{Xuyan2015} present the oscillation structure at the outer disk of only $R\gtrsim10$\,kpc, 
the mass excess stripes for young populations at $R\lesssim10$\,kpc of the southern disk 
are likely extensions of the oscillation toward the inner part of the disk.     
 
 \subsection{Stellar mass density in the $R$-$\phi$ plane} 
 \begin{figure*}
\centering
\includegraphics[width=180mm]{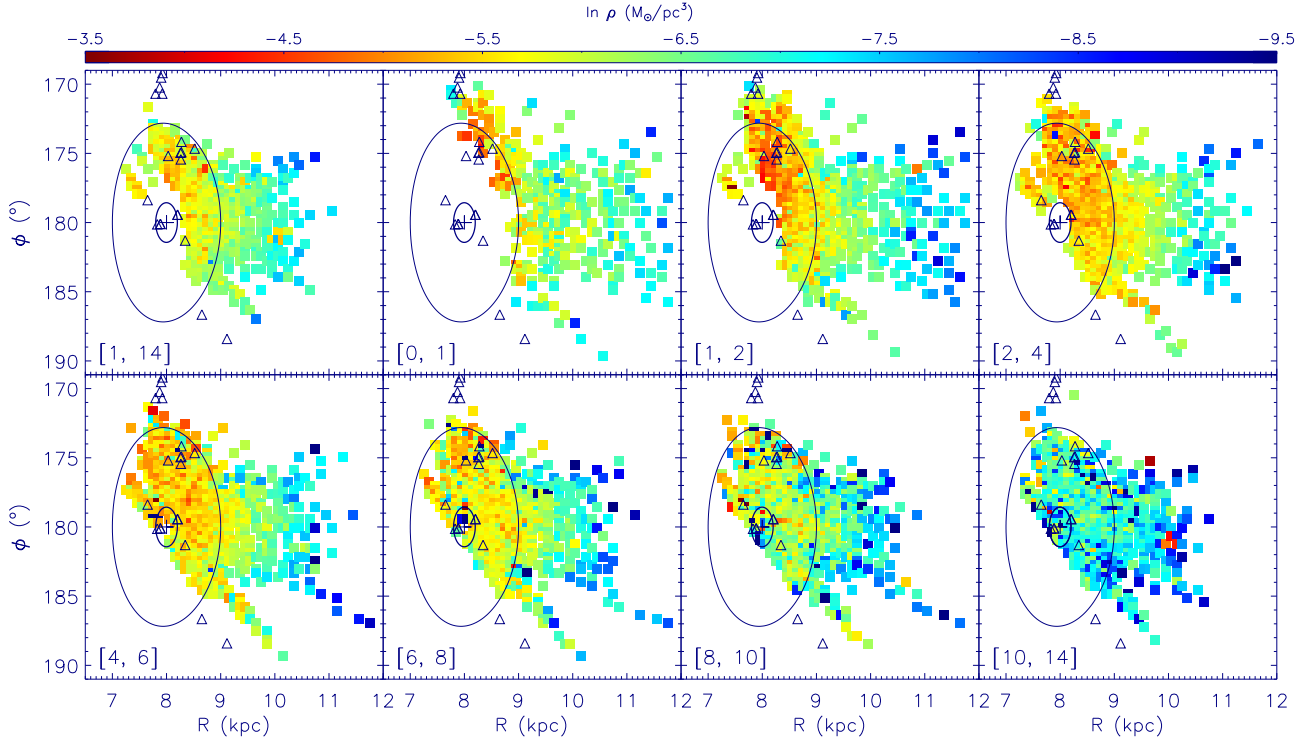}
\caption{Color-coded stellar mass density distribution in the disk $R-\phi$ plane
for the vertical slice of $|Z|<0.2$\,kpc. Different panels are results for different 
age bins, as marked at the bottom-left corner. 
For each bin, density values shown in the figure are time-averaged by dividing 
the age span of the bin. The plus indicates the
position of the Sun, while the inner and outer circle delineates respectively
a constant distance of 0.2 and 1.0\,kpc from the Sun on the disk mid-plane.
Triangles are sources of molecular masers associated with the Local arm
from \citet{Xuye2013}.}
\label{Fig16}
\end{figure*}
Fig.\,16 plots the stellar mass distribution in the disk $R$-$\phi$ plane for the 
vertical slice of $|Z|<0.2$\,kpc. The map is created by dividing the measurements within $|Z|<0.2$\,kpc 
into bins of 0.1\,kpc by 0.3$^\circ$, and average the individual measurements in each bin. 
Stellar populations of different ages exhibit different spatial coverage due to their different 
intrinsic brightness (thus different complete volume).
The population of 0--1\,Gyr shows poor coverage within 1\,kpc of the Sun as stars 
in this distance range have apparent magnitudes out of the bright limiting magnitude of the surveys, 
while the older populations reach smaller distance in the farther side due to their fainter intrinsic
brightness. Generally, the data have a good coverage of the disk within
500\,pc of the Sun for stellar populations of 1--4\,Gyr, and within 200\,pc
for populations older than 4\,Gyr.
The figure shows a significant mass excess at around the solar radius for young and 
intermediate stellar populations. In the azimuthal direction, the over-density structure 
reach a maximum distance of at least $\sim$1.2\,kpc, and it extends to larger 
Galactocentric radius ($\sim$9\,kpc) in the anti-center direction ($\phi=180^\circ$) than in the
second quadrant ($90<\phi<180^\circ$). 
The structure becomes more diffused with increasing age, but still visible for the population of 6--8\,Gyr.  
The location of the structure is consistent with the Local arm revealed
by young stellar associations and molecular gas \citep{Xuye2013}, implying that 
they are probably associated with each other. 
 
To better present the structures, in Fig.\,17 we plot the residual map after subtracting the  
fits with the double-component disk model with constant scale heights. 
Residual map after subtracting fits with the flared double-component disk model is also 
presented in the Appendix.
The residual maps shows clear patterns. For young populations ($\lesssim$4\,Gyr), 
it is clear that residuals at $R\lesssim9$\,kpc exhibit mass excesses, while they become 
under-densities at $R\gtrsim9$\,kpc, as has been seen in Figs.\,14 and 15. 
For populations older than 8\,Gyr, it seems that positions of the mass excesses in the anti-center direction 
have moved slightly outwards to $8.5\lesssim R\lesssim10.5$\,kpc. 
The mass excesses for young populations are especially prominent in the second quadrant, 
and the positions are consistent with the molecule clouds in the Local arm. 
While the mass excess patterns become fragmented and loose for the older populations. 
The 0--1\,Gyr population exhibits also some over-densities at $R\gtrsim10$\,kpc, 
which are probably signatures of the Perseus arm \citep{Xuye2006}. 
Interestingly, it is found that those over-density signatures are even more explicit for 
the vertical slice $0.2<|Z|<0.4$\,kpc (see Appendix), indicating that the over-density 
structure reach at least 200--400\,pc above the disk, which may provide constrains 
on the nature of the Perseus arm.
 \begin{figure*}
\centering
\includegraphics[width=180mm]{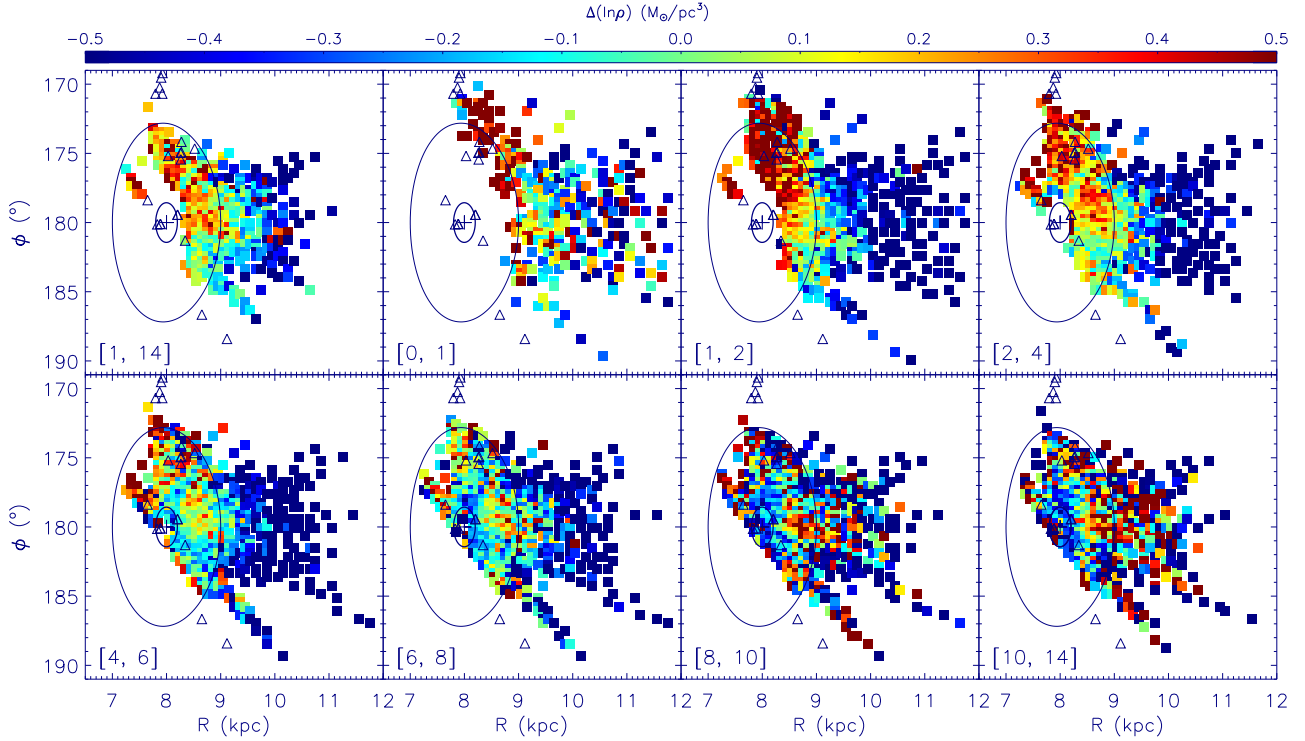}
\caption{Relative residuals of stellar mass density distribution in the disk $R-\phi$ plane for the vertical 
slice of $|Z|<0.2$\,kpc. The residuals are derived by subtracting the global fitting in the disk $R$--$Z$ 
plane with constant scale heights.}
\label{Fig17}
\end{figure*}

 \subsection{Stellar mass density at the solar radius}
 To have an estimate of the mid-plane disk stellar mass density at the solar radius, we average 
measurements within $7.8<R<8.2$\,kpc and $|Z|<50$\,pc. Since this region is not a complete 
volume for the whole stellar populations of 0--14\,Gyr due to the large spreading of absolute 
magnitudes of the MSTO-SG stars, 
we use the summation of the 1--14\,Gyr and the 0--1\,Gyr populations as a measure 
of the whole stellar populations. However, the 0--1\,Gyr population 
covers a different disk region with that of the 1--14\,Gyr population, as the former covers 
the disk region of $\sim$1\,kpc away from the solar position, while the later covers the region 
of 0.4--0.8\,kpc from position of the Sun. We therefore are forced to assume that from 0.4\,kpc 
to 0.8\,kpc in the azimuthal direction of $7.8<R<8.2$\,kpc, there is no abrupt variation 
of stellar mass density for the 0--1\,Gyr population. 
This seems to be a reasonable approximation, as we do not see strong azimuthal variation 
of stellar mass density in this region for the 1--2\,Gyr population. 

The underlying stellar mass density for the overall populations of 0--14\,Gyr 
within $7.8<R<8.2$\,kpc and $|Z|<50$\,pc is then 
\begin{equation}
\bar{\rho}_{\rm 0-14Gyr} = (1-c)\times(\bar{\rho}_{\rm 1-14Gyr}+\bar{\rho}_{\rm 0-1Gyr}),
\end{equation}
where 
\begin{equation}
\bar{\rho} = \frac{\sum_{i=1}^n\rho_iV_i}{\sum_{i=1}^nV_i}.
\end{equation}
Here $\rho_i$ is the $i_{\rm th}$ density estimate for which the central position of the distance bin 
is located in $7.8<R<8.2$\,kpc and $|Z|<50$\,pc, $V_i$ is the volume of the $i_{\rm th}$ distance 
bin, and $c$ is a factor accounting for contribution of main-sequence star contamination. 
Our measurements yield $\bar{\rho}_{\rm 1-14Gyr}=0.0662\pm0.0010$\,$M_\odot$/pc$^3$, 
$\bar{\rho}_{\rm 0-1Gyr}=0.0062\pm0.0003$\,$M_\odot$/pc$^3$. These values give a 
total stellar mass density of $0.0724\pm0.0010$\,$M_\odot$/pc$^3$ 
if we do not consider the contamination (i.e. $c=0$). However, as discussed in Section\,4, 
our measurements must have been significantly overestimated due to inevitable contamination 
from main-sequence stars, which may have contributed up to 20\% of the measured 
stellar mass density. We therefore adopt a $c$ value of 0.2 to obtain a more reasonable estimate 
of the underlying total stellar mass density. Then we have a total stellar mass density 
$\bar{\rho}_{\rm 0-14Gyr} = 0.0579\pm0.0008$\,$M_\odot$/pc$^3$.
The value further reduces to $\sim$\,0.0521$\pm$0.0007\,$M_\odot$/pc$^3$ if the Chabrier {\rm IMF} is adopted, 
as it predicts about 10\% lower stellar mass than the Kroupa {\rm IMF}.
The result does not yet include contributions from brown dwarfs, which may contribute
another 0.0015--0.002\,$M_\odot$/pc$^3$ \citep{Flynn2006, McKee2015}.
Considering a brown dwarf mass density of 0.0015\,$M_\odot$/pc$^3$, the final total stellar
mass density is then 0.0594$\pm$0.0008\,$M_\odot$/pc$^3$ ($0.0536\pm0.0007$\,$M_\odot$/pc$^3$ 
for the Chabrier ${\rm IMF}$)

These values are significantly higher than
previous estimates at the solar-neighborhood based on the Hipparcos data, 
which are 0.044\,$M_\odot$/pc$^3$ \citep{Holmberg2000}, 0.045$\pm$0.003\,$M_\odot$/pc$^3$ \citep{Chabrier2001}, 
0.042\,$M_\odot$/pc$^3$ \citep{Flynn2006}, 0.043$\pm$0.004\,$M_\odot$/pc$^3$ \citep{McKee2015}, 
and also higher than recent estimate with the Gaia DR1 by \citet{Bovy2017}, who give a value of 0.04$\pm$0.002\,$M_\odot$/pc$^3$.
Adopting a value of 0.043$\pm$0.004\,$M_\odot$/pc$^3$ for the solar-neighborhood measurement by \citet{McKee2015},
our estimate is 0.0164\,$M_\odot$/pc$^3$ higher, which is above 4 times larger than the reported error 
by \citet{McKee2015}, or 8 times larger than the report error by \citet{Bovy2017}. 
If the Chabrier IMF is used, the amount of over-density becomes 0.0106\,$M_\odot$/pc$^3$, which 
is 3 times larger than the reported error by \citet{McKee2015}, 5 times larger than the reported error by \citet{Bovy2017}. 
As all these measurements in literature suggest a value between 0.040--0.045\,$M_\odot$/pc$^3$, 
the difference between our estimates and the literature may have even larger significance than 
the quoted values. Note that \citet{Chabrier2003} have suggested a total stellar mass density 
of 0.051$\pm$0.003\,$M_\odot$/pc$^3$ in the local disk by assuming a 20 per cent contribution from the thick disk. 
Such a value is comparable to ours when the Chabrier  {\rm IMF} is adopted. However, we argue that 
a 20 per cent contribution from the thick disk at the local disk is seriously overestimated. 
Our results suggest the thick disk contributes only a few per cent mass density at the disk mid-plane, 
which is consistent with many literature results \citep[e.g.][]{Juric+2008, Chen+2017}.

We emphasize that our results are obtained at solar radius but not the `solar neighborhood'.
Our sample stars have a good coverage at 400--800\,pc away from the Sun 
in the azimuthal direction but have poor coverage within 400\,pc.  
A likely explanation of the higher density found by this work than the solar neighborhood values 
in literature is that the Sun is located in a local low stellar density region, which has a density of 
0.0164\,$M_\odot$/pc$^3$ (or 0.0106\,$M_\odot$/pc$^3$ if Chabrier IMF adopted) lower than the nearby disk. 
Such a difference must be contributed by the Local stellar arm. Our Sun is either located at the
inner boundary of the Local arm or embedded in a cavity of stars in the arm, and it needs 
to be further studied using data with improved spatial coverage to clarify which is the real case. 
Note that the literature results for the solar-neighborhood density are usually determined 
within a complex volume, which vary with different types of stars. It is thus 
difficult to make a direct comparison of our relatively well-defined volume density with the literature results. 
To test whether the difference is caused by 
the possibility that the literature results are actually averaged values in a larger volume, we  
have also examined the mean stellar mass density within $7.8<R<8.2$\,kpc and $|Z|<100$\,pc, 
and find a density of 0.0549\,$M_\odot$/pc$^3$ (0.0496\,$M_\odot$/pc$^3$ from the Chabrier  {\rm IMF}), 
which is still significantly higher than the `solar-neighborhood' values in literature. 
We also emphasize that since the stellar mass density decreases fast with increasing height above 
the disk plane, the `underlying' mid-plane density should be higher than the current estimates 
of average values within $|Z|<50$\,pc. The mid-plane density is expect to be comparable 
to the measured values without correction for contaminations of main-sequence stars (Section\,4). 

Assuming a gas density of 0.05\,$M_\odot$/pc$^3$ as widely adopted \citep{Holmberg2000, Flynn2006},
the expected mass density of baryon matter (star and gas) in the nearby disk plane within a few
hundred parsec is thus 0.109\,$M_\odot$/pc$^3$ (0.104\,$M_\odot$/pc$^3$ for Chabrier {\rm IMF}). 
Such a value is consistent well with the local total mass density yielded by stellar dynamics, 
which suggest a typical value of 0.1\,$M_\odot$/pc$^3$ \citep{Bienayme1987, Kuijken1989c, 
Pham1997, Holmberg2000, Read2014, McKee2015, Widmark2017, Kipper2018}. 
Our results thus leave little room for the existence of a meaningful amount of dark matter in the 
nearby disk mid-plane.
However, since our results show that stellar mass distribution in the local disk is highly asymmetric, 
one needs further study to better understand how the local dark matter density estimation 
has been affected by such asymmetries. 

\begin{table*}
\centering
\caption{Stellar mass density at solar radius derived with star count method.}
\label{}
\begin{tabular}{llll}
\hline
Reference  & $\Sigma^{*}$  ($M_\odot$/pc$^{-2}$)  & $\Sigma_{\rm visible}^{*}$  ($M_\odot$/pc$^{-2}$)   \\ 
                  &     visible star + remnant             & visible  star &              \\                                         
\hline
\cite{Flynn2006}  &  35.5   &   28.3   \\
\cite{Bovy2012} &  $-$ & $32\pm1$$^a$   \\
\cite{McKee2015} & $33.4\pm3$  &  $27.0\pm2.7$   \\
\cite{Mackereth2017} & $-$ & $20.0^{+2.4}_{-2.9}({\rm stat.})^{+5.0}_{-2.4}({\rm syst.})$     \\
This work$^{b}$   & $36.8\pm0.5$ & $29.1\pm0.4$  \\
 \hline
\end{tabular}
\begin{tablenotes}
\item[1] $a$: the Kroupa (2001) IMF is adopted. The value becomes $30\pm1$ if the Chabrier {\rm IMF} is adopted.
\item[2] $b$: the Kroupa (2001) IMF is adopted. The values become $33.3\pm0.5$ and $26.3\pm0.4$ if the Chabrier {\rm IMF} is adopted.
\end{tablenotes}
\end{table*}

\begin{figure}
\centering
\includegraphics[width=80mm]{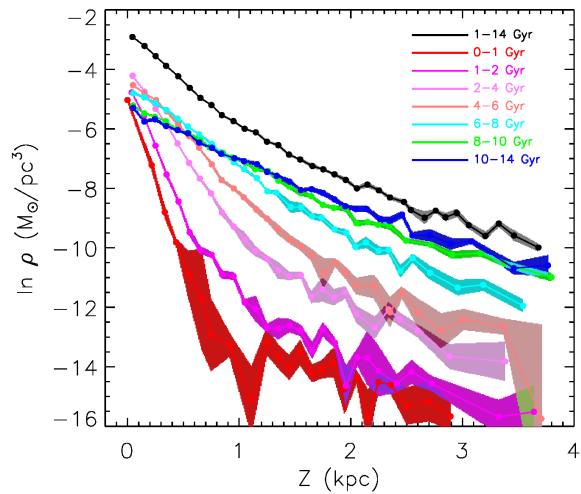}
\caption{Vertical distribution of stellar mass density for mono-age populations
in the disk radial slice $7.8<R<8.2$\,kpc. The shadow regions indicate the $1\sigma$ errors.}
\label{Fig18}
\end{figure}
\begin{figure*}
\centering
\includegraphics[width=180mm]{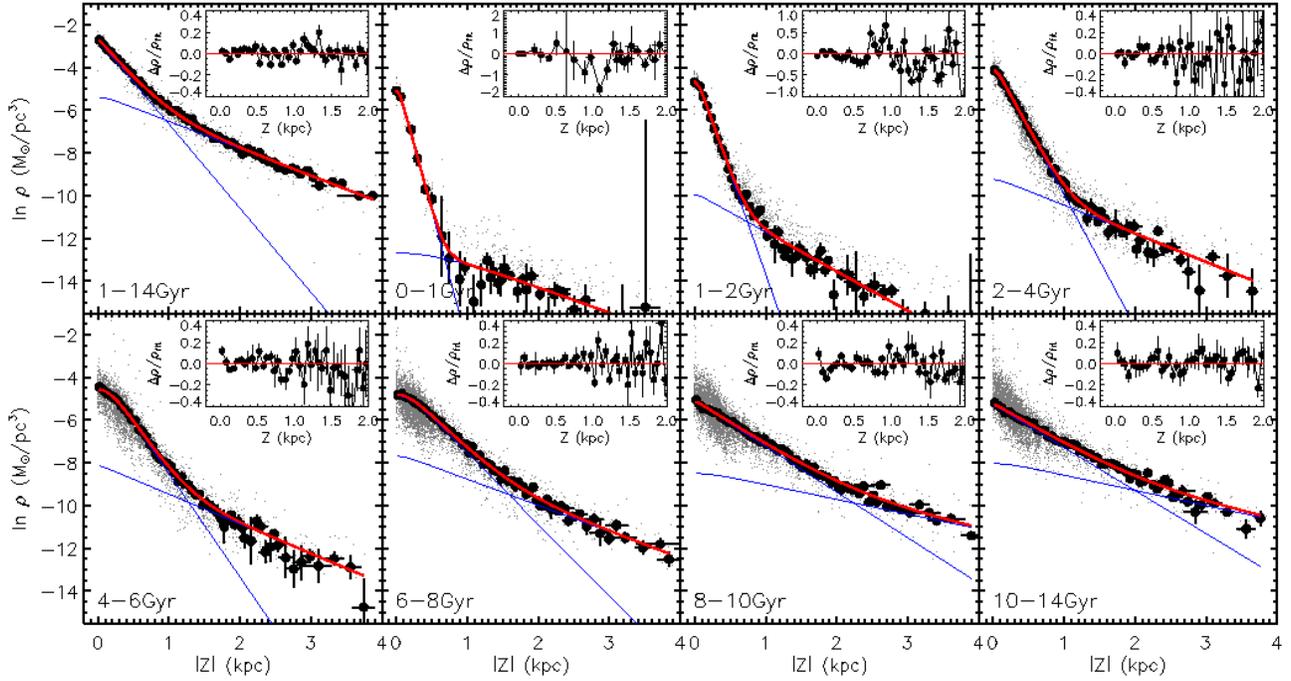} 
\caption{Fitting the vertical stellar mass distribution in the $7.8<R<8.2$\,kpc slice
with double $sech^n$ function. Different panels are results for different age
populations, as marked at the bottom-left corner. The grey dots are individual
 measurements of the stellar mass density. Note that measurements with $zero$ density 
 are not presented in the figure. The black filled circles and error bars
 are volume-weighted mean and standard errors of the mean in vertical bins 
 of 0.05\,kpc width.
 The red curve is the fit, while the blue lines are the individual
 components of the double $sech^n$ function. Residuals of the fitting are plotted at the upper-right corner.}
\label{Fig19}
\end{figure*}
Fig.\,18 shows the vertical mass distribution in the radial slice of $7.8<R<8.2$\,kpc for different
stellar populations. The figure shows a clear increasing trend of disk thickness with stellar age. 
It should be noted that as has been mentioned above, the extra component for the youngest 
populations at large heights are probably contaminations from either halo populations or thick 
disk blue stragglers whose ages are wrongly estimated. The extra component however contributes 
only a marginal ($<3$ per cent) fraction of stellar surface mass density of the youngest populations,
and will not have a significant impact on the conclusion of this paper.
We fit the vertical density distribution with a 
double $sech^n$ function \footnote{$\rho = {\rho}_1{\rm sech}^{2/n_1}\left(-\frac{n_1|Z-Z_{0,1}|}{2H_1}\right) 
+ \rho_2{\rm sech}^{2/n_2}\left(-\frac{n_2|Z-Z_{0,2}|}{2H_2}\right)$, where $Z_{0,i}$ is fixed to be 0.} (Fig.\,19) 
and integrate the function to 4\,kpc above the disk mid-plane to derive the surface mass density. 
Results of the fits are shown in Table\,5. For comparison, results of fits with a double exponential function 
are also presented in Appendix.   
The fitting yields a surface mass density of 43.1$\pm$0.5\,$M_\odot/{\rm pc}^2$ 
for the whole stellar populations by combing results of the 1--14\,Gyr and 0--1\,Gyr populations. 
After multiplying a factor of 0.82 to account for the main-sequence contamination, which may have contributed 
about 18\% (Section\,4) of the measured value, the surface stellar mass density at the solar radius becomes 
35.3$\pm$0.4\,$M_\odot/{\rm pc}^2$.
Considering that brown dwarfs may contribute another 1.5$\pm$0.3\,$M_\odot/{\rm pc}^2$ \citep{Flynn2006, McKee2015},
the total surface mass density of stellar objects and remnants is then found to be 36.8$\pm$0.5\,$M_\odot/{\rm pc}^2$.
Based on the nature of {\rm IMF}, it is found that $\sim$5\% (1.8\,$M_\odot/{\rm pc}^2$) of the surface density is in 
neutron stars and black holes, and $\sim$12\% (4.4\,$M_\odot/{\rm pc}^2$)  is in white dwarfs, 
and $\sim$79\% (29.1\,$M_\odot/{\rm pc}^2$) in the visible stars, and the remaining 4\% is in brown dwarfs. 
Our results are consistent with previous estimates based on star count method (see Table\,4),
except for that of \citet{Mackereth2017}, who report much smaller value, but note that they also report 
large systematic error due to possible systematic errors in surface gravity of their sample stars. 

Finally, we note that the sum of individual mono-age populations yields a surface mass density
of 2.8\,$M_\odot$/pc$^2$ lower than that of the 1--14\,Gyr population. Although the reason for this 
discrepancy is not fully understood, we believe it is mainly caused by the relatively large uncertainties 
of the density measurements for mono-age populations. 
Since we divide the distance bins for density measurement population by population, it is not 
surprising that the sum of mono-age populations yields sightly different mass density 
to that of the overall population. At the solar radius, the density determination is quite 
complex because many of the distance bins are located at the 
near-side boundary of the complete volume. In addition, within our selected volume of $7.8<R<8.2$\,kpc, $|Z|<50$\,pc, the 
underlying stellar density may also exhibit moderate spatial variations, and it is possible that 
the 1--14\,Gyr population actually probes the relatively high density region. 
Anyway, such a difference is not found to make a big impact on the main conclusions of this paper. 
We expect that the Gaia data will provide more insights to this discrepancy since it 
provides accurate stellar parameters for much brighter stars thus we may obtain 
improved complete volume at the solar-neighborhood. 
Note that beyond the solar radius ($R>8.0$\,kpc), where the sample stars have a good spatial coverage 
at the disk mid-plane, the sum of mono-age populations is found to yield surface 
mass density in very good agreement with that of the overall population. 

\subsection{The vertical stellar density distribution}

\begin{figure}
\centering
\includegraphics[width=85mm]{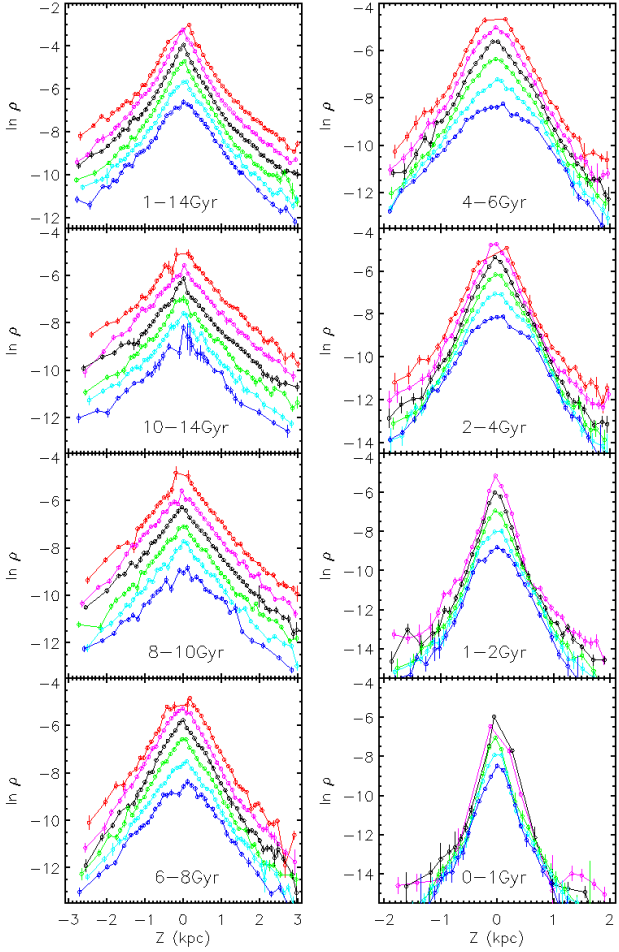}
\caption{Vertical stellar mass distribution in radial slices of 0.4\,kpc width. Different panels are 
results for populations of different ages, as marked on the figure. In each panel, from the red 
to blue are results respectively for $R$ = 7.5, 8.0, 8.5, 9.0, 9.5, 10.0\,kpc. }
\label{Fig20}
\end{figure}

\begin{figure*}
\centering
\includegraphics[width=160mm]{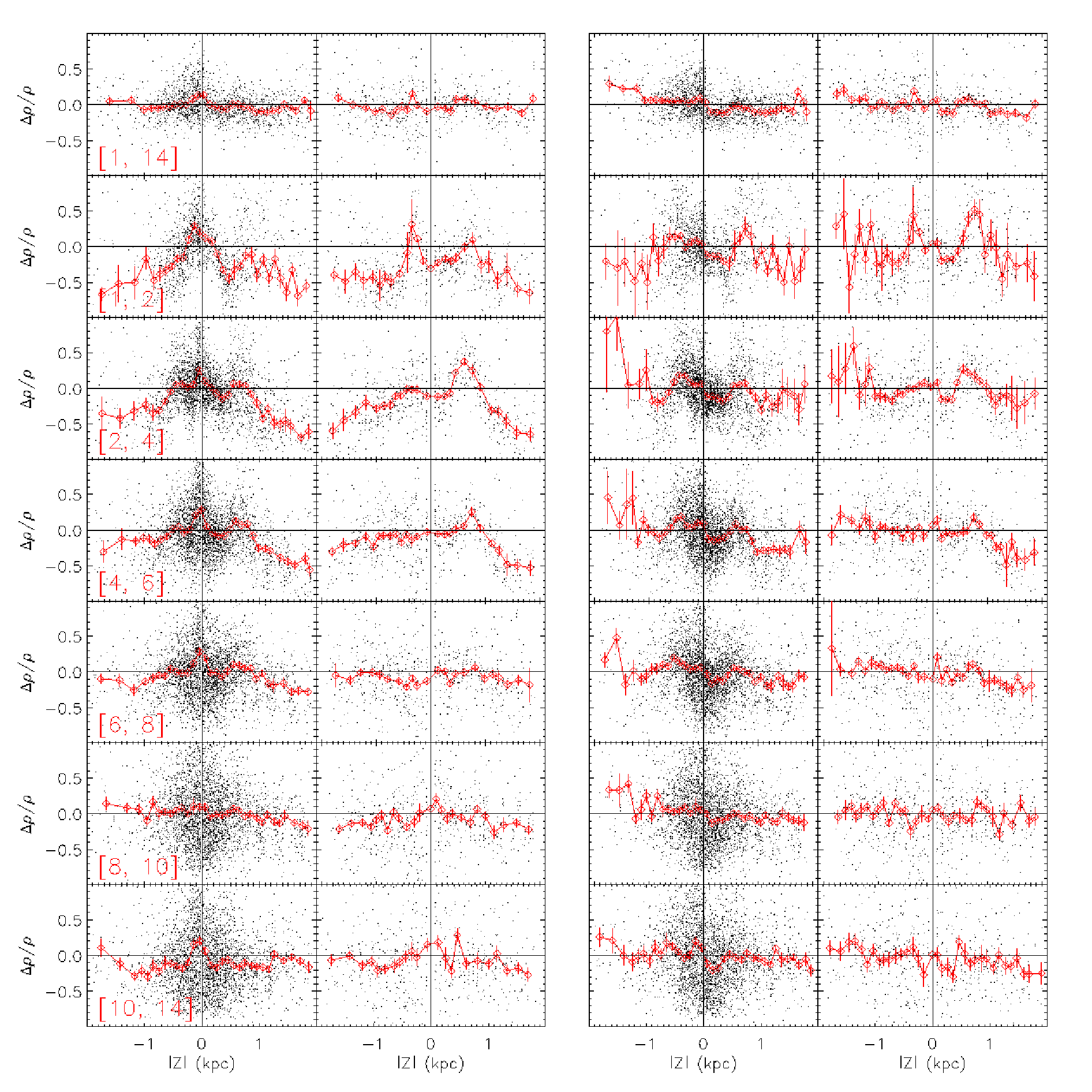}
\caption{{\em Left}: Relative residuals of stellar mass density  as a function height for two radius slices, 
$R=8.5$\,kpc and $R=9.5$\,kpc. The residuals are derived by subtracting the global fits in the disk $R$-$Z$ 
plane. The black dots are individual measurements, and the red squares with error bars are mean 
and stand error of the indvidual measurements in vertical bins; {\em Right}: Same as the left, but derived by subtracting local 
fits to the vertical mass density distribution.}
\label{Fig21}
\end{figure*}

A global fitting of the stellar density distribution in the disk $R$--$Z$ has the advantage,
in addition to derive the global structures, to reveal substructures with their true
strengths/amplitudes. While a disadvantage of the global fitting is that it can not accurately
describe the real vertical mass distribution at different radii. Here we further characterize the 
vertical mass distribution in different radial slices with 0.4\,kpc width. 

Fig.\,20 plots the vertical stellar mass distribution of different populations and at different 
radii. It shows clearly that the vertical profile become thicker with increasing age. For young 
and intermediate age populations, it is also clear that the vertical profile becomes more dumpy 
at the outer disk. The dumpy profiles clearly cannot be described with exponential functions, 
which show sharp profiles in the disk mid-plane. 
The oldest populations show sharp profiles at all radii, and the thickness does not present 
an obvious variation among those radii. This is actually why we obtain a small flaring 
strength with the global fitting in the disk $R$-$Z$ plane (Table\,3). 
However, because the old populations may have suffered serious contaminations from 
young, main-sequence stars, which may contribute a significant amount of density near 
the disk mid-plane, it is thus not clear if the sharp profiles of the old populations 
are intrinsic or just artifacts.   
While it seems quite clear that the flaring phenomenon for young and intermediate age populations 
goes parallel with a change of vertical density profile to more dumpy distribution. 
This must provide strong constrains on the origin mechanism of disk flaring. 
We suspect that such a phenomenon is possibly caused by either radial gas (star) accretion or merger events. 
For the population of 1--14\,Gyr, the sharp profiles are largely expected due to the 
superpositions of mono-age populations, which have density profiles with very different scale heights. 
Beyond the sharp and dumpy profiles, there are also visible asymmetries between the southern and 
northern part of the disk, which are especially prominent for the young and intermediate age 
populations. 

\begin{table*}
\centering
\caption{Fitting the vertical mass distribution with double $sech^n$ functions.}
\begin{tabular}{lllll}
& &  $7.8<R<8.2$\,kpc & &   \\
\end{tabular}
\begin{tabular}{lllllllll}
\hline
 Age  &  $\rho_1$   &   $\rho_2$ & $H_1$ (pc) &  $H_2$ (pc)  & $n_1$ & $n_2$ & $\chi^2_{\rm red}$ & $\Sigma_*$ \\
 \hline
  1-14  &  0.0617$\pm$0.0019  &  0.0045$\pm$0.0007  &   254$\pm$6  &   785$\pm$28  &  19.72$\pm$3.23  &   9.07$\pm$5.56  &   1.48  &   41.6$\pm$0.5  \\
   0-1  &  0.0062$\pm$0.0007  &  3.1028e-6$\pm$1.6112e-6  &    81$\pm$5  &   838$\pm$143  &   2.39$\pm$4.56  &   1.76$\pm$5.64  &   0.78  &    1.5$\pm$0.1  \\
   1-2  &  0.0096$\pm$0.0006  &  4.7183e-5$\pm$1.1913e-5  &   106$\pm$4  &   543$\pm$46  &   3.03$\pm$0.98  &  13.08$\pm$4.97  &   1.37  &    2.9$\pm$0.1  \\
   2-4  &  0.0164$\pm$0.0009  &  9.6580e-5$\pm$2.7012e-5   &   162$\pm$3  &   755$\pm$83  &   4.06$\pm$1.11  &  11.35$\pm$5.12  &   1.11  &    7.1$\pm$0.1  \\
   4-6  &  0.0102$\pm$0.0003  &  0.0003$\pm$8.5398e-5  &   206$\pm$7  &   719$\pm$87  &   1.51$\pm$0.20  &  18.20$\pm$5.06  &   1.08  &    7.7$\pm$0.1  \\
   6-8  &  0.0077$\pm$0.0004  &  0.0005$\pm$0.0002  &   298$\pm$17  &   815$\pm$126  &   2.02$\pm$0.44  &   9.11$\pm$4.92  &   1.26  &    8.1$\pm$0.1  \\ 
  8-10  &  0.0056$\pm$0.0004  &  0.0002$\pm$0.0002  &   469$\pm$36  &  1448$\pm$372  &  19.81$\pm$4.52  &   8.14$\pm$5.02  &   1.92  &    6.3$\pm$0.1  \\
 10-14  &  0.0048$\pm$0.0003  &  0.0003$\pm$0.0003  &   497$\pm$34  &  1407$\pm$203  &  19.97$\pm$4.22  &   8.22$\pm$4.85  &   1.24  &    6.1$\pm$0.1  \\
 \hline
\end{tabular}
\begin{tabular}{lllll}
& &  $8.3<R<8.7$\,kpc & &   \\
\end{tabular}
\begin{tabular}{lllllllll}
 \hline
  1-14  &  0.0477$\pm$0.0009  &  0.0024$\pm$0.0005  &   272$\pm$5  &   855$\pm$39  &  19.73$\pm$2.17  &   3.71$\pm$5.47  &   1.68  &   33.2$\pm$0.2 \\
   0-1  &  0.0081$\pm$0.0015  &  2.4691e-5$\pm$9.0011e-6  &    81$\pm$5  &   484$\pm$61  &   1.71$\pm$7.00  &  12.78$\pm$5.27  &   0.84  &    2.2$\pm$0.1  \\
   1-2  &  0.0070$\pm$0.0002  &  7.3355e-5$\pm$1.8169e-5  &   103$\pm$3  &   464$\pm$32  &   1.94$\pm$0.22  &   7.15$\pm$5.15  &   1.18  &    2.4$\pm$0.1   \\
   2-4  &  0.0129$\pm$0.0003  &  1.8466e-5$\pm$1.4687e-5  &   183$\pm$2  &   783$\pm$184  &   4.04$\pm$0.46  &   1.18$\pm$5.73  &   1.29  &    6.2$\pm$0.1  \\
   4-6  &  0.0097$\pm$0.0004  &  1.7907e-5$\pm$4.6076e-5  &   262$\pm$8  &   848$\pm$382  &   3.38$\pm$0.63  &   0.44$\pm$5.50  &   2.16  &    7.0$\pm$0.1  \\
   6-8  &  0.0079$\pm$0.0004  &  2.1899e-5$\pm$2.4046e-5  &   364$\pm$7  &  1695$\pm$280  &   6.75$\pm$4.58  &   0.84$\pm$5.54  &   1.78  &    7.0$\pm$0.1  \\
  8-10  &  0.0045$\pm$0.0003  &  0.0005$\pm$0.0002  &   417$\pm$24  &   894$\pm$190  &  19.70$\pm$2.75  &   3.52$\pm$6.41  &   1.40  &    5.2$\pm$0.1  \\
 10-14  &  0.0042$\pm$0.0003  &  0.0013$\pm$0.0003  &   259$\pm$33  &   840$\pm$51  &  18.77$\pm$3.44  &   3.60$\pm$4.84  &   1.35  &    5.1$\pm$0.1  \\
\hline
\end{tabular}
\begin{tabular}{lllll}
& &  $8.8<R<9.2$\,kpc & &   \\
\end{tabular}
\begin{tabular}{lllllllll}
 \hline
  1-14  &  0.0383$\pm$0.0008  &  0.0011$\pm$0.0003  &   308$\pm$5  &  1076$\pm$93  &  19.99$\pm$2.59  &   4.67$\pm$4.86  &   1.67  &   28.1$\pm$0.3  \\
   0-1  &  0.0044$\pm$0.0003  &  2.9377e-5$\pm$8.6060e-6  &    70$\pm$7  &   496$\pm$52  &   0.78$\pm$0.29  &   5.96$\pm$5.05  &   1.39  &    1.4$\pm$0.1  \\
   1-2  &  0.0046$\pm$0.0001  &  4.0161e-5$\pm$1.2839e-5  &   128$\pm$3  &   492$\pm$52  &   2.62$\pm$0.31  &   6.62$\pm$5.06  &   0.95  &    1.8$\pm$0.0   \\
   2-4  &  0.0091$\pm$0.0003  &  1.1131e-5$\pm$1.1531e-5  &   187$\pm$5  &  1067$\pm$318  &   1.89$\pm$0.24  &   1.39$\pm$5.82  &   1.48  &    5.5$\pm$0.1  \\
   4-6  &  0.0047$\pm$0.0004  &  0.0028$\pm$0.0001  &   150$\pm$16  &   384$\pm$319  &   0.48$\pm$0.33  &  18.93$\pm$5.26  &   1.87  &    6.2$\pm$0.1  \\
   6-8  &  0.0066$\pm$0.0005  &  1.2190e-5$\pm$2.1591e-5  &   386$\pm$9  &   837$\pm$273  &   8.32$\pm$5.09  &   0.04$\pm$5.89  &   1.61  &    6.0$\pm$0.1  \\
  8-10  &  0.0037$\pm$0.0002  &  0.0003$\pm$0.0002  &   431$\pm$27  &  1232$\pm$254  &  19.39$\pm$4.09  &   5.46$\pm$5.18  &   1.72  &    4.2$\pm$0.1  \\
 10-14  &  0.0039$\pm$0.0003  &  0.0004$\pm$0.0002  &   404$\pm$31  &  1329$\pm$220  &  19.45$\pm$4.25  &   6.69$\pm$5.12  &   1.17  &    4.4$\pm$0.1  \\
\hline
\end{tabular}
\begin{tabular}{lllll}
& &  $9.3<R<9.7$\,kpc & &   \\
\end{tabular}
\begin{tabular}{lllllllll}
 \hline
  1-14  &  0.0230$\pm$0.0014  &  0.0023$\pm$0.0006  &   300$\pm$18  &   787$\pm$100  &   4.22$\pm$2.73  &   3.97$\pm$5.92  &   1.93  &   22.6$\pm$0.3 \\
   0-1  &  0.0032$\pm$0.0001  &  3.9847e-5$\pm$1.0701e-5  &   106$\pm$3  &   414$\pm$30  &   2.41$\pm$0.52  &   7.36$\pm$4.75  &   0.68  &    1.1$\pm$0.1 \\
   1-2  &  0.0025$\pm$7.9729e-5  &  5.9241e-5$\pm$1.9654e-5  &   131$\pm$7  &   478$\pm$43  &   1.07$\pm$0.19  &  14.81$\pm$4.94  &   0.86  &    1.3$\pm$0.1 \\
   2-4  &  0.0061$\pm$0.0002  &  9.3732e-6$\pm$7.2514e-6  &   204$\pm$6  &  1940$\pm$343  &   1.36$\pm$0.16  &  11.94$\pm$6.13  &   1.10  &    4.5$\pm$0.1 \\
   4-6  &  0.0029$\pm$0.0004  &  0.0019$\pm$0.0005  &   144$\pm$22  &   442$\pm$84  &   0.29$\pm$0.25  &  18.84$\pm$4.93  &   1.22  &    4.7$\pm$0.1 \\
   6-8  &  0.0038$\pm$0.0004  &  0.0001$\pm$0.0001  &   415$\pm$23  &  1060$\pm$390  &   4.15$\pm$5.35  &   4.85$\pm$5.58  &   1.17  &    4.4$\pm$0.1 \\
  8-10  &  0.0029$\pm$0.0002  &  0.0002$\pm$0.0002  &   455$\pm$38  &  1098$\pm$196  &  19.79$\pm$3.71  &   2.46$\pm$5.11  &   1.35  &    3.4$\pm$0.1  \\
 10-14  &  0.0026$\pm$0.0003  &  0.0012$\pm$0.0003  &   244$\pm$44  &   760$\pm$48  &   9.10$\pm$5.11  &   4.39$\pm$4.82  &   1.11  &    3.8$\pm$0.1 \\
\hline
\end{tabular}
\begin{tabular}{lllll}
& &  $9.8<R<10.2$\,kpc & &   \\
\end{tabular}
\begin{tabular}{lllllllll}
 \hline
  1-14  &  0.0102$\pm$0.0022  &  0.0050$\pm$0.0005  &   230$\pm$22  &   681$\pm$141  &   0.90$\pm$6.05  &  14.20$\pm$6.17  &   2.21  &   17.3$\pm$0.6 \\
   0-1  &  0.0024$\pm$9.9234e-5  &  5.8628e-5$\pm$2.0383e-5  &   109$\pm$6  &   402$\pm$45  &   1.42$\pm$0.30  &  19.93$\pm$5.37  &   0.70  &    1.0$\pm$0.1  \\
   1-2  &  0.0017$\pm$8.6304e-5  &  6.9884e-5$\pm$2.0785e-5  &   105$\pm$16  &   468$\pm$61  &   0.38$\pm$0.19  &  19.24$\pm$5.34  &   1.02  &    1.1$\pm$0.1  \\
   2-4  &  0.0035$\pm$0.0003  &  8.6746e-5$\pm$0.0002  &   191$\pm$36  &   759$\pm$252  &   0.63$\pm$0.28  &  18.75$\pm$5.28  &   1.97  &    3.4$\pm$0.1  \\
   4-6  &  0.0028$\pm$0.0002  &  0.0002$\pm$0.0002  &   270$\pm$33  &   786$\pm$221  &   0.80$\pm$0.25  &  17.04$\pm$5.36  &   1.32  &    3.6$\pm$0.1  \\
   6-8  &  0.0011$\pm$0.0008  &  0.0013$\pm$0.0008  &   286$\pm$150  &   585$\pm$146  &   0.85$\pm$4.65  &   3.57$\pm$5.97  &   1.33  &    3.4$\pm$0.1  \\
  8-10  &  0.0005$\pm$0.0005  &  0.0011$\pm$0.0005  &   384$\pm$171  &   722$\pm$122  &   4.15$\pm$5.37  &   6.07$\pm$5.05  &   1.29  &    2.4$\pm$0.1  \\
 10-14  &  0.0023$\pm$0.0003  &  0.0006$\pm$0.0003  &   335$\pm$66  &   941$\pm$95  &  19.96$\pm$4.75  &   3.40$\pm$6.11  &   1.04  &    3.1$\pm$0.2  \\
\hline
\end{tabular}
\begin{tabular}{lllll}
& &  $10.3<R<10.7$\,kpc & &   \\
\end{tabular}
\begin{tabular}{lllllllll}
 \hline
  1-14  &  0.0084$\pm$0.0015  &  0.0024$\pm$0.0011  &   155$\pm$77  &   752$\pm$361  &   0.27$\pm$0.87  &   2.69$\pm$5.76  &   1.90  &   14.4$\pm$0.6  \\
   0-1  &  0.0019$\pm$8.2620e-5  &  4.7278e-5$\pm$1.3677e-5  &   129$\pm$7  &   439$\pm$30  &   1.29$\pm$0.25  &  17.08$\pm$5.14  &   0.69  &    0.9$\pm$0.1  \\
   1-2  &  0.0011$\pm$8.7787e-5  &  6.0059e-5$\pm$3.1413e-5  &    58$\pm$21  &   483$\pm$122  &   0.07$\pm$0.13  &   7.31$\pm$5.31  &   1.17  &    1.0$\pm$0.1   \\
   2-4  &  0.0021$\pm$0.0001  &  0.0002$\pm$0.0001  &    79$\pm$26  &   593$\pm$122  &   0.07$\pm$0.08  &   7.26$\pm$5.26  &   1.19  &    2.6$\pm$0.1  \\
   4-6  &  0.0016$\pm$0.0002  &  0.0003$\pm$0.0001  &   225$\pm$39  &   635$\pm$330  &   0.38$\pm$0.23  &   2.03$\pm$5.73  &   1.35  &    2.7$\pm$0.1  \\
   6-8  &  0.0011$\pm$0.0003  &  0.0006$\pm$0.0001  &   263$\pm$49  &   703$\pm$435  &   0.57$\pm$2.56  &   2.78$\pm$5.66  &   1.24  &    2.8$\pm$0.2  \\
  8-10  &  0.0005$\pm$0.0004  &  0.0008$\pm$0.0004  &   269$\pm$460  &   786$\pm$136  &   0.95$\pm$5.30  &   5.56$\pm$5.27  &   1.49  &    2.0$\pm$0.1  \\
 10-14  &  0.0012$\pm$0.0003  &  0.0002$\pm$0.0003  &   495$\pm$110  &  1437$\pm$275  &   2.23$\pm$5.57  &  11.95$\pm$5.14  &   1.20  &    2.3$\pm$0.2  \\
\hline
\end{tabular}
\end{table*}

We fit the vertical mass distribution in each radial slice using a double $sech^n$ 
function. Results of the fitting are presented in Table\,5. For comparison, results from fitting 
with a double exponential function are also presented in the Appendix.  
At the solar radius, scale heights of the thin disk component are found to increase from 
80 to 300\,pc as the age increases from 0--1 to 6--8\,Gyr, and become $\sim$500\,pc 
for the old populations of 8--10 and 10--14\,Gyr. However, as emphasized above, 
scale heights of these old populations may have suffered large systematic errors  
due to contaminations from the young, main sequence stars. 
The 1--14\,Gyr population has a scale height of 254$\pm$6 and 785$\pm$28 for the 
thin and thick disk, respectively. These values are slightly smaller than the global 
fitting (Table\,3).
The young and intermediate age populations have a $sech^n$ index value of about 
1.5 -- 4.0 for the thin disk, which means that their profiles are between the isothermal ($n=1$) and 
exponential ($n=\infty$), while the old populations and the 1--14\,Gyr population 
have a large index, which means that their profiles are close to exponentials. 
The thick disk component has a relatively large index in general, but the fitted values 
have large error bars.  

The derived scale heights are not always increasing with Galactocentric distance. For example, 
at $R=9.5$\,kpc the 4--6\,Gyr population has a scale height of only $144\pm22$\,pc, much smaller than 
the value 262$\pm$8\,pc at $R=8.5$\,kpc. At $R=10.5$\,kpc, the 1--2 and 2--4\,Gyr populations 
exhibit also very small scale heights. These decreases of scale heights are always happened 
with a significant decrease of the $sech^n$ index. We believe this is not due to degeneracy, but because of 
a significant change of the vertical profiles to much more dumpy distribution.
For the intermediate age population at $R\gtrsim10$\,kpc, the $sech^n$ index
usually have a value significantly smaller than 1, indicating that their vertical mass 
profiles are more dumpy than the isothermal distribution. 

Fig.\,21 plots the residuals of the fitting for the $R=8.5$ and $R=9.5$\,kpc slices. 
Also shown in Fig.\,21 are the residuals for the global fitting in the disk $R$--$Z$ plane. 
The figures show clear wave-like oscillations in the vertical mass 
distribution for almost all populations at $R=8.5$\,kpc. The oscillation has an amplitude
of $\sim$20\,per cent for the 1--14\,Gyr population, while the amplitudes for young populations 
reach $\sim$40\,per cent. Patterns of the vertical oscillations for relatively young populations 
are consistent well with that found by \citet{Widrow2012}.  
 \citet{Widrow2012} found dips at $Z\sim-1000$\,pc, $Z\sim400$\,pc and $Z\sim1200$\,pc, 
 and peaks at $Z\sim-400$\,pc and $Z\sim800$\,pc. All of these dips and peaks are exactly 
 matched with our results for young populations. At $R=9.5$\,kpc, the asymmetric features 
 for young populations are quite strong, and the most prominent features are caused by the over density stripes 
in both the northern and the southern disk. While oscillations of the old populations become 
very weak. However, given the 
large scatters of individual measurements, it is possible that any potentially 
intrinsic oscillation patterns have been smeared out artificially. 
Compared to the local fitting, the global fitting yields different residual profiles to some extent. 
The main difference is that the global fitting yields over-density in the disk mid-plane at $R=8.5$\,kpc, 
which is actually contributed by the Local stellar arm. 
As for the origin of these oscillations, it is suggested that they can be caused by external perturbations 
of dwarf galaxies \citep[e.g.][]{Widrow2012, Gomez2013}. 
It is even suggested that interaction with satellite galaxies or halo substructures can provoke 
the formation of spiral arms and bars \citep[e.g.][]{Gauthier2006, Purcell2011}.

Finally, we note that although there may exist strong degeneracy among different parameters, 
which induce large uncertainties to the structural parameters, the derived surface mass 
densities are however, largely free from degeneracy.

\subsection{Surface mass density as a function of radius}
\begin{figure}
\centering
\includegraphics[width=85mm]{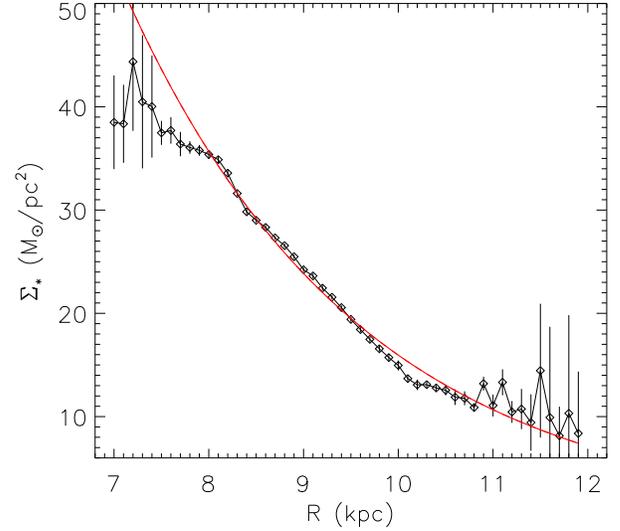}
\caption{Radial distributions of surface stellar mass density. The red curve is an exponential fit 
to the measurements. The exponential function has a scale length of $2.48\pm0.05$\,kpc, and a surface 
mass density of $35.7\pm0.3$\,$M_{\odot}$/pc$^2$ at the solar radius.}
\label{Fig22}
\end{figure}
 Fig.\,22 plots the surface stellar mass density (integrated to 4\,kpc above the disk mid-plane) 
 of the whole stellar population of 0--14\,Gyr as a function of Galactocentric radius. 
Here the effect of main sequence contamination has been corrected by multiplying 
a factor of 0.82 to the derived values. 
The figure shows a fast decreasing of surface stellar mass density with increasing radius, 
and the trend is well described by an exponential profile with scale length of
$2.48\pm0.05$\,kpc and surface density of $35.7\pm0.3$\,$M_{\odot}$/pc$^2$ at the 
solar radius ($R=8$\,kpc). 
Assuming the brown dwarf contribute another 1.5\,$M_{\odot}$/pc$^2$ at the solar radius, 
the exponential profile yields a disk total stellar mass of $3.6(\pm0.1)\times10^{10}$\,$M_\odot$. 
The value becomes $3.2(\pm0.1)\times10^{10}$\,$M_\odot$ if the Chabrier {\rm IMF} is adopted. 
The derived disk stellar mass are slightly lower than previous estimates with dynamic methods, 
which suggest a value of 3.7 -- 9.5$\times10^{10}$\,M$_{\odot}$ \citep{Bovy2013, Kafle2014, 
Licquia2015, Huang+2016, McMillan2017}.
This is partly caused by the different positions of the Sun adopted by different work. 
If we assume the Sun is located at $R=8.3$\,kpc, as adopted by \citet{Huang+2016}, 
we obtain a disk stellar mass of $4.1(\pm0.1)\times10^{10}$\,$M_\odot$ and 
$3.7(\pm0.1)\times10^{10}$\,$M_\odot$ when the Kroupa {\rm IMF} 
and the Chabrier {\rm IMF} are adopted, respectively.
Beyond the overall trend, the measurements exhibit also significant deviations 
from the exponential fit. The deviations present wave-like features, with an under density 
at $R\sim7.5$\,kpc and $R\sim10.2$\,kpc, an over-density at $R\sim8.2$\,kpc, $R\sim9.2$\,kpc 
and $R\gtrsim11.0$\,kpc. 
These features are likely contributed by the asymmetric structures shown in Section\,5.1 and 
Section\,5.2, and are probably results of disk oscillations. 
Note that for the innermost region ($R\sim7$\,kpc), the large deviations from the exponential 
fit are likely due to an underestimate of the surface mass density as a consequence of poor spatial 
coverage of the sample stars near the disk mid-plane.  

Fig.\,23 plots the radial distribution of the surface mass density for mono-age populations. 
It shows clear wave-like oscillations. Amplitudes of the oscillations are 1--2\,$M_\odot$/pc$^2$ 
for intermediate age populations, while become smaller ($\lesssim$1\,$M_\odot$/pc$^2$) for old ($>8$\,Gyr) populations.  
For intermediate age and old populations, a peak of the surface density oscillations occurs 
at $R\sim9$\,kpc, while a dip occurs at $R\sim10.5$\,kpc. 
For stellar populations younger than 2\,Gyr, the most prominent features are mass excess 
at $R\sim8.2$\,kpc, likely contributed by the Local arm. 
Such an evolution of oscillations features with age must provide crucial insights on the 
disk perturbation history. It is possibly that a perturbation event that cause the oscillations 
happened at 2--4\,Gyr ago, and stars younger than 2--4\,Gyr were formed after the perturbation, 
probably from gases suppressed (thus changed the position) by the perturbation.	 
Note however that there are some caveats for measurements in the inner disk ($R<8$\,kpc). 
As has been emphasized, because the poor spatial coverage of the sample stars near the 
disk mid-plane, the measurements at this region, especially for the young populations, 
may suffer large systematics. In most cases, extrapolation from the larger heights of the disk 
underestimates the surface mass density, but for a few cases where the fitting are failed via 
yielding unphysical (small) scale heights and (large) mid-plane density, the surface mass density 
are overestimated, as presented at $\sim$7.3\,kpc for the 1--2\,Gyr population.

Table\,6 presents the results of fits to the radial surface mass density distribution with an exponential function\footnote{$\Sigma=\Sigma_{R_\odot}e^{-(R-R_\odot)/L}$}. 
The younger populations have generally larger scale length, which increases from 2.23$\pm$0.06\,kpc 
for the 8--10\,Gyr population to 6.61$\pm$1.30\,kpc for the 0--1\,Gyr populations. While the populations of 
1--2\,Gyr and 10--14\,Gyr are exceptions. The 1--2\,Gyr population has a rather small 
scale length of 2.14\,kpc, while the 10--14\,Gyr population has a significantly larger scale length than 
the 8--10\,Gyr population. Note however that as the radial coverage of sample stars is rather limited, 
the fitting can be easily affected by the oscillation features as well as incomplete spatial 
coverage of the data in the inner disk. 
For the 0--1\,Gyr population, the very large scale length is likely an effect of the incomplete 
spatial coverage in the inner disk. If only measurements of $R>8$\,kpc are adopted, we obtain a disk scale 
length of 4.11$\pm$0.42\,kpc. 
For the 1--2\,Gyr population, the fitting has likely overestimated the background value 
at $R=8$\,kpc, as one expects that a significant part of the surface mass density is contributed by 
the Local stellar arm (Fig.\,17). For the 10--14\,Gyr, contaminations from young, main-sequence 
stars may have also a big impact on the derived parameters. Main sequence contaminations 
will cause an overestimate of the scale length if the young populations have larger scale length. 
Unexpectedly, we do not see a strong feature of the Perseus arm at the expected 
position ($\sim$11\,kpc) for the young populations. 
A possible explanation is that the Perseus arm itself has small ($<1$\,$M_\odot$/pc$^2$) 
surface density of young stars, and at the same time, it covers a wide range of $R$, from $\sim$10\,kpc to $>12$\,kpc, so that 
although it contributes the results, it looks not obvious given the small radial coverage of our sample stars. 

 \citet{Amores2017} found a contiguous increasing trend of scale length with time, from about 
 2.3\,kpc for the old ($\sim$8\,Gyr) to 3.9\,kpc for the young ($<$0.5\,Gyr) populations. In general, 
 our results show a trend consistent well with their's. Our values of scale length are also consistent well 
 with their's for populations of 0--1, 6--8 and 8--10\,Gyr. For the 2--4 and 4--6\,Gyr populations, 
 we obtain slightly larger scale length, although the differences are within the error bars. 
Note that their results are derived from the 2MASS photometric data with $80^\circ\lesssim l \lesssim280^\circ$ 
and $|b|<5.5^\circ$ only. Contribution from the flaring disk at larger heights to the surface mass density 
may yield larger scale length. 
From the radial distribution of surface number density of LAMOST red clump (RC) stars, 
\citet{Wan2017} found a scale length of 4.7$\pm$0.5\,kpc and 3.4$\pm$0.2\,kpc for the young (2.7\,Gyr) 
and old (4.6\,Gyr) RC populations, respectively. These values, especially for the younger population, 
are significantly larger than both ours and \citet{Amores2017} by 2--3$\sigma$. It is likely that these  
differences are largely consequences of different radial coverage of the sample stars, as their sample 
cover a Galactocentric distance of 9--13.5\,kpc. As has been discussed, the derivation of scale length 
are sensitive to the radial coverage of the data because the existence of radial oscillations in the 
stellar mass density.

A non-monotonic radial surface stellar density profile was declared recently by \citet{Bovy2016} and 
\citet{Mackereth2017} using red giant branch and red clump stars from the APOGEE survey. 
\citet{Bovy2016} show that radial profiles of surface mass density for mono-abundance populations 
can be described with broken exponentials, with peak radii change from $\sim$6 to $\sim$11\,kpc as 
[Fe/H] decreases from 0.3 to $-0.6$\,dex for low-alpha populations (see their Fig.\,11). 
\citet{Mackereth2017} show further that break radii of the surface mass density change from 
$\sim$8\,kpc for young population to 12\,kpc for old populations (see their Fig.\,13). Moreover, 
their results show that surface density of the old populations exhibits another break at $R\sim6$\,kpc.  
Our sample stars have smaller radial coverage, so that it is difficult to make a direction 
comparison with \citet{Bovy2016} and \citet{Mackereth2017}. While it is still possible to  
make a comparison for the young populations of $\lesssim$5\,Gyr, as \citet{Mackereth2017} 
show a break radius at $R\sim8$\,kpc for the $<3$\,Gyr population and at $R\sim10$\,kpc 
for the 3--5\,Gyr population, for which the break radius are well within the radial coverage of 
our data. Our results show no clear break at $R\sim10$\,kpc for the 2--4\,Gyr 
and 4--6\,Gyr populations. For the youngest populations, our results show a peak at $R\sim8.2$\,kpc 
due to the Local arm, whereas we believe the sudden drop of surface density at the inner disk 
are fake features due to poor spatial completeness of the sample stars near the mid-plane of the inner disk. 
We do not expect a significant, continue decreasing or flattening of the surface mass density 
in the inner disk of $R<8$\,kpc, although it is indeed possible that there is a local peak caused 
by the Local arm. We therefore tend to believe that the break exponentials,  
shown by \citet{Bovy2016} and \citet{Mackereth2017}, at least some of them, are probably 
artifacts due to either incompleteness of the data or their method to explain the data. 
We emphasize that determinations of disk profiles are easily affected by asymmetric structures. 
It is possible to explain the oscillation structures as a `broken' radial profile if the underlying density 
distribution are not well characterized due to, for instance, in completeness of the data 
or strong presumptions about the density profiles. 
Anyway, although the current work focus on the mono-age populations as a whole, it is 
interesting to have a further examination on the structure of mono-age and mono-abundance 
populations, as \citet{Mackereth2017} done, utilizing this larger database as well as the coming Gaia DR2.  

\begin{figure}
\centering
\includegraphics[width=85mm]{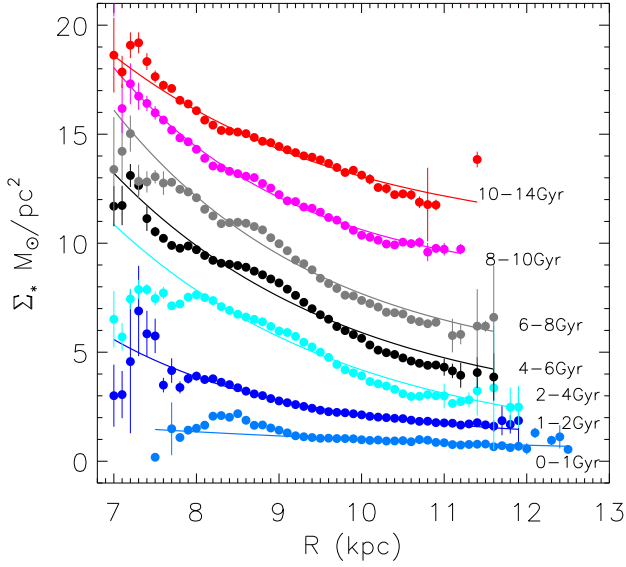}
\caption{Radial distributions of surface stellar mass density for mono-age populations. The solid 
curves are fits to the measurements with exponential function. The profiles have been shifted 
arbitrarily in the y-axis.}
\label{Fig23}
\end{figure}
\begin{table}
\caption{Fitting the radial surface stellar density distribution with exponential function.}
\centering
\label{}
\begin{tabular}{cccc}
\hline
 Age (Gyr)  &  $\Sigma_{R_\odot}^{\rm fit}$ ($M_{\odot}$/pc$^2$)    &  $L$ (kpc)   &    $\chi^2_{\rm red}$  \\
\hline                                                                                                                                                
 0--14      &   $35.7\pm0.3$$^a$ & $2.48\pm0.05$ & 2.27  \\
 0--1        &   $1.35\pm0.07$   & $6.61\pm1.30$  &  2.30 \\
 0--1$^b$    &   $1.62\pm0.07$   & $4.11\pm0.42$  &  1.32 \\
 1--2        &    $2.87\pm0.04$ & $2.14\pm0.05$ & 1.48  \\
 2--4        &    $7.34\pm0.11$ & $2.90\pm0.13$ & 2.23 \\
 4--6        &   $7.89\pm0.12$  & $2.84\pm0.13$ &  2.06 \\
 6--8        &    $8.14\pm0.11$ & $2.52\pm0.09$ & 1.65 \\
 8--10      &    $6.42\pm0.07$ & $2.23\pm0.06$ & 1.12 \\
 10--14    &    $6.09\pm0.07$  & $2.90\pm0.11$ & 0.97 \\
 \hline
\end{tabular}
\begin{tablenotes}
\item[1] $a$:  main-sequence contaminations have been corrected. 
\item[2] $b$: only measurements at $R>8$\,kpc are adopted for the fitting. 
\end{tablenotes}
\end{table}

\subsection{Star formation history of the disk}
\begin{figure}
\centering
\includegraphics[width=85mm]{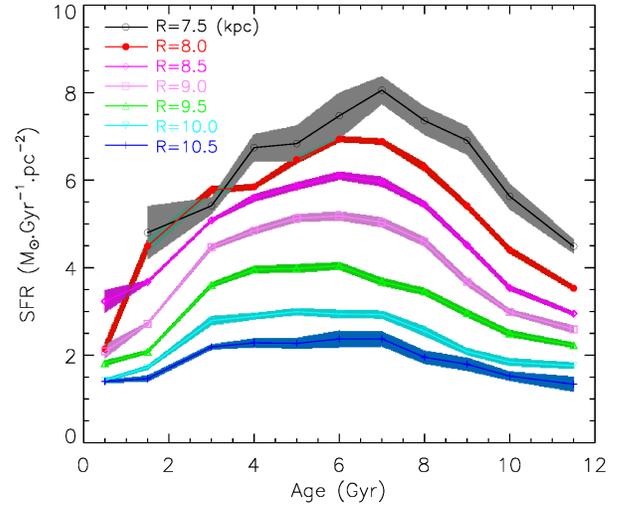}
\caption{The disk star formation history at different radii. The shadow regions indicate the 1$\sigma$ errors.}
\label{Fig24}
\end{figure}
Given the surface mass density for mono-age populations, the star formation rate ({\rm SFR}) can be
derived directly by
\begin{equation}
{\rm SFR}=\frac{\Sigma_{\rm ini}^*}{1000\times\Delta t}, 
\end{equation}
where $\Sigma_{\rm ini}^*$ is the initial stellar mass formed in a given time span $\Delta t$.
Because the accuracy of the current age estimates is not good enough to settle when
was the first disk star started to form, we simply assume the disk started to form
at 13\,Gyr ago, so that $\Delta t$ for the oldest age bin is 3.0\,Gyr.
Fig.\,24 plots the disk {\rm SFR} as a function of age at different Galactocentric annuli.
The figure shows that at the disk of $R\lesssim8.5$\,kpc, the {\rm SFR} exhibits
a peak at 6--8\,Gyr ago, and the {\rm SFR} decreases with time at younger ages. 
While the peak  {\rm SFR} shifts to about 4--6\,Gyr at the outer disk of $R\gtrsim9.0$\,kpc, 
and there is a plateau in the age range 3--7\,Gyr at $R\gtrsim10$\,kpc. 
Such a trend is consistent with the concept of an inside-out galaxy assemblage history. 
Below 3\,Gyr, the {\rm SFR} shows a steep decreasing trend with time at almost all radii, 
probably indicating that the disk may have been largely quenched from 3\,Gyr ago.   
At the solar radius, the 0--1\,Gyr population yields a recent {\rm SFR} of $2.14\pm0.15$\,$M_{\odot}$/pc$^2$/Gyr, 
about a half of that from the 1--2\,Gyr population. While it seems that such a low {\rm SFR} 
from the 0--1\,Gyr population has been underestimated, likely because of the poor spatial 
completeness of the sample stars at the solar radius. Also, as shown in Section\,4, a systematic 
overestimate of stellar ages for the very young stars may induce an underestimate (by $\sim$20\%) 
of the stellar mass formed in this young age bin. 
At the inner disk ($R\leq7.5$\,kpc), the {\rm SFR} for the youngest population is found to 
drop to a value close to zero because of the poor spatial coverage of the sample stars
near the disk mid-plane, and we have thus omitted that point in the figure. 
Note that no corrections for main-sequence contaminations have been implemented, so that the 
underlying {\rm SFR} of the disk at the early epochs must be lower than the current estimates 
derived from the old populations (Section\,4). 

We fit the radial {\rm SFR} profile with an exponential function\footnote{$\psi = \psi_{R_\odot}e^{-(R-R_\odot)/L}$}, 
and derive the disk total {\rm SFR} by integrating the function. The derived results are presented in Table\,7.
The 0--1\,Gyr population yields a recent total {\rm SFR} of 1.96$\pm$0.12\,$M_{\odot}$/yr across the whole disk, 
which has a scale length of 3.65$\pm$0.25\,kpc derived from measurements of $R>8$\,kpc. 
Note that here we have not considered contributions of the brown dwarfs when deriving the initial 
stellar mass, so that the {\rm SFR} may has been underestimated by a few per cent.
The value is in very good agreement with literature estimates for the present 
{\rm SFR} of the Milky Way, which have a typical value of 1.9$\pm$0.4\,$M_{\odot}$/yr \citep{Chomiuk2011}. 
The 1--2\,Gyr population yields however a very large total {\rm SFR} of 5.69$\pm$0.31\,$M_{\odot}$/yr, 
likely an artifact caused by incorrect scale length, which has been significantly underestimated due to 
probably effects of the Local stellar arm. 
Although the total disk {\rm SFR} may be largely uncertain due to uncertainty of the derived disk scale length,
we expect the {\rm SFR} in $7.5<R<11.5$\,kpc are much better determined. 
The {\rm SFR} in $7.5<R<11.5$\,kpc exhibits an increasing trend with time at the early epoch, 
reaching a peak value at 4--6\,Gyr ago, and then decrease with time. This is consistent with
the downsizing trend of galactic star formation history derived from extra-galaxies \citep{Heavens2004}. 

\citet{Snaith2015} derived the disk star formation history using the age-[Si/Fe] relation for a sample 
of nearby stars with high resolution spectroscopy. They found a rather high {\rm SFR} at the early 
epoch, which has produced about half of the total disk mass, and they also suggested the {\rm SFR} has 
a dip at 8--9\,Gyr ago. Our results obviously do not support their conclusions. There could be two 
major reasons to explain the conflicts, both may have contributed a significant part. One is that the underlying scale lengths 
of the old, high-[$\alpha$/Fe] disk is much smaller than the current estimates from the 10--14\,Gyr population. 
It is suggested that the high-[$\alpha$/Fe] stellar populations have scale length of $\sim$2\,kpc 
\citep{Bovy2012, Bovy2016, Mackereth2017}. In fact, we have checked our data, and find even 
smaller scale length of $\sim$1.5\,kpc for the old disk with ${\rm [\alpha/Fe]} > 0.15$. At the same time, 
the surface mass density at solar radius also reduces to be about a half the current estimates. 
This will increase the total {\rm SFR} of the oldest disk by almost a factor of 2. The other explanation 
is that \citet{Snaith2015} may have underestimated the {\rm SFR} of their thin disk populations, 
as they utilized a small and incomplete sample of stars as well as simple chemical models. A further, more detailed 
study using complete stellar samples and more realistic chemical models is certainly necessary 
to better constrain the disk {\rm SFH} with the chemical modeling approach.   

\begin{table*}
\caption{Star formation history of the Galactic disk.}
\centering
\label{}
\begin{tabular}{ccccccc}
\hline
 Age   & $\psi_{R_{\odot}}$  & $\psi_{R_{\odot}}^{\rm fit}$   &  $L$ &  $\psi_{7.5<R<11.5{\rm kpc}}$ & $\psi_{\rm tot}$  & $\chi_{\rm red}^2$  \\
 (Gyr)  &  ($M_{\odot}$/pc$^2$/Gyr)  & ($M_{\odot}$/pc$^2$/Gyr) & (kpc)  &  ($M_{\odot}$/yr) & ($M_{\odot}$/yr) &  \\
\hline                                                                                                                                                
  0--1       &  $2.14\pm0.15$   &    $2.69\pm0.11$     &    $3.65\pm0.25$     &     $0.43\pm0.01$          &   $1.96\pm0.12$      & 1.06   \\
 1--2        &   $4.49\pm0.17$  &  $4.51\pm0.06$   & $2.09\pm0.05$ & $0.57\pm0.01$  & $5.69\pm0.31$ & 1.26  \\
 2--4        &   $5.77\pm0.09$    & $5.94\pm0.10$    & $2.97\pm0.14$ & $0.88\pm0.02$ & $4.82\pm0.22$ & 2.85  \\
 4--6        &  $6.47\pm0.09$     & $6.70\pm0.11$    & $2.71\pm0.12$ & $0.96\pm0.01$ & $5.88\pm0.31$ & 2.13 \\
 6--8        &  $6.88\pm0.07$   & $6.88\pm0.07$    & $2.53\pm0.08$  & $0.95\pm0.01$ & $6.51\pm0.28$  & 1.64 \\
 8--10     &   $5.41\pm0.08$  & $5.51\pm0.05$  &  $2.29\pm0.06$   & $0.73\pm0.01$ & $5.95\pm0.26$ & 1.20 \\
 10--14   &  $3.53\pm0.07$ &  $3.56\pm0.03$  & $2.90\pm0.09$  &  $0.52\pm0.01$ & $2.95\pm0.09 $   &1.09 \\
 \hline
\end{tabular}
\end{table*}

\section{Summary}
We have carried out an unprecedented measurement and analysis of 3D stellar mass density of the Galactic 
disk within a few kilo-parsec from the Sun using 0.93 million MSTO and subgiant stars with robust age estimates. 
Our results suggest that the disk is strongly flared in the $R$-$Z$ plane for stellar 
populations of all ages younger than 10\,Gyr. 
The global structure of the disk for all populations are approximately described by a double-component flared 
disk with exponential profiles in the radial direction and $sech^n$ profiles in the vertical direction. 
For the overall populations, the thin disk component has a scale length of $2216\pm30$\,pc, and  
a scale height of 265$\pm$2\,pc at solar radius and increases with Galactocentric distance 
with a slope of 0.178$\pm$0.005. The thick disk has a scale length of 1405$\pm$25\,pc, and a 
scale height of 920$\pm$8\,pc at solar radius and increases with Galactocentric distance 
with a slope of 0.123$\pm$0.004. All populations younger than 10\,Gyr have comparable 
strengths of disk flaring. 
If we impose a constant scale height at all radius, we find the thin disk has a scale length of 
$3677\pm57$\,pc and a scale height of $300\pm2$\,pc, and the thick disk has a scale length 
of $4457\pm80$\,pc and a scale height of $981\pm12$\,pc. Our results provide 
insights to understand the large scatters in disk structure parameters presented in literature. 
The global fitting also suggests that the Sun is at 10$\pm$1\,pc above the mass-weighted disk mid-plane. 
Whereas the value changes with stellar populations from $\sim$1\,pc for the youngest population to 
$\sim$30\,pc for the old populations. 

The global fitting also suggests that the vertical density distribution for young and intermediate age 
populations of the thin disk are best described by a $sech^n$ function with index of 1--5, 
which means that the vertical profiles are between the isothermal ($n=1$) and exponential ($n=\infty$) 
distribution. While the vertical density distribution of the old or thick disk populations need a large  
index value, suggesting they are well described by exponential profiles. 
A local characterization of the vertical density distribution further suggests that the vertical profiles 
may change significantly with Galactocentric distance. The vertical density profiles of young and intermediate 
age populations at the outer disk of $R\gtrsim9.5$\,kpc become rather dumpy, which need to be 
described by $sech^n$ function with an index value even as small as $\sim$0.1. 
These dumpy profiles in the vertical direction may have tight correlations with the disk flaring, 
which may provide strong constrains on the origin mechanism of disk flaring. 
We suspect such a phenomenon is probably caused by either radial gas/star accretion or merger events.  
 Although the mono-age populations may have dumpy vertical profiles, the superposition 
 of individual populations with different scale heights result a profile that can be well 
 approximated by an exponential function for the overall populations. 

Wave-like oscillation features are seen in both the radial and vertical direction. 
In the radial direction, the surface mass density exhibits wave-like distribution, 
which is particularly prominent for young and intermediate age population, with an amplitude of 
1--2\,$M_{\odot}$/pc$^2$, while the amplitude becomes weak ($<1$\,$M_{\odot}$) for old populations. 
Positions of the peak mass of the waves also vary with age. The intermediate to 
old populations show peak mass at $R\sim9$\,kpc, while the young populations 
show peak mass at $R\sim8$\,kpc. The mass oscillations are mainly contributed 
by in-plane structures, such as the Local stellar arm at $R\sim8$\,kpc 
and over-densities at $R\sim9$\,kpc in the anti-center direction. 
The Local stellar arm is a prominent 
structure for relatively young ($\lesssim4$\,Gyr) populations, and particularly strong in the second 
quadrant. The over-densities at $R\sim9$\,kpc for intermediate age and old populations 
may be not independent structures but have some intrinsic 
relation with the Local arm given their coherence in both position and age. 
It is possible that they are originated from the same perturbations.

In the vertical direction, the oscillations cause strong asymmetric mass 
distribution for young and intermediate populations. At $R\sim8.5$\,kpc, the wave-like 
patterns are consistent well with those found by \citet{Widrow2012}. Amplitudes of 
the oscillations are 10--20\% for the overall populations, while become 30--40\% for the 
young populations. The peak mass excess of the waves at the southern disk have generally 
larger value than that of the northern disk. At $9\lesssim R\lesssim12$\,kpc, on the contrary, 
the peak mass excess of the vertical oscillations at the northern disk has larger value 
than that of the southern disk, which is consistent with the findings of \citet{Xuyan2015}, 
who suggest that there are more stars at the northern disk about 2\,kpc away from the Sun 
in the anti-center direction. Our results show that the mass excesses at both the southern and 
the northern disk occur in the form of stripes in the $R$-$Z$ plane, which may provide 
further constrains on their origin.   

By averaging stellar mass density in $7.8<R<8.2$\,kpc and $|Z|<50$\,pc, we find a disk mid-plane 
stellar mass density of 0.0594$\pm$0.0008\,$M_\odot$/pc$^3$ at the solar radius when the 
Kroupa {\rm IMF} is used to convert the mass density of MSTO-SG stars to the mass density of stellar populations of all masses. 
Such a value is 0.0164\,$M_\odot$/pc$^3$ higher than previous estimates at the solar neighborhood. 
The over density is likely contributed by the Local stellar arm, while our Sun is probably located 
 in a local low density region respect to the Local stellar arm.  
Assuming a gas density of 0.05\,$M_\odot$/pc$^3$ as widely adopted,
the expected mass density of baryon matter (star and gas) in the nearby disk within a few
hundred parsec is thus 0.109\,$M_\odot$/pc$^3$. Such a baryon matter density is consistent 
well with the local total mass density yielded by local dynamic methods. 
Our results thus leave little room for the existence of a meaningful amount of dark matter 
in the nearby disk mid-plane.
However, since our results show that stellar mass distribution in the local disk is highly asymmetric 
and non-smooth, one needs further study to better understand how the estimation of local dark matter density  
has been affected by such asymmetries. 
The Chabrier {\rm IMF} yields stellar mass density of $\sim$10\% lower, 
which predict a disk mid-plane stellar mass density of 0.0536$\pm$0.0007\,$M_\odot$/pc$^3$, 
and a total baryon mass of 0.104\,$M_\odot$/pc$^3$. 

The surface stellar mass density at the solar radius is found to be 36.8$\pm$0.5\,$M_{\odot}$/pc$^2$, 
which is consistent with literature values. The radial distribution of surface mass density 
yields a disk scale length evolving from 4.11$\pm$0.42\,kpc for the 0--1\,Gyr to 2.23$\pm$0.06\,kpc 
for the 8--10\,Gyr populations. The overall population has a disk scale length of 2.48$\pm$0.05\,kpc, 
and a disk total stellar mass of $3.6(\pm0.1)\times10^{10}$\,$M_\odot$ assuming the Sun is located 
at 8.0\,kpc away from the Galactic center, and the value becomes $4.1(\pm0.1)\times10^{10}$\,$M_\odot$ 
if the Sun is located at 8.3\,kpc away from the Galactic center.

The current work leads to a direct measure of disk star formation history. The results show that 
the disk star formation rate exhibits a peak at 6--8\,Gyr ago in the inner disk of $R\sim7.5$\,kpc, 
and the epoch of peak star formation rate decreases to 4--6\,Gyr ago at the outer disk of $R\sim10$\,kpc. 
This is consistent with the concept of inside-out disk assemblage history. 
The recent disk total {\rm SFR} is found to be $1.96\pm0.12$\,$M_{\odot}$/yr, which is in good 
agreement with literature results using different methods \citep[e.g.][]{Chomiuk2011}. 

Future studies utilizing more precise stellar ages based on the Gaia parallax will certainly improve 
the current work by significantly reducing the main-sequence contaminations, and thus to better 
characterize the disk structure and stellar mass density.

\vspace{7mm} \noindent {\bf Acknowledgments}{
We acknowledge supports from the National Natural Science Foundation of China (Grant No. 11703035), 
the National Key Basic Research Program of China (Grant No. 2014CB845700)
and the Joint Funds of the National Natural Science Foundation of China (Grant No. U1531244 and U1331120).
H.-B. Yuan is also supported by NSFC grant No. 11443006,
No. 11603002 and Beijing Normal University grant No. 310232102.
The LAMOST FELLOWSHIP is supported by Special Funding for Advanced Users,
budgeted and administrated by Center for Astronomical Mega-Science,
Chinese Academy of Sciences (CAMS).

Guoshoujing Telescope (the Large Sky Area Multi-Object Fiber
Spectroscopic Telescope LAMOST) is a National Major Scientific
Project built by the Chinese Academy of Sciences. Funding for
the project has been provided by the National Development and
Reform Commission. LAMOST is operated and managed by the National
Astronomical Observatories, Chinese Academy of Sciences.

This work has used photometric data from the Xuyi Schmidt telescope survey, 
the SDSS-III survey and the APASS survey. 
Funding for SDSS-III has been provided by the Alfred P. Sloan Foundation, 
the Participating Institutions, the National Science Foundation and the 
U.S. Department of Energy Office of Science. The SDSS-III web site is 
http://www.sdss3.org/. SDSS-III is managed by the Astrophysical Research 
Consortium for the Participating Institutions of the SDSS-III Collaboration 
including the University of Arizona, the Brazilian Participation Group, 
Brookhaven National Laboratory, Carnegie Mellon University, University of Florida, 
the French Participation Group, the German Participation Group, Harvard University, 
the Instituto de Astrofisica de Canarias, the Michigan State/Notre Dame/JINA 
Participation Group, Johns Hopkins University, Lawrence Berkeley National Laboratory, 
Max Planck Institute for Astrophysics, Max Planck Institute for Extraterrestrial Physics, 
New Mexico State University, New York University, Ohio State University, 
Pennsylvania State University, University of Portsmouth, Princeton University, 
the Spanish Participation Group, University of Tokyo, University of Utah, 
Vanderbilt University, University of Virginia, University of Washington and Yale University.
APASS is funded through grants from the Robert Martin Ayers Sciences Fund.}  

\bibliographystyle{aasjournal}
\bibliography{reference2.bib}

\newpage
\appendix

\begin{figure*}
\centering
\includegraphics[width=180mm]{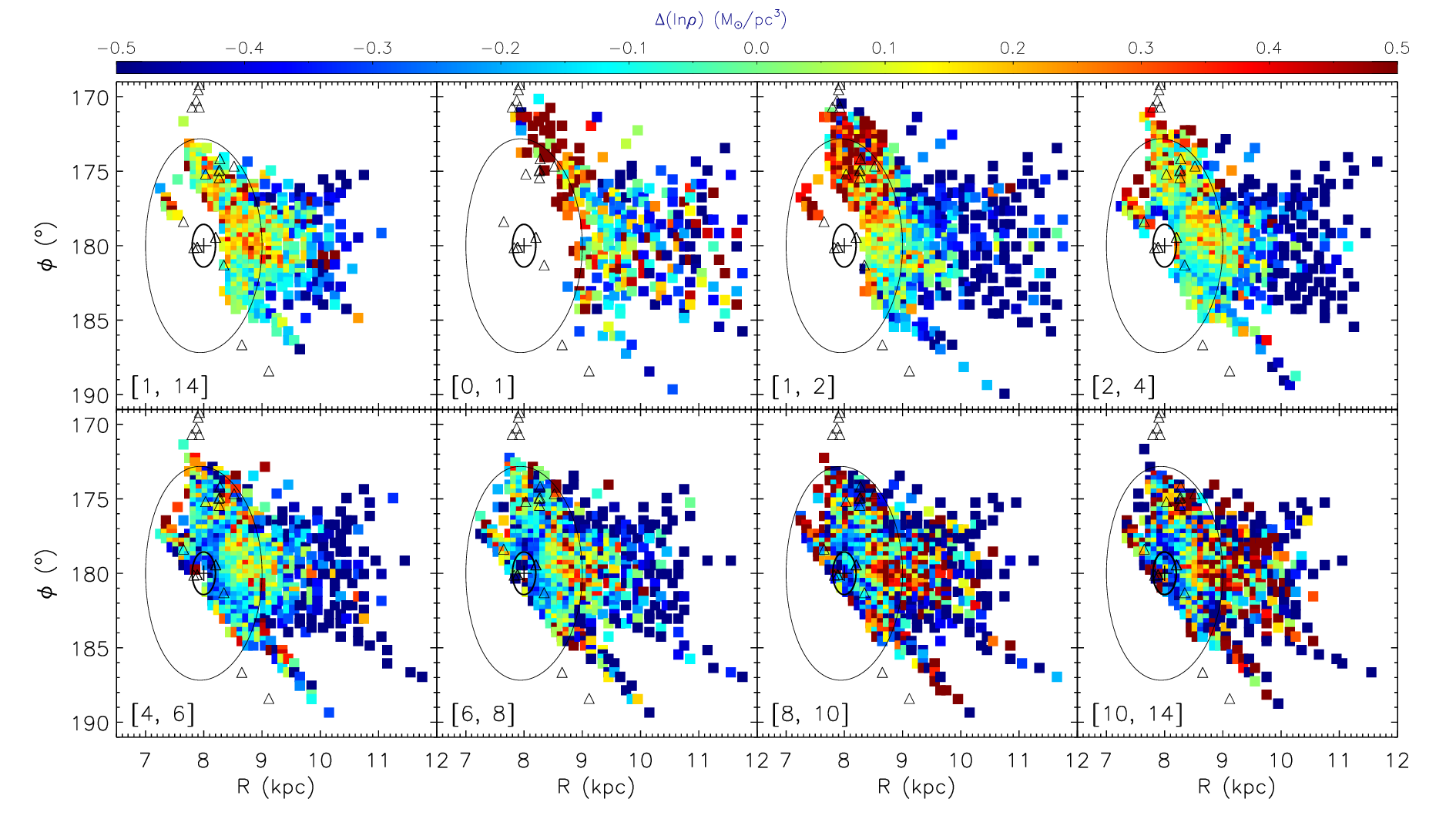}
\caption{Same as Fig.\,17, but derived by subtracting fits with the flared double-component disk model.}
\label{Fig25}
\end{figure*}

\begin{figure*}
\centering
\includegraphics[width=180mm]{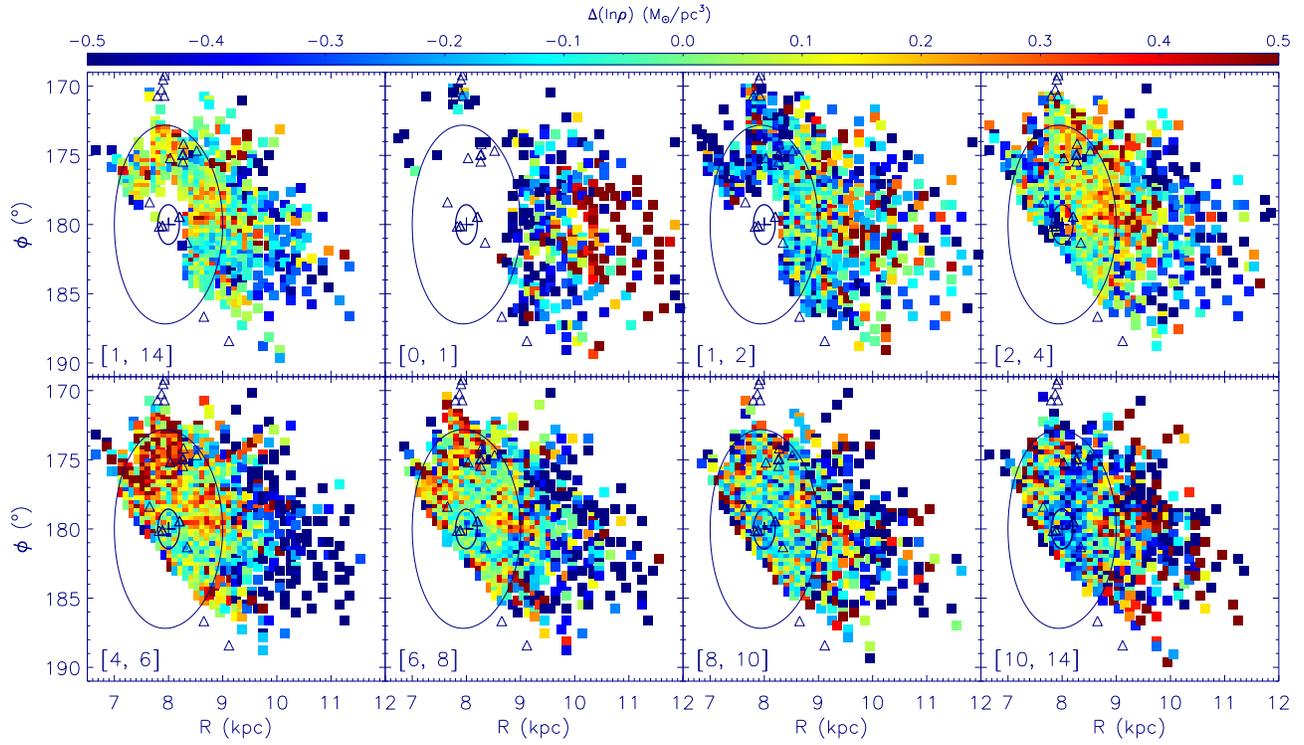}
\caption{Same as Fig.\,17, but for the vertical slice of $0.2<|Z|<0.4$\,kpc.}
\label{Fig26}
\end{figure*}

\begin{table*}
\caption{Fitting the vertical mass distribution with double exponential functions.}
\centering
\begin{tabular}{lllllll}
\hline
$7.8<R<8.2$\,kpc & & &&&& \\
\hline
 Age  &  $\rho_1$   &   $\rho_2$ & $H_1$ (pc) &  $H_2$ (pc)  & $\chi^2_{\rm min}$ & $\Sigma_*$ \\
 \hline
  1-14  &  0.0656$\pm$0.0014  &  0.0055$\pm$0.0006  &   253$\pm$5  &   771$\pm$28  &   1.47  &   41.6$\pm$0.5   \\
   0-1  &  0.0078$\pm$0.0013  &  0.0001$\pm$1.4614e-5  &    81$\pm$8  &   388$\pm$17  &   1.73  &    1.3$\pm$0.2   \\
   1-2  &  0.0122$\pm$0.0007  &  3.7309e-5$\pm$1.2009e-5  &   115$\pm$3  &   599$\pm$69  &   1.71  &    2.8$\pm$0.1  \\
   2-4  &  0.0216$\pm$0.0006  &  9.3425e-5$\pm$2.6440e-5  &   167$\pm$2  &   796$\pm$90  &   1.28  &    7.3$\pm$0.1  \\
   4-6  &  0.0160$\pm$0.0005  &  4.7716e-5$\pm$3.5533e-5  &   250$\pm$5  &  1421$\pm$583  &   1.92  &    8.2$\pm$0.2  \\
   6-8  &  0.0119$\pm$0.0003  &  7.9270e-5$\pm$4.7042e-5  &   349$\pm$7  &  1613$\pm$538  &   1.67  &    8.5$\pm$0.1  \\
  8-10  &  0.0061$\pm$0.0002  &  9.3029e-5$\pm$6.6702e-5  &   485$\pm$15  &  2688$\pm$515  &   1.87  &    6.3$\pm$0.1  \\
 10-14  &  0.0051$\pm$0.0002  &  0.0003$\pm$0.0002  &   506$\pm$28  &  1519$\pm$534  &   1.23  &    6.1$\pm$0.1  \\
 \hline
 \end{tabular}
\begin{tabular}{lllllll}
$8.3<R<8.7$\,kpc & & &&&& \\
 \hline
  1-14  &  0.0497$\pm$0.0008  &  0.0035$\pm$0.0004  &   274$\pm$5  &   855$\pm$39  &   1.67  &   33.2$\pm$0.2  \\
   0-1  &  0.0122$\pm$0.0006  &  2.0921e-5$\pm$8.0822e-6  &    89$\pm$4  &   521$\pm$57  &   0.93  &    2.2$\pm$0.1  \\
   1-2  &  0.0098$\pm$0.0004  &  4.9172e-5$\pm$1.9298e-5  &   119$\pm$3  &   548$\pm$87  &   2.33  &    2.4$\pm$0.1   \\
   2-4  &  0.0162$\pm$0.0003  &  1.9577e-5$\pm$1.0871e-5  &   193$\pm$2  &  1195$\pm$390  &   1.92  &    6.3$\pm$0.1  \\
   4-6  &  0.0127$\pm$0.0003  &  1.4869e-5$\pm$1.7665e-5  &   280$\pm$5  &  2672$\pm$642  &   2.62  &    7.2$\pm$0.1  \\
   6-8  &  0.0094$\pm$0.0002  &  2.8229e-5$\pm$1.4003e-5  &   371$\pm$6  &  2984$\pm$489  &   1.85  &    7.1$\pm$0.1  \\
  8-10  &  0.0047$\pm$0.0002  &  0.0006$\pm$0.0003  &   438$\pm$20  &   947$\pm$154  &   1.38  &    5.2$\pm$0.1  \\
 10-14  &  0.0039$\pm$0.0003  &  0.0018$\pm$0.0003  &   277$\pm$33  &   853$\pm$62  &   1.35  &    5.1$\pm$0.1  \\
 \hline
\end{tabular}
\begin{tabular}{lllllll}
$8.8<R<9.2$\,kpc & & &&&& \\
 \hline
  1-14  &  0.0404$\pm$0.0007  &  0.0014$\pm$0.0004  &   311$\pm$6  &  1106$\pm$143  &   1.66  &   28.1$\pm$0.3  \\
   0-1  &  0.0081$\pm$0.0007  &  1.8327e-5$\pm$8.2969e-6  &    98$\pm$5  &   606$\pm$95  &   2.04  &    1.6$\pm$0.1  \\
   1-2  &  0.0057$\pm$0.0002  &  1.2498e-5$\pm$1.0836e-5  &   146$\pm$3  &   800$\pm$148  &   2.10  &    1.7$\pm$0.1  \\
   2-4  &  0.0133$\pm$0.0005  &  1.0068e-5$\pm$8.8573e-6  &   211$\pm$4  &  1710$\pm$427  &   2.74  &    5.7$\pm$0.1  \\
   4-6  &  0.0108$\pm$0.0007  &  1.0536e-5$\pm$0.0005  &   294$\pm$10  &  2031$\pm$177  &   3.04  &    6.4$\pm$0.2  \\
   6-8  &  0.0076$\pm$0.0002  &  2.6779e-5$\pm$1.2143e-5  &   386$\pm$6  &  3000$\pm$512  &   1.66  &    6.0$\pm$0.1  \\
  8-10  &  0.0040$\pm$0.0002  &  0.0002$\pm$8.2142e-5  &   455$\pm$17  &  1663$\pm$504  &   1.67  &    4.2$\pm$0.1  \\
 10-14  &  0.0041$\pm$0.0002  &  0.0004$\pm$0.0002  &   409$\pm$31  &  1343$\pm$379  &   1.17  &    4.4$\pm$0.1  \\
 \hline
\end{tabular}
\begin{tabular}{lllllll}
$9.3<R<9.7$\,kpc & & &&&& \\
 \hline
 1-14  &  0.0281$\pm$0.0009  &  0.0015$\pm$0.0007  &   347$\pm$12  &   988$\pm$145  &   2.32  &   22.4$\pm$0.3   \\
   0-1  &  0.0045$\pm$0.0002  &  2.7913e-5$\pm$1.0960e-5  &   118$\pm$3  &   478$\pm$63  &   1.08  &    1.1$\pm$0.1 \\
   1-2  &  0.0038$\pm$0.0002  &  1.0589e-5$\pm$1.1534e-5  &   175$\pm$5  &   834$\pm$127  &   1.91  &    1.4$\pm$0.1  \\
   2-4  &  0.0093$\pm$0.0004  &  1.0016e-5$\pm$6.5894e-6  &   245$\pm$6  &  1659$\pm$424  &   2.48  &    4.6$\pm$0.2  \\
   4-6  &  0.0075$\pm$0.0008  &  1.2179e-5$\pm$0.0008  &   335$\pm$112  &  1759$\pm$15  &   2.04  &    5.0$\pm$0.1  \\
   6-8  &  0.0048$\pm$0.0002  &  1.8326e-5$\pm$0.0002  &   452$\pm$15  &  2726$\pm$583  &   1.31  &    4.4$\pm$0.1 \\
  8-10  &  0.0030$\pm$0.0002  &  0.0003$\pm$0.0002  &   465$\pm$32  &  1115$\pm$222  &   1.35  &    3.4$\pm$0.1  \\
 10-14  &  0.0025$\pm$0.0004  &  0.0016$\pm$0.0003  &   266$\pm$49  &   766$\pm$55  &   1.12  &    3.8$\pm$0.1 \\
 \hline
\end{tabular}
 \begin{tabular}{lllllll}
$9.8<R<10.2$\,kpc & & &&&& \\
 \hline
  1-14  &  0.0210$\pm$0.0010  &  0.0010$\pm$0.0004  &   383$\pm$15  &  1078$\pm$310  &   2.43  &   18.3$\pm$0.5  \\
   0-1  &  0.0034$\pm$0.0003  &  0.0001$\pm$2.2632e-5  &   134$\pm$5  &   346$\pm$20  &   1.65  &    1.0$\pm$0.1  \\
   1-2  &  0.0028$\pm$0.0003  &  1.0062e-5$\pm$1.7796e-5  &   204$\pm$9  &   750$\pm$125  &   2.19  &    1.2$\pm$0.1  \\
   2-4  &  0.0059$\pm$0.0022  &  1.0179e-5$\pm$0.0021  &   295$\pm$38  &  1342$\pm$54  &   3.54  &    3.5$\pm$0.2  \\
   4-6  &  0.0047$\pm$0.0016  &  0.0004$\pm$0.0016  &   390$\pm$31  &   398$\pm$41  &   2.15  &    4.0$\pm$0.2  \\
   6-8  &  0.0031$\pm$0.0012  &  0.0004$\pm$0.0012  &   511$\pm$101  &   515$\pm$52  &   1.66  &    3.6$\pm$0.1  \\
  8-10  &  0.0016$\pm$0.0005  &  0.0001$\pm$0.0005  &   651$\pm$70  &   918$\pm$168  &   1.34  &    2.3$\pm$0.1  \\
 10-14  &  0.0021$\pm$0.0004  &  0.0010$\pm$0.0003  &   336$\pm$72  &   917$\pm$98  &   1.04  &    3.1$\pm$0.2  \\
 \hline
\end{tabular}
\begin{tabular}{lllllll}
$10.3<R<10.7$\,kpc & & &&&& \\
 \hline
  1-14  &  0.0174$\pm$0.0014  &  0.0006$\pm$0.0004  &   409$\pm$24  &  1273$\pm$516  &   2.42  &   15.8$\pm$0.7 \\
   0-1  &  0.0028$\pm$0.0002  &  1.5432e-5$\pm$1.4784e-5  &   165$\pm$6  &   580$\pm$82  &   1.41  &    1.0$\pm$0.1  \\
   1-2  &  0.0026$\pm$0.0009  &  1.2106e-5$\pm$0.0008  &   221$\pm$37  &   673$\pm$95  &   2.23  &    1.2$\pm$0.1  \\
   2-4  &  0.0039$\pm$0.0014  &  0.0005$\pm$0.0014  &   330$\pm$37  &   333$\pm$44  &   2.92  &    2.9$\pm$0.2  \\
   4-6  &  0.0034$\pm$0.0014  &  0.0003$\pm$0.0014  &   433$\pm$45  &   435$\pm$68  &   2.12  &    3.2$\pm$0.2  \\
   6-8  &  0.0028$\pm$0.0009  &  1.7382e-5$\pm$0.0009  &   528$\pm$199  &  1828$\pm$77  &   1.62  &    3.0$\pm$0.2  \\
  8-10  &  0.0007$\pm$0.0003  &  0.0008$\pm$0.0011  &   512$\pm$58  &   790$\pm$369  &   1.55  &    2.0$\pm$0.3  \\
 10-14  &  0.0018$\pm$0.0005  &  4.8729e-5$\pm$0.0004  &   600$\pm$198  &  2822$\pm$363  &   1.26  &    2.4$\pm$0.2   \\
  \hline
\end{tabular}
\end{table*}

\label{lastpage}

\end{document}